\documentclass[12pt]{article}

\usepackage{axodraw}
\usepackage{cite}

\makeatletter

\def\paragraph{\@startsection{paragraph}{4}{\z@}{+2.00ex plus
 +1ex minus +.2ex}{1.5ex plus .2ex}{\it\normalsize}}

\def\section{\@startsection {section}{1}{\z@}{+3.0ex plus +1ex minus
  +.2ex}{2.3ex plus .2ex}{\normalsize\bf\boldmath}}
\def\subsection{\@startsection{subsection}{2}{\z@}{+2.5ex plus +1ex
minus +.2ex}{1.5ex plus .2ex}{\normalsize\bf\boldmath}}
\def\subsubsection{\@startsection{subsubsection}{3}{\z@}{+2.5ex plus
 +1ex minus +.2ex}{1.5ex plus .2ex}{\normalsize\it}}

\expandafter\ifx\csname mathrm\endcsname\relax\def\mathrm#1{{\rm #1}}\fi

\renewcommand{\theequation}{\thesection.\arabic{equation}}
\newcounter{saveeqn}

\@addtoreset{equation}{section}

\newcommand{\ks}{k\hspace{-0.52em}/\hspace{0.1em}}
\newcommand{\lslash}{l\hspace{-0.42em}/\hspace{0.1em}}
\newcommand{\ps}{p\hspace{-0.42em}/}%\hspace{0.1em}}
\newcommand{\qs}{q\hspace{-0.48em}/}%\hspace{0.1em}}
\newcommand{\rs}{r\hspace{-0.42em}/}%\hspace{0.1em}}
\def\bit{\begin{itemize}}
\def\eit{\end{itemize}}
\def\beq{\begin{equation}}
\def\eeq{\end{equation}}
\def\beqar{\begin{eqnarray}}
\def\eeqar{\end{eqnarray}}
\def\barr#1{\begin{array}{#1}}
\def\earr{\end{array}}
\def\bfi{\begin{figure}}
\def\efi{\end{figure}}
\def\btab{\begin{table}}
\def\etab{\end{table}}
\def\bce{\begin{center}}
\def\ece{\end{center}}
\def\nn{\nonumber}

\def\text{\textstyle}

\def\be{\beta}
\def\ga{\gamma}
\def\de{\delta}
\def\teps{\varepsilon}
\def\veps{\epsilon}
\def\la{\lambda}
\def\si{\sigma}
\def\om{\omega}

\def\De{\Delta}

\def\refeq#1{\mbox{(\ref{#1})}}

\def\refse#1{\mbox{Sect.~\ref{#1}}}
\def\refses#1{\mbox{Sects.~\ref{#1}}}
\def\refapp#1{\mbox{App.~\ref{#1}}}
\def\refapps#1{\mbox{Apps.~\ref{#1}}}
\def\citere#1{\mbox{Ref.~\cite{#1}}}
\def\citeres#1{\mbox{Refs.~\cite{#1}}}

\def\solid{\raise.9mm\hbox{\protect\rule{1.1cm}{.2mm}}}
\def\dash{\raise.9mm\hbox{\protect\rule{2mm}{.2mm}}\hspace*{1mm}}

\def\nl{\nonumber\\}

\def\nln{\nonumber\\*[-1ex]\phantom{\fbox{\rule{0em}{2ex}}}}

\newcommand{\dsl}[1]{\not \hspace{-0.7mm}#1}
\def\dsl{\mathpalette\make@slash}
\def\make@slash#1#2{\setbox\z@\hbox{$#1#2$}%
  \hbox to 0pt{\hss$#1/$\hss\kern-\wd0}\box0}

\newcommand{\order}{\mathcal{O}}

\newcommand{\lsim}
{\mathrel{\raisebox{-.3em}{$\stackrel{\displaystyle <}{\sim}$}}}
\newcommand{\gsim}
{\mathrel{\raisebox{-.3em}{$\stackrel{\displaystyle >}{\sim}$}}}
\def\asymp#1%
{\mathrel{\raisebox{-.4em}{$\widetilde{\scriptstyle #1}$}}}

\def\Nequal#1%
{\mathrel{\raisebox{-.5em}{$\stackrel{=}{\scriptstyle\rm#1}$}}}

\newcommand{\NLLA}{\mathrel{\stackrel{\mathrm{NLL}}{=}}}

\def\sgn{\mathop{\mathrm{sgn}}\nolimits}

\newcommand{\Tr}{\mathop{\mathrm{Tr}}\nolimits}

\newcommand{\GeV}{\unskip\,\mathrm{GeV}}

\newcommand{\TeV}{\unskip\,\mathrm{TeV}}

\def\mathswitchr#1{\relax\ifmmode{\mathrm{#1}}\else$\mathrm{#1}$\fi}

\newcommand{\PW}{\mathswitchr W}
\newcommand{\PZ}{\mathswitchr Z}

\newcommand{\PA}{\mathswitchr A}
\newcommand{\Pg}{\mathswitchr g}
\newcommand{\PH}{\mathswitchr H}

\newcommand{\Pb}{\mathswitchr b}

\newcommand{\Pt}{\mathswitchr t}

\newcommand{\Pep}{\mathswitchr {e^+}}
\newcommand{\Pem}{\mathswitchr {e^-}}

\newcommand{\PWpm}{\mathswitchr {W^\pm}}

\newcommand{\FA}{A}
\newcommand{\FZ}{Z}
\newcommand{\FW}{W}
\newcommand{\FWpm}{{W^\pm}}

\def\mathswitch#1{\relax\ifmmode#1\else$#1$\fi}

\newcommand{\MW}{\mathswitch {M_\PW}}

\newcommand{\MA}{\mathswitch {M_A}}
\newcommand{\MZ}{\mathswitch {M_\PZ}}
\newcommand{\MH}{\mathswitch {M_\PH}}

\newcommand{\Mt}{\mathswitch {m_\Pt}}

\newcommand{\scrs}{\scriptscriptstyle}
\newcommand{\sw}{\mathswitch {s_{\scrs\PW}}}
\newcommand{\cw}{\mathswitch {c_{\scrs\PW}}}

\newcommand{\muD}{\mu_{\mathrm{D}}}
\newcommand{\muR}{\mu_\mathrm{R}}

\newcommand{\betacoeff}[1]{b_{#1}}

\newcommand{\vev}{v} %{{\bf v}}

\def\ie{i.e.\ }
\def\eg{e.g.\ }
\def\cf{cf.\ }
\newcommand{\etal}{{\it et al.}}

\newcommand{\rd}{\mathrm{d}}

\newcommand{\ri}{\mathrm{i}}

\newcommand{\rF}{\mathrm{F}}
\newcommand{\rL}{\mathrm{L}}
\newcommand{\rR}{\mathrm{R}}

\newcommand{\SU}{\mathrm{SU}}
\newcommand{\U}{\mathrm{U}}

\newcommand{\fact}{\mathrm{F}}

\newcommand{\elm}{\mathrm{em}}
\newcommand{\sew}{\mathrm{sew}}
\newcommand{\QED}{{\mathrm{QED}}}
\newcommand{\M}{{\cal {M}}}
\newcommand{\univfact}[2]{I(#1,#2)}
\newcommand{\FF}[2]{F_{#1}^{#2}}
\newcommand{\GG}[2]{G_{#1}^{#2}}
\newcommand{\Ff}[2]{f_{#1}^{#2}}
\newcommand{\Gg}[2]{g_{#1}^{#2}}
\newcommand{\DD}[1]{D_{#1}}
\newcommand{\deDD}[1]{\Delta D_{#1}}
\let \DDsub \DD 
\let \deDDsub \deDD
\newcommand{\DDUV}[1]{\DD{#1}^{\mathrm{UV}}}
\newcommand{\gb}{g_1}
\newcommand{\gw}{g_2}

\newcommand{\mel}[2]{\M_{#1}^{#2}}
\newcommand{\melQ}[2]{\M_{#1}^{#2}(Q^2)}
\newcommand{\nmel}[2]{\tilde{\M}_{#1}^{#2}}
\newcommand{\alphaeps}{\alpha_{\veps}}

\newcommand{\eqdiagl}{&=&\hspace{-3mm}}

\newcommand{\lr}[1]{l_{#1}} 
\newcommand{\LMZW}{\lr{\mathrm{Z}}}
\newcommand{\LmuR}{\lr{\muR}}
\newcommand{\LrMI}{\lr{1}}
\newcommand{\LrMII}{\lr{2}}
\newcommand{\LrMIII}{\lr{3}}
\newcommand{\Lrij}{\lr{ij}}
\newcommand{\Lrik}{\lr{ik}}

\newcommand{\Yuk}{{\mathrm{Y}}}
\newcommand{\zYuk}{z^{\Yuk}}

\newcommand{\Lt}{\lr{\Pt}}
\newcommand{\Lrmi}{\lr{i}}
\newcommand{\Lrmj}{\lr{j}}
\newcommand{\Lrmk}{\lr{k}}
\newcommand{\LYuk}{L_\Pt}

\newcommand{\deltat}[1]{\delta_{#1,\Pt}}
\newcommand{\deltaz}[1]{\delta_{#1,0}}

\newcommand{\lambdat}{\la_{t}}%{\la_{\Pt}}

\newcommand{\MSbar}{{\overline{\mathrm{MS}}}}

\def\draftdate{\relax}
\def\mda{\relax}
\def\mua{\relax}
\def\mla{\relax}
\def\draft{
\def\thtystars{******************************}
\def\sixtystars{\thtystars\thtystars}
\typeout{}
\typeout{\sixtystars**}
\typeout{* Draft mode!
         For final version remove \protect\draft\space in source file *}
\typeout{\sixtystars**}
\typeout{}
\def\draftdate{\today}
\def\mua{\marginpar[\boldmath\hfil$\uparrow$]%
                   {\boldmath$\uparrow$\hfil}%
                    \typeout{marginpar: $\uparrow$}\ignorespaces}
\def\mda{\marginpar[\boldmath\hfil$\downarrow$]%
                   {\boldmath$\downarrow$\hfil}%
                    \typeout{marginpar: $\downarrow$}\ignorespaces}
\def\mla{\marginpar[\boldmath\hfil$\rightarrow$]%
                   {\boldmath$\leftarrow $\hfil}%
                    \typeout{marginpar: $\leftrightarrow$}\ignorespaces}
\def\Mua{\marginpar[\boldmath\hfil$\Uparrow$]%
                   {\boldmath$\Uparrow$\hfil}%
                    \typeout{marginpar: $\Uparrow$}\ignorespaces}
\def\Mda{\marginpar[\boldmath\hfil$\Downarrow$]%
                   {\boldmath$\Downarrow$\hfil}%
                    \typeout{marginpar: $\Downarrow$}\ignorespaces}
\def\Mla{\marginpar[\boldmath\hfil$\Rightarrow$]%
                   {\boldmath$\Leftarrow $\hfil}%
                    \typeout{marginpar: $\Leftrightarrow$}\ignorespaces}

\overfullrule 5pt
\oddsidemargin -15mm
\marginparwidth 29mm
}

\makeatletter

\def\eqnarray{\stepcounter{equation}\let\@currentlabel=\theequation
\global\@eqnswtrue
\global\@eqcnt\z@\tabskip\@centering\let\\=\@eqncr
$$\halign to \displaywidth\bgroup\hskip\@centering
  $\displaystyle\tabskip\z@{##}$\@eqnsel&\global\@eqcnt\@ne
  \hskip 2\arraycolsep \hfil${##}$\hfil
  &\global\@eqcnt\tw@ \hskip 2\arraycolsep $\displaystyle\tabskip\z@{##}$\hfil
   \tabskip\@centering&\llap{##}\tabskip\z@\cr}

\makeatother

\begin{document}

\newcommand{\leg}[1]{\scriptstyle{#1}}

%%%%%%%%%%%%%%%%%%%%%%  Feynman diagrams %%%%%%%%%%%%%%%%%%%%%%
\newcommand{\wblob}{
\Vertex(-15.9138,3.75675){0.8}
\Vertex(-16.3512,-0.00758122){0.8}
\Vertex(-15.9138,-3.75675){0.8}
\Line(0.,0.)(-19.5,-9.75)
\Line(0.,0.)(-19.5,9.75)
\GCirc(0.,0.){10.9008}{1}
}
\newcommand{\blob}{
\Vertex(-15.9138,3.75675){0.8}
\Vertex(-16.3512,-0.00758122){0.8}
\Vertex(-15.9138,-3.75675){0.8}
\Line(0.,0.)(-19.5,-9.75)
\Line(0.,0.)(-19.5,9.75)
\GCirc(0.,0.){10.9008}{0.5}
}
\newcommand{\factblob}{\wblob \Text(0,0)[]{\scriptsize F}}
\newcommand{\nfactblob}{\wblob \Text(0,0)[]{\scriptsize N}}

\newcommand{\diaggeneric}{
\begin{picture}(120.,104.)(-28.,-52.)
\Gluon(-53.6656,0.)(0.,0.){3.2}{5}
\Line(0.,0.)(24.,12.)
\Line(24.,12.)(48.,24.)
\Line(24.,-12.)(0.,0.)
\Line(48.,-24.)(24.,-12.)
\GCirc(4.47214,0.){17}{0.8}
%external lines
\Text(84.,42.)[l]{$\leg{i}$}
\Text(84.,-42.)[l]{$\leg{j}$}
%gauge bosons
%inflowing lines
\Text(27.668,14.6453)[br]{}
\Text(28.,-14.)[tr]{}
\end{picture}
}

\newcommand{\diagone}[4]{
\begin{picture}(120.,104.)(-28.,-52.)
\Line(0.,0.)(48.,24.)
\Line(48.,24.)(80.,40.)
\Line(48.,-24.)(0.,0.)
\Line(80.,-40.)(48.,-24.)
\Photon(48.,24.)(48.,-24.){2.4}{3.5}
\Vertex(48.,24.){2}
\Vertex(48.,-24.){2}
%external lines
\Text(84.,42.)[l]{#1}
\Text(84.,-42.)[l]{#2}
%gauge bosons
\Text(53.6656,0.)[l]{#3}
%inflowing lines
\Text(27.668,14.6453)[br]{}
\Text(28.,-14.)[tr]{}
{#4}
\end{picture}
}

\newcommand{\diagnewb}[1]{
\begin{picture}(120.,104.)(-28.,-52.)
\Line(0.,0.)(30.,15.)
\Line(30.,-15.)(0.,0.)
{#1}
\end{picture}
}

\newcommand{\diagnewc}[1]{
\begin{picture}(120.,104.)(-28.,-52.)
\Line(0.,0.)(48.,24.)
\Line(48.,24.)(80.,40.)
\Line(48.,-24.)(0.,0.)
\Line(80.,-40.)(48.,-24.)
\Photon(48.,24.)(48.,-24.){2.4}{3.5}
\Vertex(48.,24.){2}
\Vertex(48.,-24.){2}
%external lines
\Text(84.,42.)[l]{$\leg{i}$}
\Text(84.,-42.)[l]{$\leg{j}$}
%gauge bosons
\Text(53.6656,0.)[l]{$\scriptscriptstyle{V_1}$}
%inflowing lines
\Text(27.668,14.6453)[br]{}
\Text(28.,-14.)[tr]{}
{#1}
\end{picture}
}
\newcommand{\diagnewd}[1]{
\begin{picture}(120.,104.)(-28.,-52.)
\Line(0.,0.)(80,0)
\PhotonArc(25.,-8.)(26.2488,17.7447,162.255){2}{5}
\Vertex(50.,0.){2}
\Text(84.,0)[l]{$\leg{i}$}
%gauge bosons
\Text(23,28.)[l]{{$\scriptstyle{V_\mu}$}}
%inflowing lines
\Text(27.668,14.6453)[br]{}
\Text(28.,-14.)[tr]{}
{#1}
\end{picture}
}

\newcommand{\diagnewe}[1]{
\begin{picture}(120.,104.)(-28.,-52.)
\Line(0.,0.)(80,0)
\Line(0.,0.)(50,50)
\PhotonArc(0,0)(48,0,45){2}{4}
\Vertex(48.,0.){2}
\Vertex(33.941,33.941){2}
\Text(84.,0)[l]{$\leg{i}$}
\Text(53.,53.)[l]{$\leg{j}$}
%gauge bosons
\Text(50,22.)[l]{{$\scriptstyle{V_\mu}$}}
%inflowing lines
\Text(27.668,14.6453)[br]{}
\Text(28.,-14.)[tr]{}
{#1}
\end{picture}
}

\newcommand{\diagoned}[1]{
\begin{picture}(120.,104.)(-28.,-52.)
\Line(0.,0.)(50,0)
\PhotonArc(50.,-50.)(70.71,90.,135.){2}{5}
\Text(54.,0)[l]{$\leg{i}$}
%gauge bosons
\Text(54,22.)[l]{{$\scriptstyle{V_\mu}$}}
%inflowing lines
\Text(27.668,14.6453)[br]{}
\Text(28.,-14.)[tr]{}
{#1}
\end{picture}
}

\newcommand{\diagonee}[1]{
\begin{picture}(120.,104.)(-28.,-52.)
\Line(0.,0.)(60,0)
\Line(0.,0.)(50,50)
\Vertex(25,25){2}
\Photon(25,25)(60,25){2}{3}
\Text(64.,0)[l]{$\leg{i}$}
\Text(53.,53.)[l]{$\leg{j}$}
%gauge bosons
\Text(64,22.)[l]{{$\scriptstyle{V_\mu}$}}
%inflowing lines
\Text(27.668,14.6453)[br]{}
\Text(28.,-14.)[tr]{}
{#1}
\end{picture}
}

\newcommand{\diagI}[5]{
\begin{picture}(120.,104.)(-28.,-52.)
\Line(0.,0.)(28.,14.)
\Line(28.,14.)(56.,28.)
\Line(56.,28.)(80.,40.)
\Line(28.,-14.)(0.,0.)
\Line(56.,-28.)(28.,-14.)
\Line(80.,-40.)(56.,-28.)
\Photon(56.,28.)(56.,-28.){2.4}{3.5}
\Photon(28.,14.)(28.,-14.){2.4}{2}
\Vertex(56.,28.){2}
\Vertex(56.,-28.){2}
\Vertex(28.,14.){2}
\Vertex(28.,-14.){2}
%external lines
\Text(84.,42.)[l]{#1}
\Text(84.,-42.)[l]{#2}
%gauge bosons
\Text(62.6099,0.)[l]{#3}
\Text(31.305,0.)[l]{#4}
%inflowing lines
\Text(41.5019,21.9679)[br]{}
\Text(42.,-21.)[tr]{}
\Text(19.515,10.9162)[br]{}
\Text(20.,-10.)[tr]{}
{#5}
\end{picture}
}

\newcommand{\diagII}[5]{
\begin{picture}(120.,104.)(-28.,-52.)
\Line(0.,0.)(28.,14.)
\Line(28.,14.)(56.,28.)
\Line(56.,28.)(80.,40.)
\Line(28.,-14.)(0.,0.)
\Line(56.,-28.)(28.,-14.)
\Line(80.,-40.)(56.,-28.)
\Photon(56.,28.)(28.,-14.){-2.4}{3}
\Photon(56.,-28.)(28.,14.){-2.4}{3}
\Vertex(56.,28.){2}
\Vertex(56.,-28.){2}
\Vertex(28.,14.){2}
\Vertex(28.,-14.){2}
%external lines
\Text(84.,42.)[l]{#1}
\Text(84.,-42.)[l]{#2}
%gauge bosons
\Text(48.3499,8.07117)[l]{#3}
\Text(48.3499,-9.07117)[l]{#4}
%inflowing lines
\Text(41.5019,21.9679)[br]{}
\Text(42.,-21.)[tr]{}
\Text(19.515,10.9162)[br]{}
\Text(20.,-10.)[tr]{}
{#5}
\end{picture}
}

\newcommand{\diagIII}[6]{
\begin{picture}(120.,104.)(-28.,-52.)
\Line(0.,0.)(24.,12.)
\Line(24.,12.)(56.,28.)
\Line(56.,28.)(80.,40.)
\Line(49.1935,-24.5967)(0.,0.)
\Line(80.,-40.)(49.1935,-24.5967)
\Photon(56.,28.)(49.1935,0.){2.4}{2}
\Photon(24.,12.)(49.1935,0.){-2.4}{2}
\Photon(49.1935,0.)(49.1935,-24.5967){2.4}{2}
\Vertex(56.,28.){2}
\Vertex(24.,12.){2}
\Vertex(49.1935,0.){2}
\Vertex(49.1935,-24.5967){2}
%external lines
\Text(84.,42.)[l]{#1}
\Text(84.,-42.)[l]{#2}
%gauge bosons
\Text(56.5825,13.3573)[l]{#3}
\Text(56.5825,-13.3573)[l]{#4}
\Text(35.307,5.78034)[tr]{#5}
%inflowing lines
\Text(41.5019,21.9679)[br]{}
\Text(32.,-16.)[tr]{}
\Text(19.515,10.9162)[br]{}
{#6}
\end{picture}
}

\newcommand{\diagIIIYuka}[6]{
\begin{picture}(120.,104.)(-28.,-52.)
\Line(0.,0.)(24.,12.)
\Line(24.,12.)(56.,28.)
\Line(56.,28.)(80.,40.)
\Line(49.1935,-24.5967)(0.,0.)
\Line(80.,-40.)(49.1935,-24.5967)
%\Photon(56.,28.)(49.1935,0.){2.4}{2}%1
\DashLine(56.,28.)(49.1935,0.){3}%1
%\Photon(24.,12.)(49.1935,0.){-2.4}{2}%2
\DashLine(24.,12.)(49.1935,0.){3}%2
\Photon(49.1935,0.)(49.1935,-24.5967){2.4}{2}%3
%\DashLine(49.1935,0.)(49.1935,-24.5967){4}%3
\Vertex(56.,28.){2}
\Vertex(24.,12.){2}
\Vertex(49.1935,0.){2}
\Vertex(49.1935,-24.5967){2}
%external lines
\Text(84.,42.)[l]{#1}
\Text(84.,-42.)[l]{#2}
%gauge bosons
\Text(56.5825,13.3573)[l]{#3}
\Text(56.5825,-13.3573)[l]{#4}
\Text(35.307,5.78034)[tr]{#5}
%inflowing lines
\Text(41.5019,21.9679)[br]{}
\Text(32.,-16.)[tr]{}
\Text(19.515,10.9162)[br]{}
{#6}
\end{picture}
}

\newcommand{\diagIIIYukb}[6]{
\begin{picture}(120.,104.)(-28.,-52.)
\Line(0.,0.)(24.,12.)
\Line(24.,12.)(56.,28.)
\Line(56.,28.)(80.,40.)
\Line(49.1935,-24.5967)(0.,0.)
\Line(80.,-40.)(49.1935,-24.5967)
\Photon(56.,28.)(49.1935,0.){2.4}{2}%1
%\DashLine(56.,28.)(49.1935,0.){3}%1
%\Photon(24.,12.)(49.1935,0.){-2.4}{2}%2
\DashLine(24.,12.)(49.1935,0.){3}%2
\Photon(49.1935,0.)(49.1935,-24.5967){2.4}{2}%3
%\DashLine(49.1935,0.)(49.1935,-24.5967){4}%3
\Vertex(56.,28.){2}
\Vertex(24.,12.){2}
\Vertex(49.1935,0.){2}
\Vertex(49.1935,-24.5967){2}
%external lines
\Text(84.,42.)[l]{#1}
\Text(84.,-42.)[l]{#2}
%gauge bosons
\Text(56.5825,13.3573)[l]{#3}
\Text(56.5825,-13.3573)[l]{#4}
\Text(35.307,5.78034)[tr]{#5}
%inflowing lines
\Text(41.5019,21.9679)[br]{}
\Text(32.,-16.)[tr]{}
\Text(19.515,10.9162)[br]{}
{#6}
\end{picture}
}

\newcommand{\diagIIIYukc}[6]{
\begin{picture}(120.,104.)(-28.,-52.)
\Line(0.,0.)(24.,12.)
\Line(24.,12.)(56.,28.)
\Line(56.,28.)(80.,40.)
\Line(49.1935,-24.5967)(0.,0.)
\Line(80.,-40.)(49.1935,-24.5967)
%\Photon(56.,28.)(49.1935,0.){2.4}{2}%1
\DashLine(56.,28.)(49.1935,0.){3}%1
\Photon(24.,12.)(49.1935,0.){-2.4}{2}%2
%\DashLine(24.,12.)(49.1935,0.){3}%2
\Photon(49.1935,0.)(49.1935,-24.5967){2.4}{2}%3
%\DashLine(49.1935,0.)(49.1935,-24.5967){4}%3
\Vertex(56.,28.){2}
\Vertex(24.,12.){2}
\Vertex(49.1935,0.){2}
\Vertex(49.1935,-24.5967){2}
%external lines
\Text(84.,42.)[l]{#1}
\Text(84.,-42.)[l]{#2}
%gauge bosons
\Text(56.5825,13.3573)[l]{#3}
\Text(56.5825,-13.3573)[l]{#4}
\Text(35.307,5.78034)[tr]{#5}
%inflowing lines
\Text(41.5019,21.9679)[br]{}
\Text(32.,-16.)[tr]{}
\Text(19.515,10.9162)[br]{}
{#6}
\end{picture}
}

\newcommand{\diagV}[5]{
\begin{picture}(120.,104.)(-28.,-52.)
\Line(0.,0.)(16.,8.)
\Line(16.,8.)(44.,22.)
\Line(44.,22.)(64.,32.)
\Line(64.,32.)(80.,40.)
\Line(64.,-32.)(0.,0.)
\Line(80.,-40.)(64.,-32.)
\Photon(64.,32.)(64.,-32.){2.4}{3.5}
\PhotonArc(30.,15.)(15.6525,26.5651,206.565){2}{4}
\Vertex(64.,-32.){2}
\Vertex(16.,8.){2}
\Vertex(44.,22.){2}
\Vertex(64.,32.){2}
%external lines
\Text(84.,42.)[l]{#1}
\Text(84.,-42.)[l]{#2}
%gauge bosons
\Text(71.5542,0.)[l]{#3}
\Text(23.8885,35.1332)[br]{#4}
%inflowing lines
\Text(41.5019,21.9679)[br]{}
\Text(42.,-21.)[tr]{}
\Text(19.515,10.9162)[br]{}
\Text(20.,-10.)[tr]{}
{#5}
\end{picture}
}

\newcommand{\diagVYuk}[5]{
\begin{picture}(120.,104.)(-28.,-52.)
\Line(0.,0.)(16.,8.)
\Line(16.,8.)(44.,22.)
\Line(44.,22.)(64.,32.)
\Line(64.,32.)(80.,40.)
\Line(64.,-32.)(0.,0.)
\Line(80.,-40.)(64.,-32.)
\Photon(64.,32.)(64.,-32.){2.4}{3.5}
%\PhotonArc(30.,15.)(15.6525,26.5651,206.565){2}{4}
\DashCArc(30.,15.)(15.6525,26.5651,206.565){3}
\Vertex(64.,-32.){2}
\Vertex(16.,8.){2}
\Vertex(44.,22.){2}
\Vertex(64.,32.){2}
%external lines
\Text(84.,42.)[l]{#1}
\Text(84.,-42.)[l]{#2}
%gauge bosons
\Text(71.5542,0.)[l]{#3}
\Text(23.8885,35.1332)[br]{#4}
%inflowing lines
\Text(41.5019,21.9679)[br]{}
\Text(42.,-21.)[tr]{}
\Text(19.515,10.9162)[br]{}
\Text(20.,-10.)[tr]{}
{#5}
\end{picture}
}

\newcommand{\diagVII}[5]{
\begin{picture}(120.,104.)(-28.,-52.)
\Line(0.,0.)(24.,12.)
\Line(24.,12.)(40.,20.)
\Line(40.,20.)(56.,28.)
\Line(56.,28.)(80.,40.)
\Line(40.,-20.)(0.,0.)
\Line(80.,-40.)(40.,-20.)
\Photon(40.,20.)(40.,-20.){2.4}{3.5}
\PhotonArc(40.,20.)(17.8885,26.5651,206.565){2}{4}
\Vertex(40.,-20.){2}
\Vertex(24.,12.){2}
\Vertex(40.,20.){2}
\Vertex(56.,28.){2}
%external lines
\Text(84.,42.)[l]{#1}
\Text(84.,-42.)[l]{#2}
%gauge bosons
\Text(49.1935,0.)[l]{#3}
\Text(32.1994,42.9325)[br]{#4}
%inflowing lines
\Text(41.5019,21.9679)[br]{}
\Text(42.,-21.)[tr]{}
\Text(19.515,10.9162)[br]{}
\Text(20.,-10.)[tr]{}
{#5}
\end{picture}
}

\newcommand{\diagVIIYuk}[5]{
\begin{picture}(120.,104.)(-28.,-52.)
\Line(0.,0.)(24.,12.)
\Line(24.,12.)(40.,20.)
\Line(40.,20.)(56.,28.)
\Line(56.,28.)(80.,40.)
\Line(40.,-20.)(0.,0.)
\Line(80.,-40.)(40.,-20.)
\Photon(40.,20.)(40.,-20.){2.4}{3.5}
%\PhotonArc(40.,20.)(17.8885,26.5651,206.565){2}{4}
\DashCArc(40.,20.)(17.8885,26.5651,206.565){3}
\Vertex(40.,-20.){2}
\Vertex(24.,12.){2}
\Vertex(40.,20.){2}
\Vertex(56.,28.){2}
%external lines
\Text(84.,42.)[l]{#1}
\Text(84.,-42.)[l]{#2}
%gauge bosons
\Text(49.1935,0.)[l]{#3}
\Text(32.1994,42.9325)[br]{#4}
%inflowing lines
\Text(41.5019,21.9679)[br]{}
\Text(42.,-21.)[tr]{}
\Text(19.515,10.9162)[br]{}
\Text(20.,-10.)[tr]{}
{#5}
\end{picture}
}

\newcommand{\diagIX}[7]{
\begin{picture}(120.,104.)(-28.,-52.)
\Line(0.,0.)(56.,28.)
\Line(56.,28.)(80.,40.)
\Line(56.,-28.)(0.,0.)
\Line(80.,-40.)(56.,-28.)
\Photon(56.,28.)(56.,10.5064){2.4}{2.5}
\Photon(56.,-10.5064)(56.,-28.){2.4}{2.5}
\ArrowArc(56.9771,0.)(11.7466,-90.,90.)
\ArrowArc(56.9771,0.)(11.7466,90.,270.)
\Vertex(56.,-28.){2}
\Vertex(56.,28.){2}
\Vertex(56.,10.5064){2}
\Vertex(56.,-10.5064){2}
%external lines
\Text(84.,42.)[l]{#1}
\Text(84.,-42.)[l]{#2}
%gauge bosons
\Text(64.4276,21.6767)[lu]{#3}
\Text(63.3537,-24.6395)[lb]{#4}
\Text(74.2375,0.)[l]{#5}
\Text(41.1437,0.)[r]{#6}
%inflowing lines
\Text(41.5019,21.9679)[br]{}
\Text(42.,-21.)[tr]{}
\Text(19.515,10.9162)[br]{}
\Text(20.,-10.)[tr]{}
{#7}
\end{picture}
}

\newcommand{\diagself}[3]{
\begin{picture}(120.,104.)(-28.,-52.)
\Line(0.,0.)(56.,28.)
\Line(56.,28.)(80.,40.)
\Line(56.,-28.)(0.,0.)
\Line(80.,-40.)(56.,-28.)
\Photon(56.,28.)(56.,9){2.4}{2.5}
\Photon(56.,-9)(56.,-28.){2.4}{2.5}
\Vertex(56.,-28.){2}
\Vertex(56.,28.){2}
%external lines
\Text(84.,42.)[l]{$\leg{i}$}
\Text(84.,-42.)[l]{$\leg{j}$}
%gauge bosons
\Text(64.4276,21.6767)[lu]{#1}
\Text(63.3537,-24.6395)[lb]{#2}
%inflowing lines
\Text(41.5019,21.9679)[br]{}
\Text(42.,-21.)[tr]{}
\Text(19.515,10.9162)[br]{}
\Text(20.,-10.)[tr]{}
\GCirc(56,0.){9}{0.9}
{#3}
\end{picture}
}

\newcommand{\diagX}[1]{
\begin{picture}(120.,104.)(-28.,-52.)
\Line(0.,0.)(56.,28.)
\Line(56.,28.)(80.,40.)
\Line(56.,-28.)(0.,0.)
\Line(80.,-40.)(56.,-28.)
\Photon(56.,28.)(56.,10.5064){2.4}{2.5}
\Photon(56.,-10.5064)(56.,-28.){2.4}{2.5}
\PhotonArc(56.9771,0.)(11.7466,-101.459,258.541){-2}{8}
\Vertex(56.,-28.){2}
\Vertex(56.,28.){2}
\Vertex(56.,10.5064){2}
\Vertex(56.,-10.5064){2}
%external lines
\Text(84.,42.)[l]{$\leg{i}$}
\Text(84.,-42.)[l]{$\leg{j}$}
%gauge bosons
\Text(64.4276,21.6767)[lu]{$\scriptscriptstyle{V_1}$}
\Text(76.0263,0.)[l]{$\scriptscriptstyle{V_2}$}
\Text(38.4604,0.)[r]{$\scriptscriptstyle{V_3}$}
\Text(63.3537,-24.6395)[lb]{$\scriptscriptstyle{V_4}$}
%inflowing lines
\Text(41.5019,21.9679)[br]{}
\Text(42.,-21.)[tr]{}
\Text(19.515,10.9162)[br]{}
\Text(20.,-10.)[tr]{}
{#1}
\end{picture}
}

\newcommand{\diagXI}[1]{
\begin{picture}(120.,104.)(-28.,-52.)
\Line(0.,0.)(56.,28.)
\Line(56.,28.)(80.,40.)
\Line(56.,-28.)(0.,0.)
\Line(80.,-40.)(56.,-28.)
\Photon(56.,28.)(56.,10.5064){2.4}{2.5}
\Photon(56.,-10.5064)(56.,-28.){2.4}{2.5}
\DashArrowArcn(56.9771,0.)(11.7466,90.,-90.){1}
\DashArrowArcn(56.9771,0.)(11.7466,270.,90.){1}
\Vertex(56.,-28.){2}
\Vertex(56.,28.){2}
\Vertex(56.,10.5064){2}
\Vertex(56.,-10.5064){2}
%external lines
\Text(84.,42.)[l]{$\leg{i}$}
\Text(84.,-42.)[l]{$\leg{j}$}
%gauge bosons
\Text(64.4276,21.6767)[lu]{$\scriptscriptstyle{V_1}$}
\Text(73.343,1.)[l]{$\scriptstyle{u^{V_2}}$}
\Text(41.1437,1.)[r]{$\scriptstyle{u^{\bar V_3}}$}
\Text(63.3537,-24.6395)[lb]{$\scriptscriptstyle{V_4}$}
%inflowing lines
\Text(41.5019,21.9679)[br]{}
\Text(42.,-21.)[tr]{}
\Text(19.515,10.9162)[br]{}
\Text(20.,-10.)[tr]{}
{#1}
\end{picture}
}

\newcommand{\diagXII}[1]{
\begin{picture}(120.,104.)(-28.,-52.)
\Line(0.,0.)(56.,28.)
\Line(56.,28.)(80.,40.)
\Line(56.,-28.)(0.,0.)
\Line(80.,-40.)(56.,-28.)
\Photon(56.,28.)(56.,10.5064){2.4}{2.5}
\Photon(56.,-10.5064)(56.,-28.){2.4}{2.5}
\DashCArc(56.9771,0.)(11.7466,-90.,90.){3}
\DashCArc(56.9771,0.)(11.7466,90.,270.){3}
\Vertex(56.,-28.){2}
\Vertex(56.,28.){2}
\Vertex(56.,10.5064){2}
\Vertex(56.,-10.5064){2}
%external lines
\Text(84.,42.)[l]{$\leg{i}$}
\Text(84.,-42.)[l]{$\leg{j}$}
%gauge bosons
\Text(64.4276,21.6767)[lu]{$\scriptscriptstyle{V_1}$}
\Text(71.5542,0.)[l]{$\scriptscriptstyle{\Phi_{2}}$}
\Text(42.9325,0.)[r]{$\scriptscriptstyle{\Phi_{3}}$}
\Text(63.3537,-24.6395)[lb]{$\scriptscriptstyle{V_4}$}
%inflowing lines
\Text(41.5019,21.9679)[br]{}
\Text(42.,-21.)[tr]{}
\Text(19.515,10.9162)[br]{}
\Text(20.,-10.)[tr]{}
{#1}
\end{picture}
}

\newcommand{\diagXV}[1]{
\begin{picture}(120.,104.)(-28.,-52.)
\Line(0.,0.)(56.,28.)
\Line(56.,28.)(80.,40.)
\Line(56.,-28.)(0.,0.)
\Line(80.,-40.)(56.,-28.)
\Photon(56.,28.)(56.,10.5064){-2.4}{2.5}
\Photon(56.,-10.5064)(56.,-28.){-2.4}{2.5}
\PhotonArc(56.9771,0.)(11.7466,90.,270.){-2}{3.5}
\DashCArc(56.9771,0.)(11.7466,-90.,90.){3}
\Vertex(56.,-28.){2}
\Vertex(56.,28.){2}
\Vertex(56.,10.5064){2}
\Vertex(56.,-10.5064){2}
%external lines
\Text(84.,42.)[l]{$\leg{i}$}
\Text(84.,-42.)[l]{$\leg{j}$}
%gauge bosons
\Text(64.4276,21.6767)[lu]{$\scriptscriptstyle{V_1}$}
\Text(76.0263,0.)[l]{$\scriptscriptstyle{\Phi_{2}}$}
\Text(38.4604,0.)[r]{$\scriptscriptstyle{V_3}$}
\Text(63.3537,-24.6395)[lb]{$\scriptscriptstyle{V_4}$}
%inflowing lines
\Text(41.5019,21.9679)[br]{}
\Text(42.,-21.)[tr]{}
\Text(19.515,10.9162)[br]{}
\Text(20.,-10.)[tr]{}
{#1}
\end{picture}
}

\newcommand{\diagXVI}[1]{
\begin{picture}(120.,104.)(-28.,-52.)
\Line(0.,0.)(48.,24.)
\Line(48.,24.)(80.,40.)
\Line(48.,-24.)(0.,0.)
\Line(80.,-40.)(48.,-24.)
\Photon(48.,24.)(58.1378,0.){-2.4}{2.5}
\Photon(48.,-24.)(58.1378,0.){2.4}{2.5}
\PhotonArc(67.082,0.)(8.94427,-177.135,182.865){-2}{7}
\Vertex(48.,-24.){2}
\Vertex(48.,24.){2}
\Vertex(58.1378,0.){2}
%external lines
\Text(84.,42.)[l]{$\leg{i}$}
\Text(84.,-42.)[l]{$\leg{j}$}
%gauge bosons
\Text(55.9503,18.8245)[lu]{$\scriptscriptstyle{V_1}$}
\Text(80.4984,0.)[l]{$\scriptscriptstyle{V_2}$}
\Text(55.0177,-21.3975)[lb]{$\scriptscriptstyle{V_3}$}
%inflowing lines
\Text(41.5019,21.9679)[br]{}
\Text(42.,-21.)[tr]{}
\Text(19.515,10.9162)[br]{}
\Text(20.,-10.)[tr]{}
{#1}
\end{picture}
}

\newcommand{\diagXVII}[1]{
\begin{picture}(120.,104.)(-28.,-52.)
\Line(0.,0.)(48.,24.)
\Line(48.,24.)(80.,40.)
\Line(48.,-24.)(0.,0.)
\Line(80.,-40.)(48.,-24.)
\Photon(48.,24.)(58.1378,0.){-2.4}{2.5}
\Photon(48.,-24.)(58.1378,0.){2.4}{2.5}
\DashCArc(67.082,0.)(8.94427,-177.135,182.865){3}
\Vertex(48.,-24.){2}
\Vertex(48.,24.){2}
\Vertex(58.1378,0.){2}
%external lines
\Text(84.,42.)[l]{$\leg{i}$}
\Text(84.,-42.)[l]{$\leg{j}$}
%gauge bosons
\Text(55.9503,18.8245)[lu]{$\scriptscriptstyle{V_1}$}
\Text(80.4984,0.)[l]{$\scriptscriptstyle{\Phi_{2}}$}
\Text(55.0177,-21.3975)[lb]{$\scriptscriptstyle{V_3}$}
%inflowing lines
\Text(41.5019,21.9679)[br]{}
\Text(42.,-21.)[tr]{}
\Text(19.515,10.9162)[br]{}
\Text(20.,-10.)[tr]{}
{#1}
\end{picture}
}

\newcommand{\diagAZ}[1]{
\begin{picture}(120.,104.)(-28.,-52.)
\Line(0.,0.)(56.,28.)
\Line(56.,28.)(80.,40.)
\Line(56.,-28.)(0.,0.)
\Line(80.,-40.)(56.,-28.)
\Photon(56.,28.)(56.,10.5064){-2.4}{2.5}
\Photon(56.,-10.5064)(56.,-28.){-2.4}{2.5}
\GCirc(56.,0.){11}{0.8}
\Vertex(56.,-28.){2}
\Vertex(56.,28.){2}
%external lines
\Text(84.,42.)[l]{$\leg{i}$}
\Text(84.,-42.)[l]{$\leg{j}$}
%gauge bosons
\Text(64.4276,21.6767)[lu]{$\scriptscriptstyle{A}$}
\Text(76.0263,0.)[l]{}
\Text(38.4604,0.)[r]{}
\Text(63.9079,-23.1643)[lb]{$\scriptscriptstyle{Z}$}
%inflowing lines
\Text(41.5019,21.9679)[br]{}
\Text(42.,-21.)[tr]{}
\Text(19.515,10.9162)[br]{}
\Text(20.,-10.)[tr]{}
{#1}
\end{picture}
}

\newcommand{\diagXX}[6]{
\begin{picture}(120.,104.)(-28.,-52.)
\Line(0.,0.)(65.,32.5)
\Line(0.,0.)(65.,-32.5)
\Line(0.,0.)(72.6722,0.)
\Photon(55.25,27.625)(55.25,0.){1.95}{2.5}
\Photon(37.05,-18.525)(37.05,0.){-1.95}{2}
\Vertex(55.25,27.625){2}
\Vertex(37.05,-18.525){2}
\Vertex(55.25,0.){2}
\Vertex(37.05,0.){2}
%external lines
\Text(68.25,34.125)[l]{#1}
\Text(76.3058,0.)[l]{#2}
\Text(68.25,-34.125)[l]{#3}
%gauge bosons
\Text(60.4318,13.7948)[l]{#4}
\Text(42.4369,-13.018)[l]{#5}
%inflowing lines
\Text(46.5102,0.)[b]{}
\Text(32.1146,16.999)[br]{}
\Text(22.75,-11.375)[tr]{}
\Text(25.4353,0.)[b]{}
{#6}
\end{picture}
}

\newcommand{\diagXXd}[6]{
\begin{picture}(120.,104.)(-28.,-52.)
\Line(0.,0.)(65.,32.5)
\Line(0.,0.)(65.,-32.5)
\Line(0.,0.)(72.6722,0.)
\PhotonArc(30.,15.)(15.6525,26.5651,206.565){2}{4}
\Vertex(16.,8.){2}
\Vertex(44.,22.){2}
\Photon(50,-25)(50,0.){-1.95}{2}
\Vertex(50,-25){2}
\Vertex(50,0.){2}
%external lines
\Text(68.25,34.125)[l]{#1}
\Text(76.3058,0.)[l]{#2}
\Text(68.25,-34.125)[l]{#3}
%gauge bosons
\Text(10.4318,30.7948)[b]{#4}
\Text(55,-15)[l]{#5}
%inflowing lines
\Text(46.5102,0.)[b]{}
\Text(32.1146,16.999)[br]{}
\Text(22.75,-11.375)[tr]{}
\Text(25.4353,0.)[b]{}
{#6}
\end{picture}
}

\newcommand{\diagXXc}[6]{
\begin{picture}(120.,104.)(-28.,-52.)
\Line(0.,0.)(65.,32.5)
\Line(0.,0.)(65.,-32.5)
\Line(0.,0.)(72.6722,0.)
\Photon(37.05,18.525)(37.05,0.){1.95}{2}
\Photon(55.25,-27.625)(55.25,0.){-1.95}{2.5}
\Vertex(37.05,18.525){2}
\Vertex(55.25,-27.625){2}
\Vertex(37.05,0.){2}
\Vertex(55.25,0.){2}
%external lines
\Text(68.25,34.125)[l]{#1}
\Text(76.3058,0.)[l]{#2}
\Text(68.25,-34.125)[l]{#3}
%gauge bosons
\Text(60.1189,-14.1922)[l]{#4}
\Text(42.4369,10.018)[l]{#5}
%inflowing lines
\Text(46.5102,0.)[t]{}
\Text(32.1146,16.999)[br]{}
\Text(22.75,-11.375)[tr]{}
\Text(25.4353,0.)[t]{}
{#6}
\end{picture}
}

\newcommand{\diagXXI}[7]{
\begin{picture}(120.,104.)(-28.,-52.)
\Line(0.,0.)(65.,32.5)
\Line(0.,0.)(65.,-32.5)
\Line(0.,0.)(72.36,-6.72921)
\Photon(45.5,22.75)(50.3792,7.05301){1.95}{2}
\Photon(39.,-19.5)(50.3792,7.05301){-1.95}{3}
\Photon(62.9532,-5.85441)(50.3792,7.05301){1.95}{2}
\Vertex(45.5,22.75){2}
\Vertex(62.9532,-5.85441){2}
\Vertex(39.,-19.5){2}
\Vertex(50.3792,7.05301){2}
%external lines
\Text(68.25,34.125)[l]{#1}
\Text(75.978,-7.06567)[l]{#2}
\Text(68.25,-34.125)[l]{#3}
%gauge bosons
\Text(52.0477,16.1784)[l]{#4}
\Text(61.6758,3.43505)[l]{#5}
\Text(46.4209,-13.9719)[l]{#6}
%inflowing lines
\Text(26.,-13.)[tr]{}
\Text(28.9031,15.2991)[br]{}
\Text(29.0689,0.)[b]{}
{#7}
\end{picture}
}

\newcommand{\diagXXII}[7]{
\begin{picture}(120.,104.)(-28.,-52.)
\Line(0.,0.)(65.,32.5)
\Line(0.,0.)(65.,-32.5)
\Line(0.,0.)(72.36,6.72921)
\Line(0.,0.)(72.36,-6.72921)
\Photon(52.,26.)(57.888,5.38336){1.95}{2}
\Photon(52.,-26.)(57.888,-5.38336){-1.95}{2}
\Vertex(52.,26.){2}
\Vertex(52.,-26.){2}
\Vertex(57.888,5.38336){2}
\Vertex(57.888,-5.38336){2}
%external lines
\Text(68.25,34.125)[l]{#1}
\Text(75.978,7.06567)[l]{#2}
\Text(75.978,-7.06567)[l]{#3}
\Text(68.25,-34.125)[l]{#4}
%gauge bosons
\Text(59.3965,16.9633)[l]{#5}
\Text(59.3965,-16.9633)[l]{#6}
{#7}
\end{picture}
}

\newcommand{\diagonenf}[3]{
%\begin{picture}(120.,104.)(-28.,-52.)
\begin{picture}(120.,76.)(-28.,-38.)
\Line(0.,0.)(80,0)
\PhotonArc(25.,-8.)(26.2488,17.7447,162.255){2}{5}
\Vertex(50.,0.){2}
\Text(84.,0)[l]{#1}
%gauge bosons
\Text(23,29)[l]{#2}
%inflowing lines
\Text(27.668,14.6453)[br]{}
\Text(28.,-14.)[tr]{}
{#3}
\end{picture}
}

\newcommand{\diagoneselfnf}[3]{
%\begin{picture}(120.,104.)(-28.,-52.)
\begin{picture}(120.,76.)(-28.,-38.)
\Line(0.,0.)(80,0)
\PhotonArc(40.,0)(20,0,180){2}{5}
\Vertex(60.,0.){2}
\Vertex(20.,0.){2}
\Text(84.,0)[l]{#1}
\Text(40,29.)[b]{#2}
\Text(27.668,14.6453)[br]{}
\Text(28.,-14.)[tr]{}
{#3}
\end{picture}
}

\newcommand{\diagonefact}[4]{
\begin{picture}(120.,104.)(-28.,-52.)
\Line(0.,0.)(80,0)
\Line(0.,0.)(50,50)
\PhotonArc(0,0)(48,0,45){2}{4}
\Vertex(48.,0.){2}
\Vertex(33.941,33.941){2}
\Text(84.,0)[l]{#1}
\Text(53.,53.)[l]{#2}
%gauge bosons
\Text(50,22.)[l]{#3}
%inflowing lines
\Text(27.668,14.6453)[br]{}
\Text(28.,-14.)[tr]{}
{#4}
\end{picture}
}

\newcommand{\diagInf}[6]{
\begin{picture}(120.,104.)(-28.,-52.)
\Line(0.,0.)(28.,14.)
\Line(28.,14.)(56.,28.)
\Line(56.,28.)(80.,40.)
\Line(28.,-14.)(0.,0.)
\Line(56.,-28.)(28.,-14.)
\Line(80.,-40.)(56.,-28.)
\Photon(56.,28.)(56.,0.){2.4}{2}
\Photon(56.,0.)(56.,-28.){2.4}{2}
\Photon(56.,0.)(0,0){2.4}{4}
\Vertex(56.,28.){2}
\Vertex(56.,0){2}
\Vertex(56.,-28.){2}
%external lines
\Text(84.,42.)[l]{#1}
\Text(84.,-42.)[l]{#2}
%gauge bosons
\Text(62.6099,14.)[l]{#3}
\Text(62.6099,-14.)[l]{#4}
\Text(35.305,-5.)[t]{#5}
%inflowing lines
\Text(41.5019,21.9679)[br]{}
\Text(42.,-21.)[tr]{}
\Text(19.515,10.9162)[br]{}
\Text(20.,-10.)[tr]{}
{#6}
\end{picture}
}
\newcommand{\diagIInf}[5]{
\begin{picture}(120.,104.)(-28.,-52.)
\Line(0.,0.)(16.,8.)
\Line(16.,8.)(44.,22.)
\Line(44.,22.)(64.,32.)
\Line(64.,32.)(80.,40.)
\Line(64.,-32.)(0.,0.)
\Line(80.,-40.)(64.,-32.)
\Photon(64.,32.)(64.,-32.){2.4}{3.5}
\PhotonArc(17.9434,8.83368)(20.,26.2114,206.211){2}{5} 
\Vertex(35.8868,17.6674){2}
\Vertex(64.,-32.){2}
\Vertex(64.,32.){2}
%external lines
\Text(84.,42.)[l]{#1}
\Text(84.,-42.)[l]{#2}
%gauge bosons
\Text(71.5542,0.)[l]{#3}
\Text(20.8885,35.1332)[br]{#4}
%inflowing lines
\Text(41.5019,21.9679)[br]{}
\Text(42.,-21.)[tr]{}
\Text(19.515,10.9162)[br]{}
\Text(20.,-10.)[tr]{}
{#5}
\end{picture}
}

\newcommand{\diagIIInf}[5]{
\begin{picture}(120.,104.)(-28.,-52.)
\Line(0.,0.)(24.,12.)
\Line(24.,12.)(40.,20.)
\Line(40.,20.)(56.,28.)
\Line(56.,28.)(80.,40.)
\Line(40.,-20.)(0.,0.)
\Line(80.,-40.)(40.,-20.)
\Photon(40.,20.)(40.,-20.){2.4}{3.5}
\PhotonArc(44.6515,-11.4562)(46.0977,66.8127,165.61){2}{5}
\Vertex(62.8019,30.9179){2}
\Vertex(40.,-20.){2}
\Vertex(40.,20.){2}
%external lines
\Text(84.,42.)[l]{#1}
\Text(84.,-42.)[l]{#2}
%gauge bosons
\Text(49.1935,0.)[l]{#3}
\Text(30.1994,40.9325)[br]{#4}
%inflowing lines
\Text(41.5019,21.9679)[br]{}
\Text(42.,-21.)[tr]{}
\Text(19.515,10.9162)[br]{}
\Text(20.,-10.)[tr]{}
{#5}
\end{picture}
}

\newcommand{\diagIVnf}[6]{
\begin{picture}(120.,104.)(-28.,-52.)
\Line(0.,0.)(65.,32.5)
\Line(0.,0.)(65.,-32.5)
\Line(0.,0.)(72.6722,0.)
\PhotonArc(23.7198,2.76053)(23.8799,45.7845,186.638){2}{5}
\Vertex(40.3727,19.8758){2}
\Photon(50,-25)(50,0.){-1.95}{2}
\Vertex(50,-25){2}
\Vertex(50,0.){2}
%external lines
\Text(68.25,34.125)[l]{#1}
\Text(76.3058,0.)[l]{#2}
\Text(68.25,-34.125)[l]{#3}
%gauge bosons
\Text(10.4318,30.7948)[b]{#4}
\Text(55,-15)[l]{#5}
%inflowing lines
\Text(46.5102,0.)[b]{}
\Text(32.1146,16.999)[br]{}
\Text(22.75,-11.375)[tr]{}
\Text(25.4353,0.)[b]{}
{#6}
\end{picture}
}

\newcommand{\diagIwi}[3]{
%\begin{picture}(120.,104.)(-28.,-52.)
\begin{picture}(120.,64.)(-28.,-32.)
\Line(0.,0.)(50,0)
\PhotonArc(50.,-50.)(70.71,90.,135.){2}{5}
\Text(54.,0)[l]{#1}
%gauge bosons
\Text(54,22.)[l]{#2}
%inflowing lines
\Text(27.668,14.6453)[br]{}
\Text(28.,-14.)[tr]{}
{#3}
\end{picture}
}
\newcommand{\diagIIwi}[3]{
%\begin{picture}(120.,104.)(-28.,-52.)
\begin{picture}(120.,64.)(-28.,-32.)
\Line(0.,0.)(50,0)
\Photon(25.,0)(50,20){-2}{3}
\Vertex(25,0.){2}
\Text(54.,0)[l]{#1}
\Text(54,22.)[l]{#2}
%inflowing lines
\Text(27.668,14.6453)[br]{}
\Text(28.,-14.)[tr]{}
{#3}
\end{picture}
}

\newcommand{\diagIIIwi}[2]{
\begin{picture}(120.,104.)(-28.,-52.)
\Line(0.,0.)(50,0)
\Text(54.,0)[l]{#1}
%gauge bosons
%inflowing lines
\Text(27.668,14.6453)[br]{}
\Text(28.,-14.)[tr]{}
{#2}
\end{picture}
}

\newcommand{\diagIVwi}[4]{
\begin{picture}(120.,104.)(-28.,-52.)
\Line(0.,0.)(70,0)
\Line(0.,0.)(50,50)
\Photon(25,25)(70,25){2}{5}
\Vertex(25,25){2}
\Text(74.,0)[l]{#1}
\Text(53.,53.)[l]{#2}
%gauge bosons
\Text(75,25)[l]{#3}
%inflowing lines
\Text(27.668,14.6453)[br]{}
\Text(28.,-14.)[tr]{}
{#4}
\end{picture}
}

\newcommand{\subloopI}[4]{
\begin{picture}(120.,104.)(-28.,-52.)
\Line(0.,0.)(16.,8.)
\Line(16.,8.)(44.,22.)
\Line(44.,22.)(64.,32.)
\Line(64.,32.)(80.,40.)
\Line(64.,-32.)(0.,0.)
\Line(80.,-40.)(64.,-32.)
\Photon(64.,32.)(64.,-32.){2.4}{3.5}
\GCirc(35.,17.5){10}{0.8}
\Vertex(64.,-32.){2}
\Vertex(64.,32.){2}
%external lines
\Text(84.,42.)[l]{#1}
\Text(84.,-42.)[l]{#2}
%gauge bosons
\Text(71.5542,0.)[l]{#3}
%inflowing lines
\Text(41.5019,21.9679)[br]{}
\Text(42.,-21.)[tr]{}
\Text(19.515,10.9162)[br]{}
\Text(20.,-10.)[tr]{}
{#4}
\end{picture}
}

\newcommand{\subloopII}[4]{
\begin{picture}(120.,104.)(-28.,-52.)
\Line(0.,0.)(24.,12.)
\Line(24.,12.)(40.,20.)
\Line(40.,20.)(56.,28.)
\Line(56.,28.)(80.,40.)
\Line(40.,-20.)(0.,0.)
\Line(80.,-40.)(40.,-20.)
\Photon(40.,20.)(40.,-20.){2.4}{3.5}
\GCirc(40.,20){10}{0.8}
\Vertex(40.,-20.){2}
%external lines
\Text(84.,42.)[l]{#1}
\Text(84.,-42.)[l]{#2}
%gauge bosons
\Text(49.1935,0.)[l]{#3}
%inflowing lines
\Text(41.5019,21.9679)[br]{}
\Text(42.,-21.)[tr]{}
\Text(19.515,10.9162)[br]{}
\Text(20.,-10.)[tr]{}
{#4}
\end{picture}
}

\newcommand{\diagIcollin}[4]{
%\begin{picture}(100.,104.)(-15.,-52.)
\begin{picture}(100.,88.)(-15.,-44.)
\Line(0.,0.)(90,0)
\Text(94.,0)[l]{#1}
\Photon(0,0)(0,-30){2}{4}
\Text(0,-34)[t]{#2}
\Photon(55,0)(55,-30){2}{4}
\Vertex(55,0){2}
\Text(55,-34)[t]{#3}
{#4}
\end{picture}
}

\newcommand{\diagIIcollin}[6]{
%\begin{picture}(175.,104.)(-5.,-52.)
\begin{picture}(175.,88.)(-5.,-44.)
\Photon(0,0)(-15,-30){2}{4}
\Text(0,-34)[t]{#1}
\Photon(0,0)(15,-30){2}{4}
\Text(15,-34)[t]{#2}
\Line(0.,0.)(80,0)
\Text(90,0)[l]{$\dots$}
\Line(120.,0.)(170,0)
\Text(174.,0)[l]{#3}
\Photon(65,0)(65,-30){2}{4}
\Vertex(65,0){2}
\Text(65,-34)[t]{#4}
\Photon(135,0)(135,-30){2}{4}
\Vertex(135,0){2}
\Text(135,-34)[t]{#5}
{#6}
\end{picture}
}

\newcommand{\diagIIIcollin}[6]{
%\begin{picture}(175.,104.)(-5.,-52.)
%\begin{picture}(175.,88.)(-5.,-44.)
\begin{picture}(45.,88.)(-5.,-44.)
\Photon(0,0)(-15,-30){2}{4}
\Text(0,-34)[t]{#1}
\Photon(0,0)(15,-30){2}{4}
\Text(15,-34)[t]{#2}
\Line(0.,0.)(30,0)
%\Text(90,0)[l]{$\dots$}
%\Line(120.,0.)(170,0)
\Text(34.,0)[l]{#3}
%\Photon(65,0)(65,-30){2}{4}
%\Vertex(65,0){2}
%\Text(65,-34)[t]{#4}
%\Photon(135,0)(135,-30){2}{4}
%\Vertex(135,0){2}
%\Text(135,-34)[t]{#5}
{#6}
\end{picture}
}

%%%%%%%%%%%%%%%%%%%%%%  abbreviations for results   %%%%%%%%%%%%%%%%%%%%%%
\newcommand{\Frac}{\frac}
\newcommand{\Epsinv}[1]{\veps^{- #1}}
\newcommand{\Eps}[1]{\veps^{#1}}
\newcommand{\Right}{\right}
\newcommand{\EG}{\gamma_{\mathrm{E}}}

%%%%%%%%%%%%%%%%%%%%% %%%%%%%%%%%%%%%%%%%%%%%%%%%%%%%%%%%%%%%%%

\thispagestyle{empty} 

%\draft

\thispagestyle{empty}
\def\thefootnote{\fnsymbol{footnote}}
\setcounter{footnote}{1}
\null
\draftdate\hfill  PSI-PR-08-11 
\\
\strut\hfill  MPP-2008-89
\vskip 0cm
\vfill
\begin{center}
{\Large \bf
Two-loop electroweak next-to-leading logarithms \\ 
for 
processes involving heavy quarks
\par}
 \vskip 1em
{\large
{\sc A.\ Denner$^1$\footnote{Ansgar.Denner@psi.ch}, 
B.\ Jantzen$^1$\footnote{physics@bernd-jantzen.de},
and S.\ Pozzorini$^2$\footnote{pozzorin@mppmu.mpg.de} }}
\\[.5cm]
$^1$ {\it Paul Scherrer Institut\\
CH-5232 Villigen PSI, Switzerland}
\\[0.3cm]
$^2$ {\it 
Max-Planck-Institut f\"ur Physik,
F\"ohringer Ring 6\\
D-80805 M\"unchen, Germany}
\par
\end{center}\par
\vskip 1.0cm \vfill {\bf Abstract:} \par 

We derive logarithmically enhanced two-loop virtual 
electroweak corrections
for arbitrary fermion-scattering processes at the TeV scale.
This extends results previously obtained for massless fermion scattering 
to processes that involve also bottom and top quarks.
The contributions resulting from soft, collinear, and ultraviolet
singularities in the complete electroweak Standard Model are
explicitly extracted from two-loop diagrams to
next-to-leading-logarithmic accuracy including all effects associated
with symmetry breaking and Yukawa interactions.
%We find agreement
%with the resummation prescriptions that have been proposed in the
%literature based on a symmetric $\SU(2)\times\U(1)$ theory matched
%with QED at the electroweak scale and provide new next-to-leading
%contributions proportional to $\ln(\MZ^2/\MW^2)$.
\par
\vskip 1cm
\noindent
September 2008
\mla
\par
\null
\setcounter{page}{0}
\clearpage
\def\thefootnote{\arabic{footnote}}
\setcounter{footnote}{0}

\section{Introduction}
\label{se:intro}%\refse{intro}

%%%%%%%%%%%%%%%%%%%%%%%%%%%%%%%%%%%%%%%%%%%%%%%%%%
% EW corrections at TeV scale
%%%%%%%%%%%%%%%%%%%%%%%%%%%%%%%%%%%%%%%%%%%%%%%%%%
At TeV colliders, electroweak radiative corrections are strongly
enhanced by logarithms of the type $\ln(Q^2/{M_{\PW,\PZ}^2})$
\cite{Kuroda:1991wn,Denner:1995jv,Beccaria:1998qe,Beenakker:1993tt}.
These logarithmic corrections affect every reaction that involves
electroweakly interacting particles and is characterized by scattering
energies $Q\gg M_{\PW,\PZ}$.  The impact of the enhanced electroweak
corrections at high-energy colliders has been investigated both in
complete one-loop calculations and in logarithmic approximation for
several specific reactions, including gauge-boson pair production at
the ILC~\cite{Beenakker:1993tt,Gounaris:2002za} and the
LHC~\cite{Accomando:2001fn}, gauge-boson
scattering~\cite{Denner:1995jv,Accomando:2006hq}, fermion-pair
production in $\Pep\Pem$
collisions~\cite{Beccaria:1998qe,Beccaria:1999fk}, Drell--Yan
processes at the LHC~\cite{Baur:2001ze}, heavy-quark
production~\cite{Beccaria:1999xd},
single-gauge-boson plus jet production at the LHC~\cite{Maina:2004rb},
Higgs production in vector-boson fusion at the
LHC~\cite{Ciccolini:2007jr}, and three-jet production in
$\Pep\Pem$ collisions~\cite{CarloniCalame:2008qn}.  The results of
these studies demonstrate that at energies $Q\sim 1\TeV$ the size of
the electroweak corrections can reach, depending on the process, up to
tens of per cent at one loop and several per cent at two loops.

At high energies, the dominant effects can be described in a
systematic way by treating the electroweak corrections in the
asymptotic limit ${Q}/{M_{\PW,\PZ}}\to \infty$, where all masses of
order $\MW$ are formally handled as infinitesimally small parameters.
In this limit, the electroweak corrections appear as a sequence of
logarithms of the form
$\alpha^l\ln^{j}{\left({Q^2}/{M_{\PW,\PZ}^2}\right)}$, with $j \le
2l$, which diverge asymptotically.  
\mla
These logarithms have a two-fold
origin.  The renormalization of ultraviolet (UV) singularities at the
scale $\muR\lsim M_{\PW,\PZ}$ yields terms of the form
$\ln(Q^2/\muR^2)$ and, in addition, the interactions of the initial-
and final-state particles with soft and/or collinear gauge bosons give
rise to $\ln(Q^2/M^2_{\PW,\PZ})$ terms that represent mass
singularities.

As pointed out in~\citere{Ciafaloni:2000df}, soft and collinear
electroweak logarithms are present not only in those physical
observables that are exclusive with respect to real radiation of Z and
W bosons, but even in fully inclusive
observables~\cite{Ciafaloni:2000df,Baur:2006sn}.  However, there is no
need to combine corrections resulting from virtual and real Z/W bosons
in the same observable.  Thus, in this paper we will consider only
exclusive observables, which do not include real Z/W-boson emission
and contain only virtual contributions.

Since they originate from UV, soft, and collinear singularities,
electroweak logarithmic corrections have universal properties that can
be studied in a process-independent way and reveal interesting
analogies between QED, QCD, and electroweak interactions.  At one
loop, the leading logarithms (LLs) and next-to-leading logarithms
(NLLs) factorize and are described by a general formula that applies
to arbitrary Standard-Model
processes~\cite{Denner:2001jv,Pozzorini:rs}.  The properties of
electroweak logarithmic corrections beyond one loop can be
investigated by means of infrared evolution equations that describe
the electroweak interactions in terms of two regimes corresponding to
SU(2)$\times$U(1) and $\mathrm{U}_{\mathrm{em}}(1)$
symmetric gauge theories~\cite{%
  Fadin:2000bq,Melles:2001ia,Melles:2001gw,%Melles:2001mr,Melles:2001dh,
  Kuhn:2000nn,%Kuhn:2001hz,
  Kuhn:2007ca}.
This approach makes use of resummation techniques that were derived in
the context of symmetric gauge theories (QED, QCD), thereby assuming
that electroweak symmetry-breaking effects other than the splitting
into two symmetric regimes are negligible in next-to-next-to-leading
logarithmic (NNLL) approximation.  An alternative method, which also
relies on factorization and exponentiation of the logarithmic
corrections, is based on soft--collinear effective theory
\cite{Bauer:2000ew}. Using this method, $\ln(Q^2/M^2_{\PW,\PZ})$
corrections to the Sudakov form factor for massless and massive
fermions have been calculated and used to compute electroweak
corrections to four-fermion processes at the LHC
\cite{Chiu:2007yn,Chiu:2008vv}.  Alternatively, the electroweak
logarithms can be extracted explicitly from Feynman diagrams at two
loops~\cite{Melles:2000ed,Denner:2003wi,Pozzorini:2004rm,Feucht:2003yx,Jantzen:2005az,Jantzen:2006jv,Denner:2006jr}.
Calculations of this type explicitly implement all aspects of
electroweak symmetry breaking through the Feynman rules and thus
provide a strong check of the results obtained via evolution equations
or effective theories.

So far, all existing diagrammatic results are in agreement with the
resummation prescriptions. However, up to now only a small subset of
logarithms and processes has been computed explicitly at two loops:
while the LLs \cite{Melles:2000ed} and the angular-dependent subset of
the NLLs~\cite{Denner:2003wi} have been derived for arbitrary
processes, complete diagrammatic calculations at (or beyond) the NLL
level exist only for matrix elements involving massless external
fermions~\cite{Pozzorini:2004rm,Feucht:2003yx,Jantzen:2005az,Jantzen:2006jv,Denner:2006jr}.
In the literature, no explicit 2-loop NLL calculation exists for
reactions involving massive scattering particles.

In this paper we derive the two-loop NLL corrections for general
$n$-fermion processes $f_1f_2\to f_3\dots f_n$ involving an arbitrary
number of massless and massive fermions, \ie leptons, light or heavy
quarks.  We consider the limit where all kinematical invariants are of
order $Q^2\gg M^2_{\PW,\PZ}$, and the top mass---as well as the Higgs
mass---is of the same order as $M_{\PW,\PZ}$.  Apart from the top
quark all other fermions, including bottom quarks, are treated as
massless particles.
Soft and collinear singularities from virtual photons are regularized
dimensionally and arise as $\veps$-poles in $D=4-2\veps$ dimensions.
For consistency, the same power counting is applied to
$\ln(Q^2/\MW^2)$ and $1/\veps$ singularities. Thus, in NLL
approximation we include all $\veps^{-k}\ln^{j-k}(Q^2/\MW^2)$ terms
with total power $j=2,1$ at one loop and $j=4,3$ at two loops.  We
explicitly show that the photonic singularities can be factorized in a
gauge-invariant electromagnetic term, in such a way that the remaining
part of the corrections---which is finite, gauge invariant, and does
not depend on the scheme adopted to regularize photonic
singularities---contains only $\ln(Q^2/\MW^2)$ terms.  The divergences
contained in the electromagnetic term cancel if real-photon emission
is included.

We utilize the technique that we introduced in~\citere{Denner:2006jr}
to derive the NLL two-loop corrections for massless $n$-fermion
processes.  This method is based on collinear Ward identities which
permit to factorize the soft--collinear contributions from the
$n$-fermion tree-level amplitude and isolate them in
process-independent two-loop integrals.  The latter are evaluated to
NLL accuracy using an automatized algorithm based on the
sector-decomposition technique~\cite{Denner:2004iz} and,
alternatively, the method of expansion by regions combined with
Mellin--Barnes representations (see \citere{Jantzen:2006jv} and
references therein).

The treatment of $n$-fermion processes that involve massive top quarks
implies two new aspects: additional Feynman diagrams resulting from
Yukawa interactions, and top-mass terms that render the loop integrals
more involved.  The latter is the one that requires more effort from
the calculational point of view, since every topology has to be
evaluated for several different combinations of massive/massless
internal and external lines.  This part of the calculation was
automatized in the framework of the above-mentioned algorithms, thereby
doing an important step towards a complete NLL analysis of all
processes that involve massive particles.  A brief summary of the
results presented here was anticipated in \citere{Denner:2008hw}.

The paper is organized as follows.  Section \ref{se:definitions}
contains definitions and conventions used in the calculation.  In
\refse{se:singtreatment} we review the techniques used to extract UV
and mass singularities for the case of massless fermion
scattering~\cite{Denner:2006jr} and extend them to the case of massive
fermions. The one-loop counterterms and explicit results for the
renormalized one-loop amplitude are given in \refse{se:res1loop}. The
complete two-loop results are presented in \refse{se:res2loop}
including a discussion of the Yukawa-coupling contributions and the
two-loop renormalization. Section \ref{se:disc} is devoted to the
discussion of our final results.  Explicit results for the various
contributing one- and two-loop diagrams are presented in
\refapp{se:factcont}. Further appendices contain the definition of the
loop integrals (\refapp{app:loops}) and relations between them
(\refapp{se:looprelations}).
Specific results for four-particle processes involving external
fermions and gluons, including a comparison with the results of
\citeres{Chiu:2007yn,Chiu:2008vv}, can be found in
\refapp{se:2fproduction}.

\section{Definitions and conventions}
\label{se:definitions}%\refse{se:definitions}
The calculation is based on the formalism introduced in
\citere{Denner:2006jr} for massless fermion scattering.  Here we
summarize the most important conventions and introduce new definitions
that are needed to describe massive fermions.  For more details we
refer to Sect.~2 of \citere{Denner:2006jr}.

We consider a generic $n\to 0$ process involving an even number $n$ of
polarized fermionic particles,
\beq\label{genproc}%\refeq{genproc}
\varphi_{1}(p_1)\dots
\varphi_{n}(p_{n})\to 0.
\eeq
The symbols $\varphi_{i}$ represent $n/2$ antifermions and $n/2$
fermions: $\varphi_{i}={\bar f}_{\si_i}^{\kappa_i}$ for
$i=1,\dots,n/2$ and $\varphi_{i}={f}_{\si_i}^{\kappa_i}$ for
$i=n/2+1,\dots,n$.  The indices $\kappa_i=\rR,\rL$ and $\si_i$
characterize the chirality (see below) and the fermion type
($f_{\si_i}=\nu_e,e,\dots,\tau,u,d,\dots,t$), respectively.  All
external momenta are incoming and on shell, $p_k^2=m_k^2$.  Apart from
the top quarks, all other fermions (including bottom quarks) are
treated as massless particles.

The matrix element for the process \refeq{genproc} reads
\beq\label{matel}%\refeq{matel}
\M^{\varphi_{1}\dots\varphi_{n}}
=
\left[\prod_{i=1}^{n/2} {\bar v}(p_i,\kappa_i)\right]
G^{\underline{\varphi}_{1}\dots\underline{\varphi}_{n}}
(p_1,\dots,p_{n})
\left[\,\prod_{j=n/2+1}^{n} u(p_j,\kappa_j)\right]
,
\eeq
where $G^{\underline{\varphi}_{1}\dots\underline{\varphi}_{n}}$ is the
corresponding truncated Green function.  The spinors fulfil the Dirac
equation,
\beqar\label{topspinors1}%\refeq{topspinors1}
(\ps-m)u(p,\kappa)&=&0,
\qquad
(\ps+m)v(p,\kappa)=0,
\eeqar
and the argument $\kappa$ denotes their chirality.  More precisely,
the polarization states $\kappa=\rR,\rL$ are defined in such a way
that in the massless limit the spinors are eigenstates of the chiral
projectors
\beqar\label{chiralproj}%\refeq{chiralproj}
 \omega_{\rR}&=&
{\bar \omega}_\rL=
\frac{1}{2}(1+\gamma^5)
,\qquad
 \omega_{\rL}=
{\bar \omega}_\rR=
\frac{1}{2}(1-\gamma^5)
,
\eeqar
\ie 
\beqar\label{topspinor3}%\refeq{topspinor3}
\omega_{\rho} u(p,\kappa)&=&  \de_{\kappa\rho} u(p,\kappa)
+\order\left(\frac {m} {p^0}  u\right)
,
\qquad
\omega_{\rho} v(p,\kappa)=  \de_{\kappa\rho} v(p,\kappa)
+\order\left(\frac {m} {p^0}  v\right)
\eeqar
for $m/p^0\ll 1$. While massless spinors are exact eigenstates of the
chiral projectors, in the massive case the spinors are constructed by
means of helicity projectors
\beqar\label{topspinor2}%\refeq{topspinor2}
\Omega_{\rR}=
\hat \Omega_{\rL}=
\frac{1}{2}(1+\gamma^5\rs),
\qquad
\Omega_{\rL}=
\hat \Omega_{\rR}=
\frac{1}{2}(1-\gamma^5\rs)
\eeqar
where
\beq
r^\mu=\pm \frac 1 m\left(|\vec p|,p^0 \frac{\vec p}{|\vec p|}\right),
\qquad\mbox{for}\quad
\mathrm{sign}(p^0)=\pm 1,
\eeq
with $(rp) =0$, $r^2=-1$ and
\beqar\label{spinors}%\refeq{spinors}
 \Omega_{\rho}u(p,\kappa)&=&  \de_{\kappa\rho} u(p,\kappa)
,\qquad
 \hat\Omega_{\rho}v(p,\kappa)=  \de_{\kappa\rho} v(p,\kappa)
.
\eeqar
For $m/p^0\to 0$, we have $r^\mu \to p^\mu/m $, and the spinors
satisfy \refeq{topspinor3} in the high-energy limit.  Note that in the
case of antifermions, chirality and helicity play the opposite role,
and we always use chirality to label the spinors.

The amplitudes for physical scattering processes, \ie $\,2\to n-2\,$
reactions, are easily obtained from our results for $n\to0$ reactions
using crossing symmetry.

\subsection{Perturbative and asymptotic expansions}
\label{se:pertexp}%\refse{se:pertexp}
For the perturbative expansion of matrix elements we write
\beqar\label{pertserie1a}%\refeq{pertserie1a}
\M&=&
\sum_{l=0}^\infty
\mel{l}{}
,\qquad
\mel{l}{}=
\left(\frac{\alphaeps}{4\pi}\right)^l
\nmel{l}{}
,\qquad
\alphaeps=
\left(\frac{4\pi\muD^2}{ \mathrm{e}^{\gamma_{\mathrm{E}}}Q^2}\right)^{\veps}
\alpha,
\eeqar
where $\alpha=e^2/(4\pi)$ and in $D=4-2\veps$ dimensions we include in
the definition of $\alphaeps$ a normalization factor depending on
$\veps$, the scale $\muD$ of dimensional regularization, and the
characteristic energy $Q$ of the scattering process.
% and Euler's constant $\gamma_{\mathrm{E}}$.  

The electroweak corrections are evaluated in the region where all
kinematical invariants, $r_{j\ldots k}=(p_j+\ldots+p_k)^2$, are much
larger than the squared masses of the heavy particles that enter the
loops,
\beqar\label{asymptoticregion}%\refeq{asymptoticregion}
|r_{j\ldots k}|\sim Q^2 \gg \MW^2\sim\MZ^2\sim\Mt^2\sim\MH^2.
\eeqar
In this region, the electroweak corrections are dominated by
mass-singular logarithms,
\beqar\label{logsymbol}%\refeq{logsymbol}
L&=&\ln\left(\frac{Q^2}{\MW^2}\right)
,
\eeqar
and logarithms of UV origin.  Mass singularities that originate from
soft and collinear massless photons and UV singularities are
regularized dimensionally and give rise to $1/\veps$ poles.  The
amplitudes are computed as series in $L$ and $\veps$, classifying the
terms $\veps^{n} L^{m+n}$ according to the total power $m$ of
logarithms $L$ and $1/\veps$ poles.  At $l$ loops, terms with
$m=2l,2l-1,\dots$ are denoted as leading logarithms (LLs),
next-to-leading logarithms (NLLs), and so on.  The calculation is
performed in NLL approximation expanding the one- and two-loop terms
up to order $\veps^2$ and $\veps^0$, respectively,
\beqar\label{logexpansion}%\refeq{logexpansion}
\mel{1}{}
\NLLA \sum_{m=1}^{2}\sum_{n=-m}^{2}\mel{1,m,n}{}
\,\veps^{n}
L^{m+n}
,\qquad
\mel{2}{}
\NLLA \sum_{m=3}^{4}\sum_{n=-m}^{0}\mel{2,m,n}{}
\,\veps^{n}
L^{m+n}
.
\eeqar
Since the loop corrections depend on various masses,
$\MW\sim\MZ\sim\Mt\sim\MH$, and different invariants $r_{jk}$, the
coefficients $\mel{l,m,n}{}$ in \refeq{logexpansion} involve
logarithms of type\footnote{ This dependence only appears at the
  next-to-leading level, \ie for $m=2l-1$.  }
\beqar\label{anglogsymbol}%\refeq{anglogsymbol}
\lr{i}=\ln\left(\frac{M_i^2}{\MW^2}\right),
\qquad
\lr{jk}&=&\ln\left(\frac{-r_{jk}-\ri 0}{Q^2}\right)
\quad\mbox{for } j\neq k. 
\eeqar
%The logarithms of $-r_{jk}$ are extracted in the region $r_{jk}<0$,
%where the corrections are real. The imaginary parts that arise in the
%physical region can be obtained via analytic continuation replacing
%$r_{jk}$ by $r_{jk}+\ri 0$.
For convenience we also define $l_{jj}=0$ and
\beqar\label{logsymbolt}%\refeq{logsymbolt}
L_\Pt=\ln\left(\frac{Q^2}{\Mt^2}\right).
\eeqar
To distinguish terms associated
with massless and massive fermions we use the symbols
\beq\label{kronecker}%\refeq{kronecker}
  \deltat{i} = \left\{\barr{l}
    1, \; m_i= \Mt \\
    0, \; m_i=0
  \earr\right\}
,\qquad
  \deltaz{i} = \left\{\barr{l}
    1, \; m_i= 0 \\
    0, \; m_i=\Mt
  \earr\right\}.
\eeq
Mass-suppressed corrections of order $\MW^2/Q^2$ are systematically
neglected.

\subsection{Gauge and Yukawa couplings}
\label{se:gaugeandyuk}%\refeq{se:gaugeandyuk}
The generators associated with the gauge bosons $V=A,Z,\FWpm$ are
related to the weak isospin generators $T^i$, the hypercharge $Y$, and
the electromagnetic charge $Q$ through
\beqar\label{genmixing}%\refeq{genmixing}
e {I}^{W^\pm} &=& \frac{\gw }{\sqrt{2}}\left(T^1\pm\ri T^2\right)
%,\nl
,\quad\
e I^A=-eQ=
-\gw \sw T^3- \gb \cw \frac{Y}{2} 
,\nl
e I^Z&=&
\gw \cw T^3- \gb \sw \frac{Y}{2}
=\frac{\gw}{\cw}  T^3 - e\frac{\sw}{\cw}  Q
,
\eeqar
where $\cw=\cos\theta_\mathrm{w}$ and $\sw=\sin\theta_\mathrm{w}$
denote the sine and cosine of the weak mixing angle, and $\gb$ and
$\gw$ are the coupling constants associated with the U(1) and SU(2)
groups, respectively.  The generators \refeq{genmixing} obey the
commutation relations
\beqar\label{commrel}%\refeq{commrel}
e \left[I^{V_1},I^{V_2}\right]
=
\ri \gw \,
\sum_{V_3=A,Z,W^\pm}
\teps^{V_1V_2V_3} I^{\bar{V}_3}
,
\eeqar
where $\bar V$ denotes the complex conjugate of $V$ and the
$\teps$-tensor is defined in \citere{Denner:2006jr}.  The
SU(2)$\times$U(1) Casimir operator reads
\beq\label{casimir}%\refeq{casimir}
\sum_{V=A,Z,W^\pm} I^{\bar V} I^V = 
\frac{\gb^2}{e^2} \left(\frac{Y}{2}\right)^2+\frac{\gw^2}{e^2} C,
\qquad\mbox{with}\quad
C=\sum_{i=1}^3 (T^i)^2
.
\eeq
The gauge-boson masses, the vacuum expectation value $\vev$, and the
couplings fulfil
\beqar\label{massspectrum}%\refeq{massspectrum}
M_{\PWpm}&=&\frac{1}{2} \gw \vev
,\qquad
\MZ=\frac{1}{2\cw} \gw \vev
,\qquad
\cw =\frac{\MW}{\MZ}
,\qquad
\cw \gb Y_\Phi =\sw \gw,
\eeqar
where  $Y_\Phi$ is the hypercharge of the Higgs doublet.  
Leaving $Y_\Phi$ as a free parameter and identifying $e=\cw \gb$, the
Gell-Mann--Nishijima relation, which determines the hypercharges of
the fermions, reads $Q=Y/2+Y_\Phi T^3$.

The Feynman rules for the vector-boson--fermion--antifermion vertices
read
\beqar\label{chiralcouplingsdef}%\refeq{chiralcouplingsdef}
\vcenter{\hbox{
\begin{picture}(110,100)(-50,-50)
\Text(-45,5)[lb]{$V^{\mu}$}
\Text(35,30)[cb]{$\bar{f}_{\si'}$}
\Text(35,-30)[ct]{$f_{\si}$}
\Vertex(0,0){2}
\ArrowLine(0,0)(35,25)
\ArrowLine(35,-25)(0,0)
\Photon(0,0)(-45,0){2}{3}
\end{picture}}}
&=&\ri e \gamma^\mu \sum_{\kappa=\rR,\rL}
\omega_\kappa 
I^{V}_{f^\kappa_{\si'} f^\kappa_{\si}}
\,,
\eeqar
where $I^{V}_{f^\kappa_{\si'} f^\kappa_{\si}}$ denote the
SU(2)$\times$U(1) generators in the fundamental ($\kappa=\rL$) or
trivial ($\kappa=\rR$) representation.  The chiral projectors
$\omega_\kappa$ in \refeq{chiralcouplingsdef} can easily be shifted
along the fermionic lines using anticommutation relations%
\footnote{ In the high-energy NLL approximation diagrams involving
  chiral anomalies are irrelevant.  Thus we can use
  $\{\gamma^\mu,\gamma^5\}=0$ in $D=4-2\veps$ dimensions.}  until they
meet the spinor of an external fermion or antifermion and can be
eliminated using \refeq{topspinor3}.

It is convenient to adopt a notation that describes the interactions
of fermions and antifermions in a generic way.  To this end, for a
generic incoming particle $\varphi_i=f^{\kappa_i}_{\sigma}$ or $\bar
f^{\kappa_i}_{\sigma}$, we define
\beqar\label{gaugeint}%\refeq{gaugeint}
{I}^{V}_{\varphi''_i\varphi'_i}=
\left\{
\begin{array}{l@{\quad\mbox{for}\quad}l}
I^{V}_{f^{\kappa_i}_{\sigma''} f^{\kappa_i}_{\sigma'}}
& \varphi_i=f^{\kappa_i}_{\sigma}
\\ 
I^{V}_{\bar f^{\kappa_i}_{\sigma''} \bar f^{\kappa_i}_{\sigma'}}
=-I^{V}_{f^{\kappa_i}_{\sigma'} f^{\kappa_i}_{\sigma''}}
& \varphi_i=\bar f^{\kappa_i}_{\sigma}
\end{array}
\right.
.
\eeqar
With this notation the interaction of a gauge boson $V$ with incoming
fermions and antifermions yields
\beq\label{gaugeemission}%\refeq{gaugeemission}
\frac{\ri(\ps_i+\qs+m_i)}{(p_i+q)^2-m_i^2}
\ri e \gamma^\mu
I^V_{\varphi'_i\varphi_i}
u(p_i,\kappa_i)
\qquad\mbox{and}\qquad
{\bar v}(p_i,\kappa_i)
\ri e \gamma^\mu
I^V_{\varphi'_i\varphi_i}
\frac{\ri(\ps_i+\qs-m_i)}{(p_i+q)^2-m_i^2}
.
\eeq
Similar expressions are obtained for multiple gauge-boson
interactions.  Apart from the spinors and the reversed order of the
Dirac matrices, these expressions differ only
in the sign of the mass terms in the Dirac propagators%
\footnote{For antifermions the negative sign in the coupling
  \refeq{gaugeint} compensates the negative sign (of slashed momenta)
  due to the opposite fermion and momentum flow.  This yields
  $-(-\ps_i-\qs+m_i)=(\ps_i+\qs-m_i)$.}.  In practice we perform the
calculations assuming that certain legs ($i,j,\dots$) are incoming
fermions and we find that the results are independent of the sign of
the mass terms in the propagators (only squared mass terms give rise
to unsuppressed contributions).  Thus, the results are directly
applicable to fermions and antifermions.

The Feynman rules for the scalar--fermion--antifermion vertices read
\beqar\label{ycouplingsdef}%\refeq{ycouplingsdef}
\vcenter{\hbox{
\begin{picture}(110,100)(-50,-50)
\Text(-45,5)[lb]{$\Phi$}
\Text(35,30)[cb]{$\bar{f}_{\si'}$}
\Text(35,-30)[ct]{$f_{\si}$}
\Vertex(0,0){2}
\ArrowLine(0,0)(35,25)
\ArrowLine(35,-25)(0,0)
\DashLine(0,0)(-45,0){4}
\end{picture}}}
&=& \ri 
\left(
\omega_\rR G^{\Phi}_{f^\rL_{\si'} f^\rR_{\si}}
+
\omega_\rL G^{\Phi}_{f^\rR_{\si'} f^\rL_{\si}}
\right)
\,,
\eeqar
where $\Phi=H,\chi,\phi^\pm$.  Unitarity implies the following
relation between left- and right-handed coupling matrices:
\beq
G^{\Phi}_{f^\rR_{\si'} f^\rL_{\si}} 
= 
\left(G^{\Phi^+}_{f^\rL_{\si} f^\rR_{\si'}} \right)^*
.
\eeq
We consider only contributions proportional to the top-quark Yukawa
coupling,
\beq\label{topyukawa}%\refeq{topyukawa}
\lambdat=\frac{g_2 \Mt}{\sqrt{2}\MW},
\eeq
and neglect the bottom-quark mass.  The non-vanishing components of
the right-handed coupling matrix read
\beq
G^{H}_{t^\rL t^\rR}=-\frac{\lambdat}{\sqrt{2}} 
,\qquad
G^{\chi}_{t^\rL t^\rR}=\ri\frac{\lambdat}{\sqrt{2}} 
,\qquad
G^{\phi^-}_{b^\rL t^\rR}={\lambdat} 
.
\eeq
In analogy with \refeq{gaugeint}, we define
\beqar\label{yukint}%\refeq{yukint}
{G}^{\Phi}_{\varphi''_i\varphi'_i}=
\left\{
\begin{array}{l@{\quad\mbox{for}\quad}l}
G^{\Phi}_{f^{\rL}_{\sigma''} f^{\rR}_{\sigma'}}
& \varphi_i=f^{\rR}_{\sigma}
\\ 
G^{\Phi}_{{\bar f}_{\sigma''}^{\rL} {\bar f}^{\rR}_{\sigma'}}
=- G^{\Phi}_{f^{\rR}_{\sigma'} f^{\rL}_{\sigma''}}
%\right|_{\rR\leftrightarrow \rL}
& \varphi_i=\bar f^{\rR}_{\sigma}
\end{array}
\right.
,
\eeqar
and similarly for $\rR\leftrightarrow \rL$. With this notation the
interaction of a boson $\Phi$ with incoming fermions
$\varphi_i=f^{\kappa_i}_{\sigma}$ and antifermions $\varphi_i=\bar
f^{\kappa_i}_{\sigma}$ yields similar expressions as in the case of
gauge interactions \refeq{gaugeemission},
\beq\label{yukemission}%\refeq{yukemission}
\frac{\ri(\ps_i+\qs+m_i)}{(p_i+q)^2-m_i^2}
\ri 
G^{\Phi}_{\varphi'_i\varphi_i}
u(p_i,\kappa_i)
\qquad\mbox{and}\qquad
{\bar v}(p_i,\kappa_i)
\ri 
G^{\Phi}_{\varphi'_i\varphi_i}
\frac{\ri(\ps_i+\qs-m_i)}{(p_i+q)^2-m_i^2}
.
\eeq

In the case of light (anti)fermions ($\varphi_i=$lepton or quark of
the first two generations), the subsequent interaction with gauge
bosons $V_1,V_2,V_3$ yields coupling factors
$I^{V_3}_{\varphi_i'''\varphi_i''} I^{V_2}_{\varphi_i''\varphi_i'}
I^{V_1}_{\varphi_i'\varphi_i} $, where the representation of all
generators $I^{V}$ corresponds to the chirality $\kappa_i$ given by
the spinor of the external particle $\varphi_i$ [see
\refeq{gaugeint}].  For quarks of the third generation
($\varphi_i=$top or bottom), due to Yukawa interactions and top-mass
terms in the fermion propagators, also couplings with opposite
chirality contribute.  These are denoted as
\beqar\label{revchircoup}%\refeq{revchircoup}
\hat{I}^{V}_{\varphi_i''\varphi_i'}=
\left.
{I}^{V}_{\varphi_i''\varphi_i'}\right|_{\rR\leftrightarrow \rL}
,\qquad
\hat{G}^{\Phi}_{\varphi_i''\varphi_i'}=
\left.
{G}^{\Phi}_{\varphi_i''\varphi_i'}\right|_{\rR\leftrightarrow \rL}
.
\eeqar
For instance, the subsequent emission of scalars and gauge bosons
$\Phi_1,V_2,\Phi_3$ along a heavy-quark line yields coupling factors
$\hat G^{\Phi_3}_{\varphi_i'''\varphi_i''} \hat
I^{V_2}_{\varphi_i''\varphi_i'} G^{\Phi_1}_{\varphi_i'\varphi_i} $.

In our results the matrix elements \refeq{matel} are often abbreviated
as
\beq
\M\equiv \M^{\varphi_{1}\dots\varphi_{n}},
\eeq
and when they are multiplied by gauge- and Yukawa-coupling matrices we
write
\beqar\label{abbcouplings}%\refeq{abbcouplings}
\M I_k^{V_1}&=&\sum_{\varphi'_{k}}\M^{\varphi_{1}\dots\varphi'_{k}\dots\varphi_{n}} 
I^{V_1}_{\varphi'_{k}\varphi_{k}}
,\qquad
\M G_k^{\Phi_1}=\sum_{\varphi'_{k}}\M^{\varphi_{1}\dots\varphi'_{k}\dots\varphi_{n}} 
G^{\Phi_1}_{\varphi'_{k}\varphi_{k}}
,\nl
\M G_k^{\Phi_1}I_k^{V_2}&=&\sum_{\varphi'_{k},\varphi''_{k}}
\M^{\varphi_{1}\dots\varphi''_{{k}}\dots\varphi_{n}} 
G^{\Phi_1}_{\varphi''_{k}\varphi'_{k}}
I^{V_2}_{\varphi'_{k}\varphi_{k}},
\quad\mathrm{etc.}
\eeqar
Similar shorthands are used for the coupling matrices
\refeq{revchircoup}.  Global gauge invariance implies the 
relation
\beq\label{gaugeyukrelation}%\refeq{gaugeyukrelation}
\hat I^V_k G_k^{\Phi_i} - G_k^{\Phi_i} I_k^V = \sum_{\Phi_j=H,\chi,\phi^\pm}
G_k^{\Phi_j} I^V_{\Phi_j\Phi_i}
\eeq
between combinations of gauge and Yukawa couplings as well as the
charge-conservation identity
\beqar\label{chargeconservation}%\refeq{chargeconservation}
\M \sum_{k=1}^{n} I_k^{V}&=&0,
\eeqar
which is fulfilled up to mass-suppressed terms in the high-energy
limit.  In \refeq{gaugeyukrelation}, $I^V_{\Phi_j\Phi_i}$ denotes the
SU(2)$\times$U(1) generators for the Higgs doublet, which enter the
gauge couplings of the Higgs boson.

\section{Treatment of ultraviolet and mass singularities}
\label{se:singtreatment}%\refse{se:singtreatment}

Large logarithms and $1/\veps$ poles originate from UV and mass
singularities.  These contributions can be extracted from one- and
two-loop Feynman diagrams within the 't~Hooft--Feynman gauge, using
the technique introduced in \citere{Denner:2006jr}.  In
\refse{se:masslessfer} we review this method for the case of massless
fermion scattering.  The new aspects that emerge in the presence of
massive fermions are discussed in \refse{se:massivefer}.  Finally, in
\refse{se:formfactor}, we report on a calculation of the fermionic
form factor as a check of the validity of our methods.

\subsection{Massless fermions}\label{se:masslessfer}%\refeq{se:masslessfer}
Here we give a concise summary of the method presented in
\citere{Denner:2006jr}.  For a detailed discussion we refer to the
original paper.

\subsubsection{Origin of mass singularities}%
\label{se:masslessmasssing}%\refeq{se:masslessmasssing}
Mass singularities appear in loop diagrams involving soft and/or
collinear gauge bosons that couple to external legs.
At one loop, the mass singularities of the \mbox{$n$-fermion}
amplitude originate from diagrams of the type 
{ \unitlength 0.6pt\SetScale{0.6}
\beqar\label{oneloopdiag1}%\refeq{oneloopdiag1}
&&
\vcenter{\hbox{\diagonenf{$\leg{i}$}{$\scriptscriptstyle{V}$}{\blob}}}
,
\eeqar
}%
where an electroweak gauge boson, $V=A,Z,W^\pm$, couples to one of the
external fermions ($i=1,\dots,n$) and to either
(i) another external fermion or
\mla
(ii) an internal propagator.
While diagrams of type (ii) produce only single logarithms of
collinear origin, diagrams of type (i) give rise to single and double
logarithms. The latter originate from the region where the gauge-boson
momentum is soft and collinear to one of the external fermions.
At two loops, the following  five types of diagrams 
give rise to NLL mass singularities
{
\unitlength 0.6pt\SetScale{0.6}
\beq\label{twoloopdiag2}%\refeq{twoloopdiag2}
\!\!
\vcenter{\hbox{\diagIInf{$\leg{i}$}{$\leg{j}$}{$\scriptscriptstyle{V_1}$}{$\scriptscriptstyle{V_2}$}{\blob}}}
,
\vcenter{\hbox{\diagIIInf{$\leg{i}$}{$\leg{j}$}{$\scriptscriptstyle{V_1}$}{$\scriptscriptstyle{V_2}$}{\blob}}}
,
\vcenter{\hbox{\diagIVnf{$\leg{j}$}{$\leg{i}$}{$\leg{k}$}{$\scriptscriptstyle{V_2}$}{$\scriptscriptstyle{V_1}$}{\blob}}}
,
\vcenter{\hbox{\diagInf{$\leg{i}$}{$\leg{j}$}{$\scriptscriptstyle{V_1}$}{$\scriptscriptstyle{V_3}$}{$\scriptscriptstyle{V_2}$}{\blob}}}
,
\vcenter{\hbox{\diagself{$\scriptscriptstyle{V_1}$}{$\scriptscriptstyle{V_2}$}{\blob}}}
.
\eeq
}%
Here the NLL mass singularities originate from the regions where the
gauge boson $V_1$ is simultaneously soft and collinear, and the gauge
bosons $V_2$ and $V_3$ in the first four types of diagrams are soft
and/or collinear.

\subsubsection{Factorizable and non-factorizable contributions}
\label{se:factorizable}%\refse{se:factorizable}

The soft--collinear NLL contributions resulting from the diagrams
\refeq{oneloopdiag1} and \refeq{twoloopdiag2} are split into
factorizable and non-factorizable ones.  The one- and two-loop
factorizable (F) parts result from those diagrams where the virtual
gauge bosons couple only to external lines.  Including sums over gauge
bosons and external legs we have { \unitlength 0.6pt\SetScale{0.6}
\beqar\label{oneloopdiag4}%\refeq{oneloopdiag4}
\mel{1}{\fact}
&=&
\frac{1}{2}
\sum_{i=1}^{n}
\sum_{j=1 \atop j\neq i}^{n}
\sum_{V=A,Z,W^\pm}
\left[
\vcenter{\hbox{\diagone{$\leg{i}$}{$\leg{j}$}{$\scriptscriptstyle{V}$}{\factblob}}}
\right]_{q^\mu\to x p^\mu}
,
\eeqar
}%
and
{\unitlength 0.6pt \SetScale{0.6}
\beqar\label{twoloopdiag3}%\refeq{twoloopdiag3}
\lefteqn{
\mel{2}{\fact}
=
\sum_{i=1}^{n}
\sum_{j=1\atop j\neq i}^{n}
\sum_{V_{m}=A,Z,W^\pm}
\left\{
\frac{1}{2}
\left[
\vcenter{\hbox{
\diagI{$\leg{i}$}{$\leg{j}$}{$\scriptscriptstyle{V_1}$}{$\scriptscriptstyle{V_2}$}{\factblob}
}}
+\vcenter{\hbox{
\diagII{$\leg{i}$}{$\leg{j}$}{$\scriptscriptstyle{V_1}$}{$\scriptscriptstyle{V_2}$}{\factblob}
}}
\right]
\right.
}\quad&&\nl&&{}
+
\vcenter{\hbox{
\diagIII{$\leg{i}$}{$\leg{j}$}{$\scriptscriptstyle{V_1}$}{$\scriptscriptstyle{V_3}$}{$\scriptscriptstyle{V_2}$}{\factblob}}}
+\vcenter{\hbox{\diagV{$\leg{i}$}{$\leg{j}$}{$\scriptscriptstyle{V_1}$}{$\scriptscriptstyle{V_2}$}
{\factblob}}}
+\vcenter{\hbox{\diagVII{$\leg{i}$}{$\leg{j}$}{$\scriptscriptstyle{V_1}$}{$\scriptscriptstyle{V_2}$}{\factblob}}}
+\frac{1}{2}
\vcenter{\hbox{\diagself{$\scriptscriptstyle{V_1}$}{$\scriptscriptstyle{V_2}$}{\factblob}}}
\nl&&{}
+
\sum_{k=1\atop k\neq i,j}^{n}
\left[
\vcenter{\hbox{\diagXX{$\leg{j}$}{$\leg{i}$}{$\leg{k}$}{$\scriptscriptstyle{V_1}$}{$\scriptscriptstyle{V_2}$}{\factblob}}}
\right.
+\frac{1}{6}
\left.
\vcenter{\hbox{\diagXXI{$\leg{j}$}{$\leg{i}$}{$\leg{k}$}{$\scriptscriptstyle{V_2}$}{$\scriptscriptstyle{V_1}$}{$\scriptscriptstyle{V_3}$}{\factblob}}}
\right]
+\frac{1}{8}
\sum_{k=1\atop k\neq i,j}^{n}
\sum_{l=1\atop l\neq i,j,k}^{n}
\left.
\vcenter{\hbox{\diagXXII{$\leg{i}$}{$\leg{j}$}{$\leg{k}$}{$\leg{l}$}{$\scriptscriptstyle{V_1}$}{$\scriptscriptstyle{V_2}$}{\factblob}}}
\right\}_{q^\mu \to x p^\mu}
\!\!.\nln
\eeqar
}%
The limit $q^\mu\to x p^\mu$ in \refeq{oneloopdiag4} and
\refeq{twoloopdiag3} indicates that the above diagrams are evaluated
in the approximation where each of the four-momenta $q^\mu$ of the
various gauge bosons is collinear to one of the momenta $p^\mu$ of the
external legs or soft.  Where relevant, also the contributions of hard
regions are taken into account (see \refse{se:masslessferUV}).  The
label F in the $n$-fermion tree subdiagrams in \refeq{oneloopdiag4}
and \refeq{twoloopdiag3} indicates that, by definition, the
factorizable contributions include only those parts of the above
diagrams that are obtained by performing the loop integration with the
momenta $q_m$ of the gauge bosons $V_m$ set to zero in the $n$-fermion
tree subdiagrams.

These contributions are called factorizable since their logarithmic
soft--collinear singularities factorize from the $n$-fermion tree
amplitude (see below).  The remaining contributions are called
non-factorizable (NF).  They comprise those diagrams of type
\refeq{oneloopdiag1} and \refeq{twoloopdiag2} that involve gauge
bosons $V_i$ coupling to internal (hard) propagators, and the
non-factorizable parts of the diagrams in \refeq{oneloopdiag4} and
\refeq{twoloopdiag3}.  In \citere{Denner:2006jr}, using collinear Ward
identities, it was explicitly shown that all non-factorizable one- and
two-loop contributions cancel at the amplitude level.

\subsubsection{Soft--collinear approximation}%
\label{se:scapprox}%\refeq{se:scapprox}
The soft--collinear singularities are extracted from the above Feynman
diagrams using the following soft--collinear approximation for the
interactions of virtual gauge bosons $V_1 \dots V_n$ with an incoming
(anti)fermion line $i$
\beqar\label{collapp1}%\refeq{collapp1}
\lim_{q_k^\mu\to x_k p_i^\mu }
\vcenter{\hbox{
\unitlength 0.6pt \SetScale{0.6}
\diagIIcollin{$\scriptstyle{\bar V_n^{\mu_n}}$\footnotesize{\dots}
$\scriptstyle{\bar V_1^{\mu_1}}$}{}{$\scriptstyle{i}$}%
{$\scriptstyle{V_n^{\mu_n}}$}{$\scriptstyle{V_1^{\mu_1}}$}{\blob}
}}
&=&
\vcenter{\hbox{
\unitlength 0.6pt \SetScale{0.6}
\diagIIIcollin{$\scriptstyle{\bar V_n^{\mu_n}}$\footnotesize{\dots}
$\scriptstyle{\bar V_1^{\mu_1}}$}{}{$\scriptstyle{i}$}%
{$\scriptstyle{V_n^{\mu_n}}$}{$\scriptstyle{V_1^{\mu_1}}$}{\blob}
}}
\hspace{-2mm}
\times
\frac{-2 e I^{V_n}_i
(p_i+\tilde q_n)^{\mu_n}
}{(p_i+\tilde q_n)^2}
\cdots
\frac{-2 e I^{V_1}_i
(p_i+q_1)^{\mu_1}
}{(p_i+q_1)^2}
.
\nln
\eeqar
Here $q_k$ are the loop momenta of the gauge bosons $V_k$, and $\tilde
q_k=q_1+\dots+q_k$.  In practice, the coupling of each soft/collinear
gauge boson gives rise to a factor $-2e
I^{V_k}_i(p_i+\tilde{q_k})^{\mu}$.  When applied to the factorizable
diagrams in \refeq{oneloopdiag4}--\refeq{twoloopdiag3}, this
approximation removes all Dirac matrices occuring along the fermionic
lines and yields factorized expressions of the form
\beqar\label{factorizedform}%\refeq{factorizedform}
\mel{1,2}{\fact}&=& \mel{0}{}\, K_{1,2},
\eeqar
where the NLL corrections $ K_{1,2}$, involving 
coupling factors and logarithmically divergent loop integrals,
factorize from the \mbox{$n$-fermion} tree amplitude $\mel{0}{}$.

As discussed in \citere{Denner:2006jr}, the soft--collinear
approximation \refeq{collapp1} provides a correct description of the
gauge-boson--fermion couplings in all soft/collinear regions that are
relevant for an NLL analysis at one and two loops, with the following
exception: The soft--collinear approximation is not applicable to
topologies of type {
\beqar
\label{oneloopsubd1}%\refeq{oneloopsubd1}
\unitlength 0.6pt\SetScale{0.6}
\vcenter{\hbox{\diagself{$\scriptscriptstyle{V}$}{$\scriptscriptstyle{V'}$}{\factblob}}}
,
\vcenter{\hbox{
\unitlength 0.6pt\SetScale{0.6}
\subloopII{$\leg{i}$}{$\leg{j}$}{$\scriptscriptstyle{V}$}{\factblob}}}
,
\vcenter{\hbox{
\unitlength 0.6pt\SetScale{0.6}
\subloopI{$\leg{i}$}{$\leg{j}$}{$\scriptscriptstyle{V}$}{\factblob}}}
,
\eeqar
}%
which contain UV singularities associated with a subdiagram with
characteristic scale ${\mu_{\mathrm{loop}}^2}\ll Q^2$ (see next
section) or give rise to power singularities of $\order(1/M^2)$.  For
these diagrams, the soft--collinear approximation \refeq{collapp1} can
be applied only to the vertices that occur outside the one-loop
subdiagrams that are depicted as grey blobs in \refeq{oneloopsubd1},
whereas for the vertices and propagators inside the one-loop
subdiagrams we have to apply the usual Feynman rules.

To bring these diagrams in the factorized form \refeq{factorizedform},
we have to simplify the Dirac matrices along the fermionic lines where
the soft--collinear approximation is not applicable.  To this end we
utilize trace projectors that reduce these Dirac matrices to scalar
quantities.  In practice, for the second and third diagram in
\refeq{oneloopsubd1}, the Dirac matrices along the line $i$ are
eliminated using~\cite{Denner:2006jr}
\beqar\label{projection1}%\refeq{projection1}
X u(p_i,\kappa_i)
=
\sum_\rho \omega_\rho \Pi_{ij}\left(\omega_{\rho} X\right)
u(p_i,\kappa_i)
=
\Pi_{ij}\left(\omega_{\kappa_i} X\right)
u(p_i,\kappa_i)
,
\eeqar
where $X$ represents the diagram without 
$\vcenter{\hbox{
\unitlength 0.6pt\SetScale{0.6}
\begin{picture}(25.,12.)(-15.,-6.)
{\factblob}
\end{picture}
}}$
and external spinors, and the projector is defined as
\beqar\label{projectiondef}%\refeq{projectiondef}
\Pi_{ij}\left(\Gamma\right)
=
\frac{1}{2 p_i p_j}
\Tr\left[
\Gamma
\ps_i\ps_j
\right]
\eeqar
if $m_i=m_j=0$.
For a detailed discussion we refer to
Sect.~3.2 of \citere{Denner:2006jr}.

\subsubsection{Treatment of ultraviolet singularities}%
\label{se:masslessferUV}%\refse{se:masslessferUV}

In addition to logarithmic soft--collinear singularities, also the
logarithms originating from UV singularities need to be taken into
account.  Moreover, the soft--collinear approximation \refeq{collapp1}
can give rise to fake UV logarithms that must carefully be avoided or
subtracted.

Since UV singularities produce only a single logarithm per loop, in
NLL approximation we need to consider only UV-divergent diagrams of
one-loop order and their insertion in one-loop diagrams
\refeq{oneloopdiag4} involving leading soft--collinear singularities.
The UV logarithms originate from the UV cancellations between bare
diagrams and counterterms as
\beq\label{uvsing1}%\refeq{uvsing1}
\left(\frac{\muD^2}{Q^2}\right)^{\veps}
\Biggl[
\underbrace{\frac{1}{\veps}
\left(\frac{Q^2}{\mu_{\mathrm{loop}}^2}\right)^{\veps}}_{\mbox{bare diagrams}}
-
\underbrace{\frac{1}{\veps}
\left(\frac{Q^2}{\mu_{\mathrm{R}}^2}\right)^{\veps}}_{\mbox{counterterms}}
\Biggr]
=
\ln\left(\frac{\mu_{\mathrm{R}}^2}{\mu_{\mathrm{loop}}^2}\right)
+\order(\veps),
\eeq
where $\mu_{\mathrm{loop}}$ is the characteristic scale of the
UV-singular loop diagram, $\mu_{\mathrm{R}}$ is the renormalization
scale, and the term $\left({\muD^2}/{Q^2}\right)^{\veps}$---that we
always absorb in $\alphaeps$ [see \refeq{pertserie1a}]---is
factorized.  In our calculation, the UV poles of bare loop diagrams
and counterterms are removed by means of a minimal subtraction at the
scale $Q^2$.  As a result, the UV divergent terms take the form
\beq\label{uvsing2}%\refeq{uvsing2}
\left(\frac{\muD^2}{Q^2}\right)^{\veps}
\Biggl\{
\underbrace{\frac{1}{\veps}
\left[
\left(\frac{Q^2}{\mu_{\mathrm{loop}}^2}\right)^{\veps}
-1\right]}_{\mbox{bare diagrams}}
-
\underbrace{\frac{1}{\veps}
\left[
\left(\frac{Q^2}{\mu_{\mathrm{R}}^2}\right)^{\veps}-1\right]}_{\mbox{counterterms}}
\Biggr\}
,
\eeq
which is obviously equivalent to \refeq{uvsing1}.  The advantage of
this subtraction is that for all bare (sub)diagrams with
$\mu_{\mathrm{loop}}^2\sim Q^2$ the UV singularities
${\veps}^{-1}[(Q^2/{\mu_{\mathrm{loop}}^2})^{\veps}-1]$ do not produce
large logarithms and are thus negligible.  This holds also for fake UV
singularities resulting from the soft--collinear approximation.  Thus,
apart from the counterterms, we must consider only those (subtracted)
UV contributions that originate from bare (sub)diagrams with
${\mu_{\mathrm{loop}}^2}\ll Q^2$.  At one loop this condition is never
realized in the high-energy limit \refeq{asymptoticregion}, and in
practice we need to consider only the two-loop UV contributions which
result from the diagrams in \refeq{oneloopdiag4} through insertion of
UV-divergent subdiagrams in the lines that are not hard
(${\mu_{\mathrm{loop}}^2}\ll Q^2$).  These contributions correspond to
the diagrams depicted in \refeq{oneloopsubd1}, for which we
use---instead of the soft--collinear approximation---the projectors
\refeq{projection1}--\refeq{projectiondef}, thereby ensuring a correct
description of the UV regions.

\subsection{Massive fermions}\label{se:massivefer}%\refse{se:massivefer}
The method outlined in \refse{se:masslessfer} is applicable also to
processes involving external top and/or bottom quarks.  However, in
the presence of massive fermions, two new aspects must be taken into
account: mass terms in the top-quark propagators and new diagrams
resulting from Yukawa interactions.

\subsubsection{Top-mass terms}\label{se:massterms}%\refse{se:massterms}
Let us recall that, in the high-energy limit \refeq{asymptoticregion},
the top-quark mass is treated as a small parameter and terms of
$\order (\mathrm{NLL}\times \Mt/Q)$ are systematically neglected.  In
this approximation, only the following types of $\Mt$-terms must be
considered:
\begin{itemize}
\item[(i)] Yukawa couplings proportional to $\Mt/\MW$;
\item[(ii)] $\Mt$-terms  acting as collinear regulators in the
denominator of propagators;
\item[(iii)] $\Mt$-terms in the numerator of loop integrals of $\order
  (\mathrm{NLL}/\Mt)$ or $\order (\mathrm{NLL}/M_{\PW,\PZ,\PH})$, \ie
  integrals that involve power singularities.
\end{itemize}
The majority of the $\Mt$-terms occuring in one- and two-loop diagrams
does not belong to these categories and can be set to zero.  In
particular, by explicit inspection of the relevant one- and two-loop
integrals, we find that $\Mt$-terms of type (iii) occur only inside
the one-loop insertions in the diagrams \refeq{oneloopsubd1}.  Apart
from these special cases all other $\Mt$-terms in the numerators of
loop integrals can be set to zero.

\subsubsection{NLL contributions from gauge interactions}
\label{se:NLLfromgauge}%\refse{se:NLLfromgauge}

As in the case of massless fermion scattering, the diagrams
\refeq{oneloopdiag1}--\refeq{twoloopdiag2} with external-leg
interactions of soft/collinear gauge bosons produce logarithmic mass
singularities.  These diagrams are split into factorizable and
non-factorizable parts as discussed in \refse{se:factorizable} and, in
the presence of massive fermions, the Ward identities that are
responsible for the cancellation of the non-factorizable
contributions~\cite{Denner:2006jr} receive only negligible corrections
of $\order (\Mt/Q)$.  Thus, as in the case of massless fermions, the
non-factorizable parts do not contribute and we can restrict ourselves
to the calculation of the factorizable parts \refeq{oneloopdiag4} and
\refeq{twoloopdiag3}.

The soft--collinear approximation \refeq{collapp1} can easily be
adapted to massive fermions by including the $\Mt$-terms in the
denominator of the top-quark propagators and leaving the numerator as
in the massless case.  In principle $\Mt$-terms modify also the
numerator of \refeq{collapp1}.  However, as observed above, such terms
are relevant only in the one-loop insertion diagrams in
\refeq{oneloopsubd1}.  Here, as discussed in \refse{se:scapprox}, we
employ the usual Feynman rules---and not the soft--collinear
approximation---due to the presence of UV divergences or power
singularities of $\order(1/M^2)$.

To bring the diagrams of type \refeq{oneloopsubd1} in the factorized
form \refeq{factorizedform}, we utilize projections analogous to
\refeq{projection1}.  In the case of massive fermions, using the two
projectors
\beqar\label{projectiondef2}%\refeq{projectiondef2}
\Pi_{ij}\left(\Gamma\right)
=
\frac{2 p_i p_j}{(2 p_i p_j)^2-4m_i^2m_j^2}
\Tr\left[
\Gamma
(\ps_i+m_i)
\left(\ps_j-\frac{m_im_j^2}{p_ip_j}\right)
\right]             
\eeqar
and
\beqar\label{projectiondef3}%\refeq{projectiondef3}
\tilde\Pi_{ij}\left(\Gamma\right)
=
\frac{2 p_i p_j}{(2 p_i p_j)^2-4m_i^2m_j^2}
\Tr\left[
\Gamma
(\ps_i+m_i)
\left(1-\frac{\ps_i\ps_j}{p_ip_j}\right)
\right]
\eeqar
we obtain
\beqar\label{projection2}%\refeq{projection2}
X u(p_i,\kappa_i)
&=&
\sum_\rho \omega_\rho\left[
\Pi_{ij}\left(\omega_\rho X\right)
+
\tilde\Pi_{ij}\left(\omega_\rho X\right)\ps_j
\right]u(p_i,\kappa_i).
\eeqar
Here, in contrast to the massless case, the Dirac matrices are not
projected out completely and a term proportional to
$\tilde\Pi_{ij}(\omega_\rho X) \ps_j$ remains, which cannot be cast in
the factorized form \refeq{factorizedform}. However, after explicit
evaluation of the loop integration, we find that these
$\tilde\Pi_{ij}$-terms are suppressed in the high-energy limit and
only the factorizable terms associated with the projector $\Pi_{ij}$
contribute.

The projectors \refeq{projectiondef2}--\refeq{projection2} are
applicable in the case where particle $i$ is an incoming fermion.  For
antifermions similar projectors can be constructed, which differ only
in the sign of the $m_i$-terms.  As discussed in
\refse{se:gaugeandyuk}, this difference is irrelevant since only
squared mass terms produce NLL contributions. Thus, the results
derived for fermions are directly applicable to antifermions.

All logarithms of UV origin---also in the presence of Yukawa
interactions---are treated by means of a minimal subtraction at the
scale $Q^2$ as explained in \refse{se:masslessferUV}.

\subsubsection{NLL contributions from Yukawa interactions}
\label{se:NLLfromYukawa}%\refse{se:NLLfromYukawa}

The interaction of external bottom or top legs with scalar bosons is
suppressed in all soft/collinear regions. Thus, Yukawa interactions
can produce NLL contributions only through the counterterms (see
\refse{se:1loopren}) and the topologies of type \refeq{oneloopsubd1},
which contain UV singularities or power singularities of
$\order(1/M^2)$.
In addition to the diagrams of type \refeq{oneloopsubd1} that are
present in the massless case, the following new bare diagrams must be
taken into account: {\unitlength 0.6pt \SetScale{0.6}
\beqar\label{twoloopdiagYuk}%\refeq{twoloopdiagYuk}
\mel{2}{\mathrm{Yuk}}
&=&
\sum_{i=1}^{n}
\sum_{j=1\atop j\neq i}^{n}
\sum_{V_{m}=A,Z,W^\pm}
\sum_{\Phi_{l}=H,\chi,\phi^\pm}
\left\{
\vcenter{\hbox{\diagVYuk{$\leg{i}$}{$\leg{j}$}{$\scriptscriptstyle{V_1}$}{$\scriptscriptstyle{\Phi_2}$}
{\factblob}}}
+\vcenter{\hbox{\diagVIIYuk{$\leg{i}$}{$\leg{j}$}{$\scriptscriptstyle{V_1}$}{$\scriptscriptstyle{\Phi_2}$}{\factblob}}}
\right.\nl&&{} \left.
+
\vcenter{\hbox{
\diagIIIYuka{$\leg{i}$}{$\leg{j}$}{$\scriptscriptstyle{\Phi_1}$}{$\scriptscriptstyle{V_3}$}{$\scriptscriptstyle{\Phi_2}$}{\factblob}}}
+
\vcenter{\hbox{
\diagIIIYukb{$\leg{i}$}{$\leg{j}$}{$\scriptscriptstyle{V_1}$}{$\scriptscriptstyle{V_3}$}{$\scriptscriptstyle{\Phi_2}$}{\factblob}}}
+
\vcenter{\hbox{
\diagIIIYukc{$\leg{i}$}{$\leg{j}$}{$\scriptscriptstyle{\Phi_1}$}{$\scriptscriptstyle{V_3}$}{$\scriptscriptstyle{V_2}$}{\factblob}}}
\right\}
%_{q^\mu \to x p^\mu}
.\eeqar
}%
By explicit inspection we find that the last two diagrams in
\refeq{twoloopdiagYuk} are suppressed in NLL accuracy since the
$VV\Phi$ couplings of $\order(M_{\PW,\PZ})$ are not compensated by
$1/M$ terms from the loop integrals.  The NLL terms resulting from the
first three diagrams in \refeq{twoloopdiagYuk} are worked out in
\refse{se:res2loopYukawa}.

\subsection{Form-factor checks}
\label{se:formfactor}%\refse{se:formfactor}

When evaluating the factorizable contributions of arbitrary
\mbox{$n$-fermion} processes, we have to eliminate the Dirac matrices
along the fermionic lines in order to separate the loop integrals from
the tree amplitude~$\mel{0}{}$. For this purpose we use the
soft--collinear approximation for gauge interactions presented in
\refse{se:scapprox} and the fact that the Yukawa interactions are
suppressed in all soft/collinear regions (see
\refse{se:NLLfromYukawa}).

We have checked the validity of this procedure for the case of the
form factor which couples a fermion--antifermion pair to an external
Abelian field.  In our approach, the radiative corrections to this
form factor are given by the factorizable contributions
\refeq{oneloopdiag4}, \refeq{twoloopdiag3}, and \refeq{twoloopdiagYuk}
with $n=2$ external fermions (in this case the diagrams with three and
four legs in \refeq{twoloopdiag3} do not contribute).  Alternatively,
we have calculated the one- and two-loop corrections to the form
factor in the high-energy limit including all Feynman diagrams,
without neglecting \mbox{$\Mt$-terms} in the numerator, and employing
projection techniques (see \eg \citere{Bernreuther:2004ih}) instead of
the soft--collinear approximation.  After performing the minimal
subtraction of the UV divergences as explained in
\refse{se:masslessferUV}, the results of the complete form-factor
calculation agree with the ones resulting from the factorizable
contributions \refeq{oneloopdiag4}, \refeq{twoloopdiag3}, and
\refeq{twoloopdiagYuk} in soft--collinear approximation.  We have also
verified by explicit evaluation that, as expected from the discussion
in \refse{se:NLLfromYukawa}, the form-factor diagrams with Yukawa
interactions that are not included in \refeq{twoloopdiagYuk} are
either suppressed or vanish in NLL accuracy.
Two of these additional Yukawa diagrams%
\footnote{These are the one-loop diagram \refeq{diagram0} with the
 gauge boson $V_1$ replaced by a scalar boson and the two-loop
 diagram \refeq{diagram1} with the gauge boson $V_2$ replaced by a
 scalar boson.}
yield non-vanishing NLL contributions which are, however, of pure UV
origin and are completely removed by the minimal UV subtraction.

\section{One-loop results}
\label{se:res1loop}%\refse{se:res1loop}

The one-loop amplitude gets contributions from the factorizable
one-loop diagrams \refeq{oneloopdiag4} and from counterterms,
\beqar\label{onelooprena}%\refeq{onelooprena}
\nmel{1}{}
&=&
\nmel{1}{\mathrm{F}} 
+
\nmel{1}{\mathrm{CT}} 
.
\eeqar
The results for the loop diagrams are summarized in
\refapp{se:oneloop}.  Here we discuss the one-loop renormalization and
list the final results for the renormalized one-loop amplitude.  As
shown in \citere{Denner:2001jv} the one-loop NLL corrections
factorize, \ie they can be expressed through correction factors that
multiply the Born amplitude. Moreover, they can be split in a
symmetric-electroweak part, an electromagnetic part, which in
particular contains all soft/collinear singularities associated with
photons, and an $\MZ$-dependent part that results from the difference
in the W- and Z-boson masses.  This splitting permits to separate the
soft/collinear singularities resulting from photons in a
gauge-invariant way and is very important in view of the discussion of
the two-loop contributions.

\subsection{One-loop renormalization}
\label{se:1loopren}%\refse{se:1loopren}

As discussed in \citere{Denner:2006jr}, mass renormalization is not
relevant in NLL approximation.  Thus, counterterm contributions result
only from the renormalization of the electroweak gauge couplings and
the top-quark Yukawa coupling,
\beqar\label{PRpertserie1}%\refeq{PRpertserie1}
g_{k,0}
&=&
g_k+\sum_{l=1}^\infty
\left(\frac{\alphaeps}{4\pi}\right)^l
\de g^{(l)}_k
,\quad
e_0
=
e+\sum_{l=1}^\infty
\left(\frac{\alphaeps}{4\pi}\right)^l
\de e^{(l)}
,\quad
\lambda_{t,0}
=
\lambdat+\sum_{l=1}^\infty
\left(\frac{\alphaeps}{4\pi}\right)^l
\de \lambdat^{(l)}
,\qquad
\eeqar
and from the renormalization constants associated with the wave
functions of the external fermions $i=1,\dots,n$,
\beqar\label{WFpertserie1}%\refeq{WFpertserie1}
Z_i
&=&
1+\sum_{l=1}^\infty
\left(\frac{\alphaeps}{4\pi}\right)^l
\de Z^{(l)}_i.
\eeqar

All couplings are renormalized in the $\overline{\mathrm{MS}}$ scheme,
but we moreover subtract the UV singularities both in the bare and the
counterterm contributions as explained in \refse{se:masslessferUV}.
The counterterm for the Yukawa coupling can be determined from the
divergent parts of the counterterms to $g_2$, $\MW$ and
$\Mt$ using tree-level relations. Assuming that the renormalization scale%
\footnote{We do not identify the renormalization scale $\muR$ and the
  scale of dimensional regularization $\muD$.}  $\muR$ is of the order
of or larger than $\MW$, we find for the counterterms
\beqar\label{subcouplingren}%\refeq{subcouplingren}
\de g_{k}^{(1)} &\NLLA&
-\frac{g_k}{2}
\frac{1}{\veps}
\betacoeff{k}^{(1)}
\left[\left(\frac{Q^2}{\muR^2}\right)^\veps-1\right]
,\qquad
\de e^{(1)} \NLLA
-\frac{e}{2}
\frac{1}{\veps}
\betacoeff{e}^{(1)}
\left[\left(\frac{Q^2}{\muR^2}\right)^\veps-1\right],\nl
\de \lambdat^{(1)} &\NLLA&
-\frac{\lambdat}{2}
\frac{1}{\veps}
\betacoeff{\lambdat}^{(1)}
\left[\left(\frac{Q^2}{\muR^2}\right)^\veps-1\right]
,
\eeqar 
where the dependence on the factor $(Q^2/\muR)^\veps$ is due to the
normalization of the expansion parameter $\alphaeps$ in
\refeq{PRpertserie1}, and the one-loop $\be$-function coefficients in
the electroweak Standard Model ($Y_\Phi=1$) are given by
\beq\label{betacoeffres}%\refeq{betacoeffres}
\betacoeff{1}^{(1)}=-\frac{41}{6\cw^2}
,\qquad
\betacoeff{2}^{(1)}=\frac{19}{6\sw^2}
,\qquad
\betacoeff{e}^{(1)}=-\frac{11}{3}
,\qquad
\betacoeff{\lambdat}^{(1)}=
\frac{9}{4\sw^2}+\frac{17}{12\cw^2}-\frac{9\Mt^2}{4\sw^2\MW^2}
.
\eeq
For later convenience we also define the QED $\be$-function
coefficient, which is determined by the light-fermion contributions
only,
\beqar\label{eq:betacoeffQED}%\refeq{eq:betacoeffQED}
\label{eq:betacoeffQEDtop}%\refeq{eq:betacoeffQEDtop}
\betacoeff{\QED}^{(1)}=
-\frac{4}{3}\sum_{f\neq t}N_{\mathrm{c}}^f Q_f^2=-\frac{80}{9},
\eeqar
where $N_{\mathrm{c}}^f$ represents the colour factor, \ie
$N_{\mathrm{c}}^f=1$ for leptons and $N_{\mathrm{c}}^f=3$ for quarks.
%More details about the $\be$-function coefficients can be found in
%App.~C of \citere{Denner:2006jr} or in
%\citeres{Pozzorini:rs,Pozzorini:2004rm}.
The renormalization of the mixing parameters $\cw$ and $\sw$ can be
determined via \refeq{massspectrum} from the renormalization of the
coupling constants. 

The contribution of the counterterms \refeq{subcouplingren} can easily
be absorbed into the Born amplitude
\beq\label{bornrenormalization}%\refeq{bornrenormalization}
\melQ{0}{}\equiv
\mel{0}{}
\bigg|_{g_k=g_k(Q^2),\; e=e(Q^2),\; \lambdat=\lambdat(Q^2)}
\eeq
via the running couplings
\beqar\label{bornparamren}%\refeq{bornparamren}
g_{k}^2(Q^2)
&\NLLA&
g_{k}^2(\muR^2) \left\{ 1
-\frac{\alphaeps}{4\pi}\betacoeff{k}^{(1)}
\frac{1}{\veps}\left[\left(\frac{Q^2}{\muR^2}\right)^\veps-1\right
]
\right\}
=g_{k}^2(\muR^2) \left\{ 1
-\frac{\alphaeps}{4\pi}\betacoeff{k}^{(1)}
\ln\left(\frac{Q^2}{\muR^2}\right) + \order(\veps)
\right\}
,\nl
e^2(Q^2)
&\NLLA&
e^2(\muR^2) \left\{ 1
-\frac{\alphaeps}{4\pi}\betacoeff{e}^{(1)}
\frac{1}{\veps}\left[\left(\frac{Q^2}{\muR^2}\right)^\veps-1\right
]
\right\}
=e^2(\muR^2) \left\{ 1
-\frac{\alphaeps}{4\pi}\betacoeff{e}^{(1)}
\ln\left(\frac{Q^2}{\muR^2}\right) + \order(\veps)
\right\}
,\nl
\lambdat^2(Q^2)
&\NLLA&
\lambdat^2(\muR^2)
\left\{ 1
-\frac{\alphaeps}{4\pi}
\betacoeff{\lambdat}^{(1)}
\frac{1}{\veps}\left[\left(\frac{Q^2}{\muR^2}\right)^\veps-1\right
]
\right\}
=\lambdat^2(\muR^2) \left\{ 1
-\frac{\alphaeps}{4\pi}\betacoeff{\lambdat}^{(1)}
\ln\left(\frac{Q^2}{\muR^2}\right) + \order(\veps)
\right\}
.\nln
\eeqar
For practical applications one can use $\MSbar$ input parameters at
the scale $\muR=\MZ$, or alternatively the on-shell input parameters
\mla
$\alpha(\MZ)$, $\MZ$ and $\MW$, or the $G_\mu$ input scheme.  These
different schemes are based on input parameters at the electroweak
scale and are thus equivalent in NLL approximation.  In the following
we express all one- and two-loop results in terms of the Born
amplitude at the scale $Q^2$.  The notation $\melQ{0}{}$ emphasizes
the fact that the Born amplitude implicitly depends on logarithms
$\ln(Q^2/\muR^2)$ via the running of the couplings.

In this setup, the only one-loop NLL counterterm contribution arises
from the on-shell wave-function renormalization constants $\de
Z^{(1)}_{i}$ for the fermionic external legs.  All legs receive
contributions from massive weak bosons, whereas the photonic
contribution to the counterterm for massless external fermions
vanishes owing to a cancellation between UV and mass singularities
within dimensional regularization.  The legs with top or bottom quarks
additionally get Yukawa contributions.  After subtraction of the UV
poles we find
\beqar\label{subWFRC}%\refeq{subWFRC}
\de Z^{(1)}_{i}&\NLLA&
-
\frac{1}{\veps}
\left\{
\sum_{V=Z,W^\pm} 
I_i^{\bar{V}}
I_i^{{V}}
\left[\left(\frac{Q^2}{M_V^2}\right)^\veps-1\right]
+
I_i^{A}
I_i^{{A}}
\left[3 \deltat{i} \left(\frac{Q^2}{\Mt^2}\right)^\veps-1\right]
\rule{0pt}{4ex}\right.
\nn\\*&&\left.\rule{0pt}{4ex}{}
{}+
\zYuk_i \,
\frac{\lambdat^2}{2e^2} 
\left[\left(\frac{Q^2}{\Mt^2}\right)^\veps-1\right]
\right\}
.
\eeqar
The Kronecker symbol $\deltat{i}$ is defined in \refeq{kronecker}.
Compared with the massless case two extra terms appear, one related to
the massive top quark in photonic diagrams and one originating from
the Yukawa couplings.  The Yukawa factors~$\zYuk_i$ are obtained from
\beqar
\sum_{\Phi=H,\chi,\phi^\pm}
\hat{G}_i^{\Phi^+} G_i^{\Phi}
&=&
\zYuk_i \lambdat^2 %\frac{\gw^2 \Mt^2}{2\MW^2}
\eeqar
and read
\beq\label{Yukawafactors}%\refeq{Yukawafactors}
\zYuk_i= \left\{\barr{l}
1, \mbox{ for left-handed third-generation quarks }
    \varphi_i = t^\rL, \bar t^\rL, b^\rL, \bar b^\rL,
\nl
2, \mbox{ for right-handed top quarks},
    \varphi_i = t^\rR, \bar t^\rR,
\nl
0, \mbox{ otherwise}.
\earr\right.
\eeq
Finally, the one-loop counterterm for a process with $n$ external
fermions in NLL approximation is obtained as
\mda
\beqar\label{WFren}%\refeq{WFren}
\nmel{1}{\mathrm{CT}}
 &=&
\melQ{0}{}\,
\sum_{i=1}^{n}
\frac{1}{2}
\de Z^{(1)}_{i} .
\eeqar
\mua

\subsection{Renormalized one-loop amplitude}
\label{se:oneloopres}%\refeq{se:oneloopres}

Inserting the results from \refeq{onelooprepunr}, \refeq{onelooprepu} and
\refeq{subWFRC}--\refeq{WFren} into \refeq{onelooprena}, we can write
for the renormalized one-loop matrix element for a process with $n$
external fermions 
\beqar\label{onelooprep}%\refeq{onelooprep}
\nmel{1}{}&\NLLA&
\melQ{0}{}\,
\left[
\FF{1}{\sew}
+\Delta \FF{1}{\elm}
+\Delta \FF{1}{\PZ}
\right].
\eeqar
Here the corrections are split into a symmetric-electroweak (sew) part,
\beqar\label{onelooprepa1}%\refeq{onelooprepa1}
\FF{1}{\sew}
&\NLLA&
-\frac{1}{2}\sum_{i=1}^{n}\sum_{j=1\atop j\neq i}^{n}
\sum_{V=A,Z,W^\pm} 
I_i^{\bar{V}}I_j^{{V}}\,\univfact{\veps}{\MW;p_i,p_j}
-\frac{\lambdat^2}{4e^2}
\, C(\Mt)
\sum_{i=1}^{n} \zYuk_i
,\qquad
\eeqar
which is obtained by setting the masses of all gauge bosons, $\PA,\PZ$
and $\PW^\pm$, equal to $\MW$ everywhere, an electromagnetic (em) part
\beqar\label{onelooprepa2}%\refeq{onelooprepa2}
\Delta \FF{1}{\elm}
&=&
-\frac{1}{2}\sum_{i=1}^{n}\sum_{j=1\atop j\neq i}^{n}
I_i^{A}I_j^{{A}}\,\Delta\univfact{\veps}{0;p_i,p_j}
,
\eeqar
resulting from the mass gap between the W boson and the massless photon,
and an $\MZ$-dependent part
\beqar\label{onelooprepa3}%\refeq{onelooprepa3}
\label{onelooprepalast}%\refeq{onelooprepalast}
\Delta \FF{1}{\PZ}
&=&
-\frac{1}{2}\sum_{i=1}^{n}\sum_{j=1\atop j\neq i}^{n}
I_i^{Z}I_j^{{Z}}\,\Delta\univfact{\veps}{\MZ;p_i,p_j}
,
\eeqar
describing the effect that results from the difference between $\MW$
and $\MZ$.
The functions
\beq
\Delta\univfact{\veps}{m;p_i,p_j}=
\univfact{\veps}{m;p_i,p_j}-
\univfact{\veps}{\MW;p_i,p_j}
,
\eeq
which are associated with the exchange of gauge bosons of mass $m$,
describe the effect resulting from the difference between $m$ and
$\MW$.

We have expressed the large logarithms in the Yukawa contribution
\beq
C(\Mt)\NLLA L_\Pt + \frac{1}{2}L_\Pt^2\Eps{} +\frac{1}{6}L_\Pt^3\Eps{2} +
\order(\Eps{3}),
\eeq
which originates only from the counterterm \refeq{subWFRC}, through
their natural scale~$\Mt$ in $L_\Pt=\ln(Q^2/\Mt^2)$, while all the
gauge-interaction contributions are written in terms of
$L=\ln(Q^2/\MW^2)$ as usual.  In order to build squared one-loop
expressions to order $\veps^0$, which enter the two-loop amplitude, we
need to expand all one-loop terms up to order $\veps^2$.  For the
functions $I$ and $\Delta I$ we obtain
\beqar\label{Ifunc}%\refeq{Ifunc}
\univfact{\veps}{\MW;p_i,p_j}&\NLLA&
-L^2
-\frac{2}{3}L^3\Eps{}
-\frac{1}{4}L^4\Eps{2}
+({3}-2\Lrij)
\left(L
+\frac{1}{2}L^2\Eps{}
+\frac{1}{6}L^3\Eps{2}
\right)
+\order(\veps^3)
,\nl
\Delta\univfact{\veps}{\MZ;p_i,p_j}&\NLLA&
%\univfact{\veps}{\MW;p_i,p_j}+
\LMZW\left(
2L
+2L^2\Eps{}
+L^3\Eps{2}
\right)
+\order(\veps^3)
,\nl
\Delta\univfact{\veps}{0;p_i,p_j}&\NLLA&
\left(\deltaz{i}+\deltaz{j}\right)
  \biggl(
  -\Epsinv{2} + \frac{1}{2}L^2 + \frac{1}{3}L^3\Eps{}
  + \frac{1}{8}L^4\Eps{2}
  \biggr)
+ \left(\deltat{i}+\deltat{j}\right)
  \biggl(
  L\Epsinv{1}
  + L^2
\nn\\*&&{}
  + \frac{1}{2}L^3\Eps{}
  + \frac{1}{6}L^4\Eps{2}
  \biggr)
+ \biggl[ 2\Lrij - \frac{3}{2}\left(\deltaz{i}+\deltaz{j}\right)
  - \deltat{i}(1+\Lrmi)
\nn\\*&&{}
  - \deltat{j}(1+\Lrmj) \biggr]
  \biggl(
  \Epsinv{1}
  +L
  +\frac{1}{2}L^2\Eps{}
  +\frac{1}{6}L^3\Eps{2}
  \biggr)
+ \order(\Eps{3})
.
\eeqar
Only the function $\Delta\univfact{\veps}{0;p_i,p_j}$, which
incorporates the interactions with massless photons, depends on the
fermion masses $m_{i}$ through the symbols $\deltaz{i}$ and
$\deltat{i}$ defined in \refeq{kronecker}.  The functions
$\univfact{\veps}{\MW;p_i,p_j}$ and
$\Delta\univfact{\veps}{\MZ;p_i,p_j}$, which correspond to the
exchange of massive gauge bosons, are independent of the fermion
masses and agree with the results for massless fermions in
\citere{Denner:2006jr}.

The functions $I$ and $\Delta I$ are symmetric with respect to an
exchange of the external legs $i,j$, and apart from the
angular-dependent $\Lrij$-terms, the dependence on $p_i$ and $p_j$ can
be separated:
\beqar\label{Iijsplit}%\refeq{Iijsplit}
  \univfact{\veps}{m;p_i,p_j} \Big|_{\Lrij=0} =
  \frac{1}{2} \Bigl[
    \univfact{\veps}{m;p_i,p_i} + \univfact{\veps}{m;p_j,p_j}
  \Bigr]
,
\eeqar
where $l_{ii}=l_{jj}=0$ is understood.  Under the sums over $i,j$ in
\refeq{onelooprep}--\refeq{onelooprepa3} we can replace
$\univfact{\veps}{m;p_j,p_j} \to \univfact{\veps}{m;p_i,p_i}$, use the
charge-conservation identity \refeq{chargeconservation} and write
\beqar\label{chargeconservationsum}%\refeq{chargeconservationsum}
\melQ{0}{}\,
\sum_{i=1}^{n}\sum_{j=1\atop j\neq i}^{n}
I_i^{\bar{V}}I_j^{{V}}\,
\univfact{\veps}{M_V;p_i,p_j}
\bigg|_{\Lrij=0}
&=&
-
\melQ{0}{}\,
\sum_{i=1}^{n}
I_i^{{V}} I_i^{\bar{V}}\,
\univfact{\veps}{M_V;p_i,p_i}
%\bigg|_{l_{ii}=0}
,\nln
\eeqar
and similarly for $\Delta I$. This relation turns out to be useful
later.

\section{Two-loop results}
\label{se:res2loop}%\refse{se:res2loop}

The renormalized two-loop matrix element gets contributions from the
factorizable two-loop diagrams \refeq{twoloopdiag3} and
\refeq{twoloopdiagYuk}, from wave-function renormalization, and from
parameter renormalization,
\beqar\label{twoloopcomb}%\refeq{twoloopcomb}
\nmel{2}{}
&=&
\nmel{2}{\mathrm{F}}
+\mel{2}{\mathrm{Yuk}}
+\nmel{2}{\mathrm{WF}}
+\nmel{2}{\mathrm{PR}}
.
\eeqar
The results for the two-loop diagrams \refeq{twoloopdiag3}, which
involve only gauge interactions, are summarized in
\refapp{se:twoloop}.  Here we first show that the Yukawa contribution
of the factorizable two-loop diagrams in \refeq{twoloopdiagYuk}
vanishes in NLL accuracy.  Then we list the two-loop renormalization
contributions and finally combine the factorizable contributions from
two-loop diagrams with gauge interactions and the renormalization
contributions into the complete renormalized two-loop amplitude.

As discussed in \refse{se:singtreatment}, the NLL corrections
factorize, \ie they can be expressed through correction factors that
multiply the Born amplitude. Moreover, as we show, the two-loop
correction factors can be expressed entirely in terms of one-loop
quantities.

\subsection{Two-loop Yukawa contributions}
\label{se:res2loopYukawa}%\refse{se:res2loopYukawa}

As explained in \refse{se:NLLfromYukawa}, the only non-suppressed NLL
contributions involving Yukawa interactions, apart from counterterms,
arise from the first three diagrams in \refeq{twoloopdiagYuk},
{\unitlength 0.6pt \SetScale{0.6}
\beqar\label{twoloopcontYuk}%\refeq{twoloopcontYuk}
\mel{2}{\mathrm{Yuk}}
&\NLLA&
\sum_{i=1}^{n}
\sum_{j=1\atop j\neq i}^{n}
\sum_{V_{1}=A,Z,W^\pm}
\sum_{\Phi_{2}=H,\chi,\phi^\pm}
\left\{
\vcenter{\hbox{\diagVYuk{$\leg{i}$}{$\leg{j}$}{$\scriptscriptstyle{V_1}$}{$\scriptscriptstyle{\Phi_2}$}
{\factblob}}}
+\vcenter{\hbox{\diagVIIYuk{$\leg{i}$}{$\leg{j}$}{$\scriptscriptstyle{V_1}$}{$\scriptscriptstyle{\Phi_2}$}{\factblob}}}
\right.
\nn\\*&&\left.{}
+\sum_{\Phi_{3}=H,\chi,\phi^\pm}
\vcenter{\hbox{
\diagIIIYuka{$\leg{i}$}{$\leg{j}$}{$\scriptscriptstyle{\Phi_3}$}{$\scriptscriptstyle{V_1}$}{$\scriptscriptstyle{\Phi_2}$}{\factblob}}}
\right\}
.\eeqar
}%
The evaluation of these three diagrams is presented in
\refse{se:Yukawatwoloop}.  In NLL approximation we find that, up to a
minus sign, the integral functions associated with the three
individual diagrams equal each other, and combining all diagrams we
get
\beqar\label{twoloopresYuk}%\refeq{twoloopresYuk}
\mel{2}{\mathrm{Yuk}}
&\NLLA&
-\frac{1}{e^2}
\mel{0}{}
\sum_{i=1}^{n}
\sum_{j=1\atop j\neq i}^{n}
\sum_{V_{1}=A,Z,W^\pm}
\sum_{\Phi_{2}=H,\chi,\phi^\pm}
\DDsub{\mathrm{Y}}(M_{V_1};p_i,p_j)
\,
I_j^{\bar{V}_1}
\hat{G}_i^{\Phi_{2}^+}
\nn\\*&&{}\times
\left\{
G_i^{\Phi_{2}} I_i^{{V}_1}
- \hat{I}_i^{{V}_1} G_i^{\Phi_{2}}
+ \sum_{\Phi_{3}=H,\chi,\phi^\pm}
  G_i^{\Phi_{3}} I^{V_1}_{\Phi_{3} \Phi_{2}}
\right\}
= 0
,
\eeqar
where the result for the function $\DDsub{\mathrm{Y}}$ is given in
\mla
\refeq{idiagPhisub}.  Owing to global gauge invariance the combination
of gauge and Yukawa couplings in the curly brackets of
\refeq{twoloopresYuk} vanishes [\cf \refeq{gaugeyukrelation}], so the
only two-loop NLL contributions from Yukawa interactions originate
from the wave-function counterterms discussed in \refse{se:2loopren}.
This observation confirms the prediction of
\citeres{Chiu:2007yn,Chiu:2008vv}, where in soft--collinear effective
theory at scales below $Q$ scalar particles contribute only via
wave-function renormalization.

\subsection{Two-loop renormalization}
\label{se:2loopren}%\refse{se:2loopren}

At two loops, the mass renormalization leads to non-suppressed
logarithmic terms only through the insertion of the one-loop mass
counterterms in the one-loop logarithmic corrections.  However, these
contributions are of NNLL order and can thus be neglected in NLL
approximation.  In this approximation also the purely two-loop
counterterms that are associated with the renormalization of the
external-fermion wave functions and the couplings, \ie $\de
Z^{(2)}_i$, $\de g_k^{(2)}$, $\de e^{(2)}$ and $\de\lambdat^{(2)}$, do
not contribute.

The only NLL two-loop counterterm contributions are those that result
from the combination of the one-loop amplitude with the one-loop
counterterms $\de Z^{(1)}_i$, $\de g_k^{(1)}$, $\de e^{(1)}$ and
$\de\lambdat^{(1)}$.
The wave-function counterterms yield
\beqar\label{twoloopWF}%\refeq{twoloopWF}
\nmel{2}{\mathrm{WF}}  &=&
\nmel{1}{\rF} \sum_{i=1}^{n}
\frac{1}{2}
\de Z^{(1)}_{i}.
\eeqar
The unrenormalized one-loop amplitude $\nmel{1}{\rF}$ and the
wave-function renormalization constants $\de Z^{(1)}_{i}$ are given in
\refeq{onelooprepunr} and \refeq{subWFRC}, respectively.  In NLL
approximation only the LL part of $\nmel{1}{\rF}$ contributes to
\refeq{twoloopWF}. At this level of accuracy, the unrenormalized
amplitude $\nmel{1}{\rF}$ and the renormalized amplitude $\nmel{1}{}$
\refeq{onelooprep} are equal, and we can use
\refeq{chargeconservationsum} in order to write $\nmel{1}{\rF}$ and
$\de Z^{(1)}_{i}$ such that all gauge-group generators $I_i^V$ appear
only in terms of the Casimir operators $\sum_{V=A,Z,W^\pm}
I_i^{\bar{V}}I_i^{{V}}$ and $I_j^A I_j^A$, which commute with each
other.  This enables us to combine the counterterms
$\nmel{2}{\mathrm{WF}}$ \refeq{twoloopWF} with the unrenormalized
result $\nmel{2}{\rF}$ \refeq{twoloopresultunr} into the form
presented in \refse{se:twoloopres}.

The remaining NLL two-loop counterterms result from the insertion of
the one-loop coupling-constant counterterms \refeq{subcouplingren} in
the one-loop amplitude \refeq{onelooprep} and read
\beqar\label{twoloopPR}%\refeq{twoloopPR}
e^2 \nmel{2}{\mathrm{PR}}
&\NLLA&
-\frac{1}{2\veps}
\left[\left(\frac{Q^2}{\muR^2}\right)^\veps-1\right]
\melQ{0}{}\,
\sum_{i=1}^{n}
\Biggl\{
\left[
\betacoeff{1}^{(1)} \gb^2 \left(\frac{Y_i}{2}\right)^2
+\betacoeff{2}^{(1)} \gw^2 C_{i}
\right]
\univfact{\veps}{\MW;p_i,p_i}
\nl&&{}
+\betacoeff{e}^{(1)} e^2 Q_i^2\,
\Delta\univfact{\veps}{0;p_i,p_i}
\Biggr\}.
\eeqar
Again only the LL parts of the one-loop amplitude contribute to
$\nmel{2}{\mathrm{PR}}$, so \refeq{chargeconservationsum} has been
used to arrive at the form \refeq{twoloopPR}, and the Yukawa terms in
$\FF{1}{\sew}$ as well as the corresponding counterterms
$\de\lambdat^{(1)}$ are irrelevant.

\subsection{Renormalized two-loop amplitude}
\label{se:twoloopres}%\refse{se:twoloopres}

Using the results of \refapps{se:factcont} and \ref{se:looprelations},
we find that the renormalized two-loop matrix element can be written
as
\beqar\label{twoloopresult}%\refeq{twoloopresult}
\nmel{2}{}&\NLLA&
\melQ{0}{}\,
\Biggl\{
\frac{1}{2}\left[\FF{1}{\sew}\right]^2
+\FF{1}{\sew} \Delta \FF{1}{\elm}
+\FF{1}{\sew} \Delta \FF{1}{\PZ}
+\frac{1}{2}\left[\Delta \FF{1}{\elm}\right]^2
+\Delta \FF{1}{\PZ} \Delta \FF{1}{\elm}
\nl&&{}
+\GG{2}{\sew}
+\Delta \GG{2}{\elm}
\Biggr\}
,
\eeqar
in terms of the one-loop correction factors defined in
\refeq{onelooprepa1}--\refeq{onelooprepalast} and the additional
two-loop terms
\beqar\label{betaterms}%\refeq{betaterms}
e^2\GG{2}{\sew}
&=& \frac{1}{2}\sum_{i=1}^{n}
\left[
\betacoeff{1}^{(1)} \gb^2 \left(\frac{Y_i}{2}\right)^2
+\betacoeff{2}^{(1)} \gw^2 C_{i}
\right]
J(\veps,\MW,\muR^2;p_i,p_i)
,\nl
\Delta \GG{2}{\elm}
&=&
\frac{1}{2}\sum_{i=1}^{n}
Q_i^2 \,
\Biggl\{
\betacoeff{e}^{(1)}
\left[
\Delta J(\veps,0,\muR^2;p_i,p_i)
-\Delta J(\veps,0,\MW^2;p_i,p_i)
\right]
\nn\\*&&{}
+\betacoeff{\QED}^{(1)}\,
\Delta J(\veps,0,\MW^2;p_i,p_i)
\Biggr\}
.
\eeqar
The two-loop functions $J$ and $\Delta J$ are defined in
\refeq{Jterms} through the one-loop functions $I$ and $\Delta I$, so
the entire two-loop amplitude is expressed in terms of one-loop
\mla
quantities.  The relevant $J$-functions read explicitly
\beqar
J(\veps,\MW,\muR^2;p_i,p_i) &\NLLA&
  \frac{1}{3}L^3 - \LmuR L^2 +\order(\veps),
\nl
\De J(\veps,0,\MW^2;p_i,p_i) &\NLLA&
  \deltaz{i} \left(
    \frac{3}{2}\Epsinv{3} + 2L\Epsinv{2} + L^2\Epsinv{1} \right)
\nn\\*&&{}
  - \deltat{i}
    \left( L\Epsinv{2} + 2L^2\Epsinv{1} + 2L^3 \right)
  +\order(\veps),
\nl
&&\hspace*{-4.5cm}
\De J(\veps,0,\muR^2;p_i,p_i) - \De J(\veps,0,\MW^2;p_i,p_i) \NLLA
\LmuR\,\Biggl\{
  \deltaz{i} \,\biggl[
  -2\Epsinv{2}+\Epsinv{1}(\LmuR-2L)
\nn\\*&&{}
  + \LmuR L -\frac{1}{3}\LmuR^2
  \biggr]
  + \deltat{i}
    \left( 2L\Epsinv{1} + 4L^2 - \LmuR L \right)
\Biggr\}
+\order(\veps),
\qquad
\eeqar
where 
\beq
\LmuR=\ln\left(\frac{\muR^2}{\MW^2}\right).
\eeq
Note that the terms $\GG{2}{\sew}$ and $\Delta \GG{2}{\elm}$ in
\refeq{betaterms} only involve NLLs.

In order to be able to express \refeq{twoloopresult} in terms of the
one-loop operators \refeq{onelooprepa1}--\refeq{onelooprepalast} it is
crucial that terms up to order $\Eps{2}$ are included in the latter.

The coefficients $\betacoeff{e}^{(1)}$ and $\betacoeff{\QED}^{(1)}$
describe the running of the electromagnetic coupling above and below
the electroweak scale, respectively.  The former receives
contributions from all charged fermions and bosons, whereas the latter
receives contributions only from light fermions, \ie all charged
leptons and quarks apart from the top quark.

The couplings that enter the one- and two-loop correction factors%
\footnote{ These are the coupling $\alpha$ in the perturbative
  expansion \refeq{pertserie1a} and the couplings $\gb$, $\gw$, $e$
  and $\lambdat$ that appear explicitly in \refeq{onelooprepa1},
  \refeq{betaterms} and enter implicitly in
  \refeq{onelooprepa1}--\refeq{onelooprepalast} through the dependence
  of the generators \refeq{genmixing} on the couplings and the mixing
  parameters $\cw$ and $\sw$.  } are renormalized at the general scale
$\muR \gsim \MW$.  The renormalization of the coupling constants
$g_1$, $g_2$, $e$, and $\lambdat$ in the lowest-order matrix element
$\melQ{0}{}$ in \refeq{onelooprep} and \refeq{twoloopresult} is
discussed in \refse{se:1loopren}.  The $\muR$-dependence of
$\melQ{0}{}$ is implicitly defined by \refeq{bornparamren}, and the
dependence of the one- and two-loop correction factors on $\muR$ is
described by the terms \refeq{betaterms}.  The contributions
\refeq{betaterms} originate from combinations of UV and mass
singularities.  We observe that the term proportional to
$\betacoeff{e}^{(1)}$ vanishes for $\muR=\MW$. Instead, the terms
proportional to $\betacoeff{1}^{(1)}$, $\betacoeff{2}^{(1)}$, and
$\betacoeff{\QED}^{(1)}$ cannot be eliminated through an appropriate
choice of the renormalization scale.  This reflects the fact that such
two-loop terms do not originate exclusively from the running of the
couplings in the one-loop amplitude.

Combining the Born amplitude with the one- and two-loop NLL corrections 
we can write
\beqar\label{factresult1}%\refeq{factresult1}
\M&\NLLA&
\melQ{0}{}\,
F^{\sew}\,
F^{\PZ}\,
F^{\elm},
\eeqar
where we observe a factorization of the symmetric-electroweak
contributions,
\beqar\label{Fsew}
F^{\sew}&\NLLA&
1+
\frac{\alphaeps}{4 \pi}
\FF{1}{\sew}
+\left(\frac{\alphaeps}{4 \pi}\right)^2
\left[
\frac{1}{2}\left(\FF{1}{\sew}\right)^2
+\GG{2}{\sew}
\right]
,
\eeqar
the terms resulting from the difference between $\MW$ and $\MZ$,
\beqar
F^{\PZ}
&\NLLA&
1+
\frac{\alphaeps}{4 \pi}
\Delta \FF{1}{\PZ}
,
\eeqar
and the electromagnetic terms resulting from the mass gap between the
photon and the W~boson,
\beqar\label{eq:Felmnll}%\refeq{eq:Felmnll}
F^{\elm}
&\NLLA&
1+
\frac{\alphaeps}{4 \pi}
\Delta \FF{1}{\elm}
+\left(\frac{\alphaeps}{4 \pi}\right)^2
\left[
\frac{1}{2}\left(
\Delta \FF{1}{\elm}
\right)^2
+\Delta \GG{2}{\elm}
\right]
.
\eeqar
We also observe that the symmetric-electroweak and electromagnetic
terms are consistent with the exponentiated expressions
\beqar\label{expresult1}%\refeq{expresult1}
F^{\sew}&\NLLA&
\exp\left[
\frac{\alphaeps}{4 \pi}
\FF{1}{\sew}
+\left(\frac{\alphaeps}{4 \pi}\right)^2
\GG{2}{\sew}
\right]
,\nl
F^{\elm}&\NLLA&
\exp\left[
\frac{\alphaeps}{4 \pi}
\Delta \FF{1}{\elm}
+\left(\frac{\alphaeps}{4 \pi}\right)^2
\Delta \GG{2}{\elm}
\right].
\eeqar
In particular, these two contributions exponentiate separately.  This
double-exponentia\-ting structure is indicated by the ordering of the
non-commuting one-loop operators $\FF{1}{\sew}$ and $\Delta
\FF{1}{\elm}$ in the interference term $\FF{1}{\sew} \Delta
\FF{1}{\elm}$ in our result \refeq{twoloopresult}.  The commutator of
these two operators yields a non-vanishing NLL two-loop contribution.

Note that the $\mathcal{O}(\alpha^2)$ LL contributions in
\refeq{expresult1} are entirely given by the exponentiation of the
one-loop term.  On the other hand, at the NLL level the presence of
the $\mathcal{O}(\alpha^2)$ terms $G_2^{\mathrm{sew}}$ and $\Delta
G_2^{\mathrm{em}}$ in our result \refeq{expresult1} seems to spoil
exponentiation.  However, this is an artifact of the fixed-order
expansion of the argument of the exponential and must not be
interpreted as a breaking of exponentiation.  This can easily be seen
in the framework of evolution equations, were the
$\mathcal{O}(\alpha^2)$ terms in \refeq{expresult1} naturally emerge
from the running of the coupling associated with the one-loop
contribution.  To illustrate this feature we restrict ourselves to a
gauge theory with a simple gauge group and consider a form factor.  In
this case the structure of our result $F^{\mathrm{sew}}$ in
\refeq{expresult1} is easily obtained from the manifestly
exponentiated expression (15) in \citere{Jantzen:2005az},
\begin{equation}
\label{evolution}%\refeq{evolution}
\mathcal{F}=F_0\,\exp
\left\{
\int_{\MW^2}^{Q^2}
\frac{\mathrm{d}x}{x}
\left[
\int_{\MW^2}^{x}
\frac{\mathrm{d}x'}{x'}
\gamma(\alpha(x'))
+
\zeta(\alpha(x))
+
\xi(\alpha(\MW^2))
\right]
\right\}
.
\end{equation}
The NLL approximation requires the one-loop values of the various
anomalous dimensions as well as the one-loop running of $\alpha$ in
$\gamma(\alpha)$,
\begin{eqnarray}
\gamma(\alpha(x'))
&\NLLA&\frac{\alpha(x')}{4\pi}\gamma^{(1)}
\NLLA\frac{\alpha(\muR^2)}{4\pi}
\left[
1-\frac{\alpha(\muR^2)}{4\pi}b^{(1)}\ln\left(\frac{x'}{\muR^2}\right)
\right]\gamma^{(1)}
,\nonumber\\
\zeta(\alpha(x))
&\NLLA&
\frac{\alpha(\muR^2)}{4\pi}\zeta^{(1)}
,\qquad
\xi(\alpha(\MW^2))
\NLLA\frac{\alpha(\muR^2)}{4\pi}\xi^{(1)}.
\end{eqnarray}
Inserting these expressions in (\ref{evolution}) one obtains
\begin{eqnarray}
\label{evolution2}%\refeq{evolution2}
\mathcal{F}&\NLLA&F_0\,\exp
\Biggl\{
\frac{\alpha(\muR^2)}{4\pi}
\left[
\frac{\gamma^{(1)}}{2} L^2 
+ \left(\zeta^{(1)}+\xi^{(1)}\right) L
\right]
\nonumber\\
&&{}
-
\left(\frac{\alpha(\muR^2)}{4\pi}\right)^2
\frac{\gamma^{(1)}}{2}
b^{(1)}
\left(
\frac{1}{3} L^3 
- l_{\muR} L^2 
\right)
\Biggr\}
,
\end{eqnarray}
and one can easily verify that the two-loop term appearing in the
argument of the exponential, i.e.\ the term proportional to the
$\beta$-function coefficient $b^{(1)}$, corresponds to the term
$G_2^{\mathrm{sew}}$ in our result.

The one- and two-loop corrections
\refeq{onelooprep}--\refeq{onelooprepalast} and
\refeq{twoloopresult}--\refeq{betaterms} contain various combinations
\mla
of weak-isospin matrices $I_i^V$, which are in general non-commuting
and non-diagonal. These matrices have to be applied to the Born
amplitude $\melQ{0}{}$ according to the
definition~\refeq{abbcouplings}.  In order to express the results in a
form which is more easily applicable to a specific process, it is
useful to split the integrals $\univfact{\veps}{M_V;p_i,p_j}$ and
$\De\univfact{\veps}{M_V;p_i,p_j}$ in
\refeq{onelooprepa1}--\refeq{onelooprepa3} into an angular-dependent
part involving logarithms~$\Lrij$ and an angular-independent part.
This permits to eliminate the sum over~$j$ for the angular-independent
parts of \refeq{onelooprepa1}--\refeq{onelooprepa3} using
\refeq{chargeconservationsum}.  One can easily see that the
angular-independent part of \refeq{onelooprepa1} gives rise to the
Casimir operator \refeq{casimir}.  After these simplifications, all
operators that are associated with the angular-independent parts can
be replaced by the corresponding eigenvalues, and the one- and
two-loop results can be written as
\beqar\label{factresult1f}%\refeq{factresult1f}
\M&\NLLA&
\melQ{0}{}\, 
f^{\sew}\,
f^{\PZ}\,
f^{\elm},
\eeqar
where the electromagnetic terms read
\beqar%\label{eq:Felmnll}%\refeq{eq:Felmnll}
f^{\elm}
&\NLLA&
1+
\frac{\alphaeps}{4 \pi}
\Delta \Ff{1}{\elm}
+\left(\frac{\alphaeps}{4 \pi}\right)^2
\left[
\frac{1}{2}\left(
\Delta \Ff{1}{\elm}
\right)^2
+\Delta \Gg{2}{\elm}
\right]
\eeqar
with
\beqar%\label{eigenvalues2}%\refeq{eigenvalues2}
\Delta \Ff{1}{\elm}
&\NLLA&
\sum_{i=1}^n \Biggl\{
\deltaz{i} \biggl(
  -\Epsinv{2} + \frac{1}{2}L^2 + \frac{1}{3}L^3\Eps{}
  + \frac{1}{8}L^4\Eps{2}
  \biggr)
+ \deltat{i} \biggl(
  L\Epsinv{1}
  + L^2
  + \frac{1}{2}L^3\Eps{}
  + \frac{1}{6}L^4\Eps{2}
  \biggr)
\nn\\*&&{}
- \biggl[ \frac{3}{2}\deltaz{i} + \deltat{i}(1+\Lrmi) \biggr]
  \biggl(
  \Epsinv{1}
  +L
  +\frac{1}{2}L^2\Eps{}
  +\frac{1}{6}L^3\Eps{2}
  \biggr)
\Biggr\}
  \, q_i^2
\nl&&
-\left( 
\Epsinv{1}
+L + \frac{1}{2}L^2\Eps{} + \frac{1}{6}L^3\Eps{2} \right)
  \sum_{i=1}^n \sum_{j=1\atop j\neq i}^n l_{ij} q_i q_j
+ \order(\veps^3)
,\nl
\Delta \Gg{2}{\elm}
&\NLLA&
\sum_{i=1}^n \Biggl\{
\LmuR\,\Biggl[
  \deltaz{i} \,\Biggl(
  -\Epsinv{2}
  - \left(L - \frac{1}{2}\LmuR\right) \Epsinv{1}
  + \frac{1}{2}\LmuR L -\frac{1}{6}\LmuR^2
  \Biggr)
\nn\\*&&{}
  + \deltat{i}
    \left( L\Epsinv{1} + 2L^2 - \frac{1}{2}\LmuR L \right)
\Biggr] \,
\betacoeff{e}^{(1)} 
+
\Biggl[
  \deltaz{i} \left(
    \frac{3}{4}\Epsinv{3} + L\Epsinv{2} + \frac{1}{2}L^2\Epsinv{1} \right)
\nn\\*&&{}
  - \deltat{i}
    \left( \frac{1}{2}L\Epsinv{2} + L^2\Epsinv{1} + L^3 \right)
\Biggr] \,
\betacoeff{\QED}^{(1)} 
\Biggr\} \, q_i^2
+ \order(\veps)
.
\eeqar
%where $\LmuR=\ln{(\muR^2/\MW^2)}$, and $c_{i}$, $t^3_{i}$, $y_{i}$,
%$q_{i}$, represent the eigenvalues of the operators $C_{i}$,
%$T^3_{i}$, $Y_{i}$, and $Q_{i}$, respectively.  
For the term resulting from the difference between $\MW$ and $\MZ$ we get
\beqar
f^{\PZ}
&\NLLA&
1+
\frac{\alphaeps}{4 \pi}
\Delta \Ff{1}{\PZ}
\eeqar
with 
\beqar%\label{eigenvalues2}%\refeq{eigenvalues2}
\Delta \Ff{1}{\PZ}
&\NLLA& \left( L + L^2\Eps{} + \frac{1}{2}L^3\Eps{2} \right) \LMZW
  \sum_{i=1}^n \left(\frac{\gw}{e} \cw t^3_i - \frac{\gb}{e} \sw \frac{y_i}{2}\right)^2
+ \order(\veps^3)
,
\eeqar
and the  symmetric-electroweak contributions yield
\beqar\label{fsew}%\refeq{fsew}
f^{\sew}&\NLLA&
1+
\frac{\alphaeps}{4 \pi}
\Ff{1}{\sew}
+\left(\frac{\alphaeps}{4 \pi}\right)^2
\left[
\frac{1}{2}\left(\Ff{1}{\sew}\right)^2
+\Gg{2}{\sew}
\right]
\eeqar
with
\beqar\label{eigenvalues2}%\refeq{eigenvalues2}
\Ff{1}{\sew}
&\NLLA& 
-\left(
  \frac{1}{2}L^2
  +\frac{1}{3}L^3\Eps{}
  +\frac{1}{8}L^4\Eps{2}
  -\frac{3}{2}L
  -\frac{3}{4}L^2\Eps{}
  -\frac{1}{4}L^3\Eps{2}
  \right)
  \sum_{i=1}^n \left[ \frac{\gb^2}{e^2}\left(\frac{y_i}{2}\right)^2 
    + \frac{\gw^2}{e^2} c_{i} \right]
\nn\\*&&{}
 +\left( L + \frac{1}{2}L^2\Eps{} + \frac{1}{6}L^3\Eps{2} \right)
 \mathcal{K}^{\mathrm{ad}}_{1}
 -\frac{\lambdat^2}{4e^2}\,
 \left(L_\Pt + \frac{1}{2}L_\Pt^2\Eps{} +\frac{1}{6}L_\Pt^3\Eps{2}\right)
 \sum_{i=1}^{n} \zYuk_i
+ \order(\veps^3)
,\nl
\Gg{2}{\sew}
&\NLLA& 
\left(
\frac{1}{6}L^3-\frac{1}{2}\LmuR L^2
\right)
\sum_{i=1}^n\left[
 \betacoeff{1}^{(1)} \frac{\gb^2}{e^2}
  \left(\frac{y_i}{2}\right)^2
+ \betacoeff{2}^{(1)} \frac{\gw^2}{e^2}
c_{i}
\right]
+ \order(\veps)
.
\eeqar
In the above equations
$\LmuR=\ln{(\muR^2/\MW^2)}$, and $c_{i}$, $t^3_{i}$, $y_{i}$,
$q_{i}$ represent the eigenvalues of the operators $C_{i}$,
$T^3_{i}$, $Y_{i}$, and $Q_{i}$, respectively.

The only matrix-valued expression is the angular-dependent part of the
symmetric-electroweak contribution $\Ff{1}{\sew}$ in
\refeq{eigenvalues2},
\beqar\label{angdepmat}%\refeq{angdepmat}
\mathcal{K}^{\mathrm{ad}}_{1}
&=&\sum_{i=1}^n \sum_{j=1\atop j\neq i}^n l_{ij}  \sum_{V=A,Z,W^\pm} I_i^{\bar{V}}I_j^{{V}}
.
\eeqar
The two-loop corrections involve terms proportional to
$\mathcal{K}^{\mathrm{ad}}_{1}$ and
$[\mathcal{K}^{\mathrm{ad}}_{1}]^2$.  However, the latter are of NNLL
order and thus negligible in NLL approximation.  The combination of
the matrix \refeq{angdepmat} with the Born amplitude,
\beqar\label{angdepmat2}%\refeq{angdepmat2}
\melQ{0}{}\,
\mathcal{K}^{\mathrm{ad}}_{1}
&=&\sum_{i=1}^n \sum_{j=1\atop j\neq i}^n l_{ij}  \sum_{V=A,Z,W^\pm} 
\mel{0}{\varphi_1\dots\varphi'_i\dots\varphi'_j\dots\varphi_n}
I_{\varphi'_i\varphi_i}^{\bar{V}}I_{\varphi'_j\varphi_j}^{{V}}
\nl
&=&\sum_{i=1}^n \sum_{j=1\atop j\neq i}^n l_{ij}  
\left\{
\mel{0}{\varphi_1\dots\varphi_i\dots\varphi_j\dots\varphi_n}
\left[
%q_iq_j+
%\left(\frac{t^3_i-\sw^2 q_i}{\sw\cw}\right)
%\left(\frac{t^3_j-\sw^2 q_j}{\sw\cw}\right)
\frac{\gb^2}{e^2} \frac{y_i y_j}{4}+
\frac{\gw^2}{e^2} t^3_i t^3_j
%\left(\frac{\gw t^3_i-e\sw q_i}{e\cw}\right)
%\left(\frac{\gw t^3_j-e\sw q_j}{e\cw}\right)
\right]
\right.\nl&&\left.{}
+\sum_{V=W^\pm} 
\mel{0}{\varphi_1\dots\varphi'_i\dots\varphi'_j\dots\varphi_n}
I_{\varphi'_i\varphi_i}^{\bar{V}}I_{\varphi'_j\varphi_j}^{{V}}
\right\},
\eeqar
requires the evaluation of matrix elements involving SU(2)-transformed
external fermions $\varphi'_i,\varphi'_j$, \ie isospin partners of the
fermions $\varphi_i,\varphi_j$.
Explicit results for four-particle processes are presented in \refapp{se:2fproduction}.

\section{Discussion and Conclusion}
\label{se:disc}%\refse{se:disc}

We have studied the asymptotic high-energy behaviour of virtual
electroweak corrections to arbitrary fermionic processes in the
Standard Model.
The present analysis extends results previously obtained for
massless fermion scattering \cite{Denner:2006jr} to processes that
involve also bottom and top quarks.
By explicit evaluation of all relevant Feynman diagrams, we have
derived a general formula that describes one- and two-loop logarithmic
contributions of the form $\ln(Q^2/\MW^2)$.
Such logarithmic terms---which dominate the electroweak corrections at
TeV colliders---originate from ultraviolet and mass (soft/collinear)
singularities in the asymptotic regime where all kinematical
invariants are at an energy scale $Q^2\gg\MW^2$.
All masses of the heavy particles have been assumed to be of the same
order $\MW\sim\MZ\sim\MH\sim\Mt$ but not equal, and all light
fermions---including bottom quarks---have been treated as massless
particles.
We have included all leading (LLs) and next-to-leading (NLLs)
logarithms.

The calculation has been performed in the complete
spontaneously broken electroweak Standard Model using the
't~Hooft--Feynman gauge.
The fermionic wave functions are renormalized on shell, and
coupling-constant renormalization is performed in the
$\overline{\mathrm{MS}}$ scheme, but can be generalized easily.
Employing the method developed in \citere{Denner:2006jr}, we have
reduced all NLL contributions to factorizable diagrams.  In this way
the process-dependent part of the calculation is isolated in a generic
tree-level amplitude, which is multiplied by process-independent
factors consisting of loop integrals and gauge and Yukawa couplings.
Technically this is achieved by means of collinear Ward identities and
a soft--collinear approximation.  All relevant contributions
originating from ultraviolet singularities are consistently taken into
account in every step of the calculation.  In particular, as discussed
in \refse{se:masslessferUV}, the logarithms of ultraviolet origin are
isolated in a few bare diagrams [see \refeq{oneloopsubd1} and
\refeq{twoloopdiagYuk}] and counterterms (see \refses{se:1loopren}
and \ref{se:2loopren}) by means of minimal subtractions of the
ultraviolet singularities and an appropriate choice of the subtraction
scale.  While the ultraviolet logarithms associated with the
renormalization of the coupling parameters in the factorized
tree-level amplitude are easily absorbed into running couplings, the
process-independent correction factors receive additional
contributions of ultraviolet origin which result from the bare
two-loop diagrams of type \refeq{oneloopsubd1}, the wave-function
renormalization [see \refeq{WFren} and \refeq{twoloopWF}], and the
renormalization of the couplings in the soft--collinear corrections
[see \refeq{twoloopPR}].

Since the NLL electroweak corrections originate only from the
electroweak interactions of the external legs, our general results for
$n$-fermion processes are also applicable to processes involving
external fermions and gluons and, more generally, to hard reactions
that involve $n$ fermions plus an arbitrary number of
SU(2)$\times$U(1) singlets as external particles.
Additional legs associated with external electroweak singlets can only
enter the hard part (F) of the factorizable diagrams
\refeq{oneloopdiag4}, \refeq{twoloopdiag3} and \refeq{twoloopdiagYuk}.
This modifies only the process-dependent hard amplitude $\M_0$, which
is always factorized in our derivations, while the NLL correction
factors receive contributions only from fermionic external legs and do
not depend on additional external singlets.

All two-loop integrals have been solved by two independent methods in
NLL approximation. One makes use of sector decomposition, the other
uses the strategy of regions.  Explicit results have been given for
all contributing factorizable Feynman diagrams.

The presence of soft/collinear singularities originating from virtual
photons and their interplay, at two loops, with logarithmic
corrections resulting from massive particles is one of the most
delicate aspects of the problem.  In order to isolate the finite
$\ln(Q^2/\MW^2)$ terms in a meaningful way, one has to separate the
photonic divergences in a gauge-invariant contribution that can be
cancelled against real-photon corrections.
To this end, we have split the corrections into a finite
symmetric-electroweak part, which is constructed by setting the masses
of all gauge bosons equal to $\MW$, and remaining subtracted parts,
which describe the effects resulting from the $\gamma$--W and Z--W
mass differences.
By combining all one- and two-loop diagrams we found that these three
contributions factorize as described in
\refeq{factresult1}--\refeq{eq:Felmnll}: the term associated with the
$\gamma$--W mass splitting ($F^{\elm}$) depends only on the masses and
charges of the external fermions and behaves as a pure QED correction
subtracted at photon mass $\MA=\MW$.  The term resulting from the Z--W
splitting ($F^{\PZ}$) is proportional to $\ln(\MZ^2/\MW^2)$ and
depends only on the external-leg Z-boson couplings.  Finally, the
contribution constructed by setting $M_A=\MZ=\MW$ in all loop diagrams
($F^{\sew}$) turns out to be independent of symmetry-breaking effects
such as mixing or couplings proportional to the vacuum expectation
value. This contribution, which contains only finite $\ln(Q^2/\MW^2)$
terms, behaves as in a symmetric SU(2)$\times$U(1) theory where mass
singularities are regularized by a common mass parameter.
Moreover we find that the electromagnetic and the
symmetric-electroweak
parts exponentiate as described in \refeq{expresult1}: the
corresponding two-loop contributions can be written as the
second-order terms of exponentials of the one-loop contributions plus
additional terms that are proportional to the one-loop $\be$-function
coefficients.

These results agree with the resummations that have been proposed in
the literature and confirm---for fermion scattering processes---the
assumption that the asymptotic high-energy behaviour of electroweak
interactions at two loops can be described by a symmetric and unmixed
$\SU(2)\times\U(1)$ theory matched with QED at the electroweak scale.
Indeed, apart from the terms involving $\ln(\MZ^2/\MW^2)$, in the
final result we observe a cancellation of all effects associated with
symmetry breaking.
%, \ie gauge-boson mixing, the gap between $\MZ$ and $\MW$, and
%couplings proportional to the vacuum expectation value.  
We have explicitly checked that, upon separation of the QED
singularities, our results are consistent with the predictions of
\citere{Melles:2001ia} and \citeres{Chiu:2007yn,Chiu:2008vv}.

As an application of our results for general \mbox{$n$-fermion}
processes, we present in \refapp{se:2fproduction} explicit expressions
for the case of four-particle processes involving four fermions or two
fermions and two gluons.
In general, our process-independent results
can be applied to any reaction with external fermions and gluons as
long as all kinematical invariants are large.  We plan to extend these
results to processes involving external gauge bosons and Higgs bosons.

\begin{appendix}

\section{Loop integrals of factorizable contributions}
\label{se:factcont}%\refapp{se:factcont}

In this appendix, we present explicit results for the loop integrals
of the one- and two-loop factorizable contributions defined in
\refse{se:factorizable} and the Yukawa contributions in
\refse{se:NLLfromYukawa}.  These are evaluated within the
't~Hooft--Feynman gauge, where the masses of the Faddeev--Popov ghosts
$u^A,u^\FZ,u^\FWpm$ and would-be Goldstone bosons $\chi,\phi^\pm$ read
$M_{u^A}=\MA=0$, $M_{\chi}=M_{u^\FZ}=\MZ$, and
$M_{\phi^\pm}=M_{u^\FW}=\MW$.
Using the soft--collinear approximation 
and the projectors
introduced in
\refses{se:scapprox}
and \ref{se:NLLfromgauge}, we express the factorizable contributions
resulting from individual diagrams as products of the
\mbox{$n$-fermion} Born amplitude with matrix-valued coupling factors
and loop integrals.

The loop integrals associated with the various diagrams are denoted
with symbols of the type $\DD{h}(m_1,\dots,m_n;p_i,p_j,\ldots)$.  The
definition of these integrals is provided in \refapp{app:loops}.  They
depend on various internal masses $m_1, m_2, \ldots$ and, through the
external momenta $p_i, p_j, \ldots$, on the kinematical invariants
$r_{ij}$ and the masses $m_i^2=p_i^2$.  The symbols $m_k$ are always
used to denote generic mass parameters, which can assume the values
$m_k=\MW,\MZ,\Mt,\MH$ or $m_k=0$.  Instead we use the symbols $M_k$ to
denote non-zero masses, \ie $M_k=\MW,\MZ,\Mt,\MH$.  The integrals are
often singular when certain mass parameters tend to zero, and the
cases where such parameters are zero or non-zero need to be treated
separately.  We also define subtracted functions
\beq\label{eq:subtractedintegral}%\refeq{eq:subtractedintegral}
\deDD{h}(m_1,\dots,m_n;p_i,\ldots)=
\DD{h}(m_1,\dots,m_n;p_i,\ldots)-
\DD{h}(\MW,\dots,\MW;p_i,\ldots),
\eeq
where the integral with all internal mass parameters equal to $\MW$ is
subtracted.

The integrals have been computed in NLL accuracy, and the result is
expanded in $\veps$ up to $\order(\veps^2)$ at one loop and
$\order(\veps^0)$ at two loops.  The UV poles have been eliminated by
means of a minimal subtraction as explained in
\refse{se:masslessferUV} such that the presented results are UV
finite.  The integrals have been evaluated separately for all physical
combinations of gauge-boson and fermion masses on internal and
external lines.  All loop integrals have been solved and cross-checked
using two independent methods: an automatized algorithm based on the
sector-decomposition technique \cite{Denner:2004iz} and the method of
expansion by regions combined with Mellin--Barnes representations (see
\citere{Jantzen:2006jv} and references therein).

The one-loop diagrams are treated in \refse{se:oneloop}, the two-loop
diagrams involving gauge interactions in \refse{se:twoloop}, and the
diagrams involving Yukawa interactions in \refse{se:Yukawatwoloop}.

\subsection{One-loop diagrams}
\label{se:oneloop}%\refse{se:oneloop}
The one-loop factorizable contributions 
\refeq{oneloopdiag4} originate only from one type of
diagram,%
\footnote{The $l$-loop diagrams depicted in this appendix are
  understood without factors $(\alphaeps/4\pi)^l$.}
\beqar\label{diagram0}%\refeq{diagram0}
\nmel{1}{ij}
\eqdiagl
\vcenter{\hbox{
\unitlength 0.6pt \SetScale{0.6}
\diagone{$\leg{i}$}{$\leg{j}$}{$\scriptscriptstyle{V_1}$}{\factblob}}}
%\eqdiagr
\NLLA
-
\mel{0}{}
\sum_{V_1=A,Z, W^\pm
} 
I_i^{\bar{V}_1}
I_j^{{V}_1}
\DD{0}(M_{V_1};p_i,p_j).
\eeqar
In NLL accuracy, the representations of the generators
$I_i^{\bar{V}_1}$ and $I_j^{{V}_1}$ correspond to the chiralities
given by the spinors of the external particles $i$ and $j$,
respectively.  The loop integral $\DD{0}$ is defined in
\refeq{defint0} and to NLL accuracy yields
\beqar\label{idiag0subt}%\refeq{idiag0subt}
\DDsub{0}(M_1;p_i,p_j) &\NLLA&
-L^2
-\frac{2}{3}L^3\Eps{}
-\frac{1}{4}L^4\Eps{2}
+ \left(4-2\Lrij\right) \left(
  L
  +\frac{1}{2}L^2\Eps{}
  +\frac{1}{6}L^3\Eps{2}
  \right)
\nn\\*&&{}
+\LrMI
\left(
2L
+2L^2\Eps{}
+L^3\Eps{2}
\right)
,\nl
\DDsub{0}(0;p_i,p_j) &\NLLA&
2\Lrij \Epsinv{1}
-\left(\deltaz{i}+\deltaz{j}\right)
  \left(\Epsinv{2} + 2\Epsinv{1}\right)
+ \Biggl\{ \deltat{i} \Biggl[
  L\Epsinv{1} + \frac{1}{2}L^2 + \frac{1}{6}L^3\Eps{}
\nn\\*&&{}
  + \frac{1}{24}L^4\Eps{2}
  - \Lrmi \Epsinv{1}
  + \left(2 - \Lrmi\right)
  \left(
  L
  +\frac{1}{2}L^2\Eps{}
  +\frac{1}{6}L^3\Eps{2}
  \right)
  \Biggr]
  + (i \leftrightarrow j) \Biggr\}
,\nln
\eeqar
where the UV singularities
\beqar\label{idiag0uv}%\refeq{idiag0uv}
\DDUV{0}(m_1;p_i,p_j)
\NLLA
4\Epsinv{1}
\eeqar
have been subtracted.  The shorthands
$L,\Lrmi,\Lrij,\deltat{i},\deltaz{i}$ are defined in
\refse{se:pertexp}.

Summing over all external legs, we find for the 
factorizable one-loop contributions \refeq{oneloopdiag4}
\beqar\label{onelooprepunr}%\refeq{onelooprepunr}
\nmel{1}{\rF}&\NLLA&
\mel{0}{}
\left[
\FF{1}{\rF,\sew}
+\Delta \FF{1}{\rF,\elm}
+\Delta \FF{1}{\rF,\PZ}
\right]
\eeqar
with
\beqar\label{onelooprepu}%\refeq{onelooprepu}
\FF{1}{\rF,\sew}
&=&
-\frac{1}{2}\sum_{i=1}^{n}\sum_{j=1\atop j\neq i}^{n}
\sum_{V=A,Z,W^\pm} 
I_i^{\bar{V}}I_j^{{V}}\,\DDsub{0}(\MW;p_i,p_j)
,\nl
\Delta \FF{1}{\rF,\elm}
&=&
-\frac{1}{2}\sum_{i=1}^{n}\sum_{j=1\atop j\neq i}^{n}
I_i^{A}I_j^{{A}}\,\deDDsub{0}(0;p_i,p_j)
,\nl
\Delta \FF{1}{\rF,\PZ}
&=&
-\frac{1}{2}\sum_{i=1}^{n}\sum_{j=1\atop j\neq i}^{n}
I_i^{Z}I_j^{{Z}}\,\deDDsub{0}(\MZ;p_i,p_j)
,
\eeqar
and
$\deDDsub{0}(m;p_i,p_j)$ defined in
\refeq{eq:subtractedintegral}.

\subsection{Two-loop diagrams involving gauge interactions}
\label{se:twoloop}%\refapp{se:twoloop}
The two-loop NLL factorizable terms \refeq{twoloopdiag3} involve
fourteen different types of diagrams. The diagrams 1--3, 12 and 14
in this section give rise to LLs and NLLs, whereas all other diagrams
yield only NLLs.

% Diagram 1
\subsubsection*{Diagram 1}
\vspace*{-3ex}
\beqar\label{diagram1}%\refeq{diagram1}
\nmel{2}{1,ij}
\eqdiagl
\vcenter{\hbox{
\unitlength 0.6pt \SetScale{0.6}
\diagI{$\leg{i}$}{$\leg{j}$}{$\scriptscriptstyle{V_1}$}{$\scriptscriptstyle{V_2}$}{\factblob}
}}
\NLLA
\mel{0}{}
\sum_{V_1,V_2=A,Z,W^\pm} 
I_i^{\bar{V}_2}
I_i^{\bar{V}_1}
I_j^{{V}_2}
I_j^{{V}_1}
\DD{1}(M_{V_1},M_{V_2};p_i,p_j)
,\qquad
\eeqar
where the loop integral $\DD{1}$ is defined in \refeq{defint1} 
and yields
\beqar\label{idiag1sub}%\refeq{idiag1sub}
\DDsub{1}(M_1,m_2;p_i,p_j) &\NLLA&
\frac{1}{6}L^4
-\frac{2}{3}\left(2-\Lrij+\LrMI\right)L^3
,
\nl
\DDsub{1}(0,M_2;p_i,p_j) &\NLLA&
\left(\deltaz{i} + \deltaz{j}\right)
  \Biggl[
  L^2\Epsinv{2} + \frac{4}{3}L^3\Epsinv{1} + L^4
  - \LrMII \left( 2L\Epsinv{2} + 4L^2\Epsinv{1} + 4L^3 \right)
\nn\\*&&{}
  - \left(4 - 2\Lrij\right)
  \left( L\Epsinv{2} + L^2\Epsinv{1} + \frac{2}{3}L^3 \right)
  \Biggr]
\nn\\&&{}
+ \Biggl\{ \deltat{i} \Biggl[
  -L^3\Epsinv{1} - \frac{23}{12}L^4
  + \left(4 - 3\Lrij + 2\LrMII + \Lrmi\right) L^2\Epsinv{1}
\nn\\*&&{}
  + \left(\frac{20}{3} - \frac{17}{3}\Lrij + \frac{16}{3}\LrMII
    + \frac{7}{3}\Lrmi\right) L^3
  \Biggr]
  + (i \leftrightarrow j) \Biggr\}
,
\nl
\DDsub{1}(0,0;p_i,p_j) &\NLLA&
\deltaz{i} \deltaz{j} \left[
  \Epsinv{4}
  + \left(4 - 2\Lrij\right) \Epsinv{3}
  \right]
+ \Biggl\{ \deltat{i} \deltaz{j} \Biggl[
  \frac{1}{3}\Epsinv{4} - \frac{2}{3}L\Epsinv{3}
  - \frac{1}{3}L^2\Epsinv{2}
\nn\\*&&{}
  - \frac{1}{9}L^3\Epsinv{1}
  - \frac{1}{36}L^4
  + \left(\frac{4}{3} - \frac{4}{3}\Lrij + \frac{2}{3}\Lrmi\right)
    \Epsinv{3}
\nn\\*&&{}
  - \left(4 - \Lrij - \Lrmi\right)
    \left(
    \frac{2}{3}L\Epsinv{2} + \frac{1}{3}L^2\Epsinv{1}
    + \frac{1}{9}L^3
    \right)
  \Biggr]
  + (i \leftrightarrow j) \Biggr\}
\nn\\&&{}
+ \deltat{i} \deltat{j} \Biggl[
  2L^2\Epsinv{2} + \frac{10}{3}L^3\Epsinv{1} + \frac{7}{2}L^4
  + 4\Lrij L\Epsinv{2}
  + \left(8 + 6\Lrij\right)L^2\Epsinv{1}
\nn\\*&&{}
  + \left(\frac{40}{3}+\frac{22}{3}\Lrij\right)L^3
  - \left(\Lrmi + \Lrmj\right)
    \left( 2L\Epsinv{2} + 5L^2\Epsinv{1} + 7L^3 \right)
  \Biggr]
.
\eeqar
Here the UV singularities 
\beqar\label{idiag1sing}%\refeq{idiag1sing}
\DDUV{1}(M_1,m_2;p_i,p_j) &\NLLA&
-4 L^{2}\Epsinv{1}
-\frac{8}{3} L^{3}
,\nl
\DDUV{1}(0,m_2;p_i,p_j) &\NLLA&
- 4\left(\deltaz{i}+\deltaz{j}\right) \Epsinv{3}
+ \left(\deltat{i}+\deltat{j}\right)
  \left( 4L\Epsinv{2} + 2L^2\Epsinv{1} + \frac{2}{3}L^3 \right)
\nln
\eeqar
have been subtracted.

% Diagram 2
\subsubsection*{Diagram 2}
\vspace*{-3ex}
\beqar\label{diagram2}%\refeq{diagram2}
\nmel{2}{2,ij}
\eqdiagl
\vcenter{\hbox{
\unitlength 0.6pt \SetScale{0.6}
\diagII{$\leg{i}$}{$\leg{j}$}{$\scriptscriptstyle{V_1}$}{$\scriptscriptstyle{V_2}$}{\factblob}
}}
\NLLA
\mel{0}{}
\sum_{V_1,V_2
=A,Z,W^\pm} 
I_i^{\bar{V}_2}
I_i^{\bar{V}_1}
I_j^{{V}_1}
I_j^{{V}_2}
\DD{2}(M_{V_1},M_{V_2};p_i,p_j)
,\qquad
\eeqar
where the loop integral $\DD{2}$ is defined in \refeq{defint2}.
This integral is free of UV singularities and yields
\beqar\label{idiag2}%\refeq{idiag2}
\DD{2}(M_1,M_2;p_i,p_j) &\NLLA&
\frac{1}{3}L^4
-\frac{2}{3}\left(4-2\Lrij+\LrMI+\LrMII\right)L^3
,\nl
\DD{2}(0,M_2;p_i,p_j) &\NLLA&
\deltaz{i} \Biggl[
  -\frac{2}{3}L^3\Epsinv{1}
  - \frac{5}{6}L^4
  + \left(4 - 2\Lrij\right)
    \left(L^2\Epsinv{1} + L^3\right)
\nn\\*&&{}
  + \LrMII \left(2L^2\Epsinv{1}
    + \frac{10}{3}L^3\right)
\Biggr]
+ \deltat{i} \Biggl[
  \frac{2}{3}L^4
  -\left(4 - \frac{8}{3}\Lrij + 2\LrMII + \frac{2}{3}\Lrmi\right) L^3
  \Biggr]
,\nl
\DD{2}(M_1,0;p_i,p_j)
&=& \DD{2}(0,M_1;p_j,p_i)
,\nl
\DD{2}(0,0;p_i,p_j) &\NLLA&
\deltaz{i} \deltaz{j} \left[
  \Epsinv{4}
  + \left(4-2\Lrij\right) \Epsinv{3}
  \right]
+ \Biggl\{ \deltat{i} \deltaz{j} \Biggl[
  \frac{1}{6}\Epsinv{4} - \frac{1}{3}L\Epsinv{3}
  + \frac{1}{3}L^2\Epsinv{2}
\nn\\*&&{}
  + \frac{4}{9}L^3\Epsinv{1} + \frac{5}{18}L^4
  + \left( 2 - 2\Lrij + \Lrmi \right)
    \left(\frac{1}{3}\Epsinv{3} - \frac{2}{3}L\Epsinv{2} \right)
\nn\\*&&{}
  + \left( \frac{4}{3} + \frac{2}{3}\Lrij - \frac{4}{3}\Lrmi \right)
    L^2\Epsinv{1}
  + \left( \frac{16}{9} + \frac{2}{9}\Lrij - \frac{10}{9}\Lrmi \right) L^3
  \Biggr]
  + (i \leftrightarrow j) \Biggr\}
\nl&&{}
+ \delta_{i,\Pt} \delta_{j,\Pt} \Biggl[
  -\frac{4}{3}L^3\Epsinv{1}
  - \frac{7}{3}L^4
  - 4\Lrij L^2\Epsinv{1}
  - \left(\frac{16}{3} + \frac{20}{3}\Lrij\right) L^3
\nn\\*&&{}
  + \left(\Lrmi+\Lrmj\right)
    \left(2L^2\Epsinv{1} + \frac{14}{3}L^3\right)
  \Biggr]
.
\eeqar

% Diagram 3
\subsubsection*{Diagram 3}
\vspace*{-3ex}
\beqar\label{diagram3}%\refeq{diagram3}
\nmel{2}{3,ij}
\eqdiagl
\vcenter{\hbox{
\unitlength 0.6pt \SetScale{0.6}
\diagIII{$\leg{i}$}{$\leg{j}$}{$\scriptscriptstyle{V_1}$}{$\scriptscriptstyle{V_3}$}{$\scriptscriptstyle{V_2}$}{\factblob}
}}
\nl&\NLLA&
-\ri 
\frac{\gw}{e}
\mel{0}{}
\sum_{V_1,V_2,V_3
=A,Z,W^\pm} 
\teps^{V_1 V_2 V_3}
I_i^{\bar{V}_2}
I_i^{\bar{V}_1}
I_j^{\bar{V}_3}
\DD{3}(M_{V_1},M_{V_2},M_{V_3};p_i,p_j)
,\quad
\eeqar
where the $\teps$-tensor is defined in \citere{Denner:2006jr} [see
also \refeq{commrel}].  The loop integral $\DD{3}$ is defined in
\refeq{defint3} and yields
\beqar\label{idiag3sub}%\refeq{idiag3sub}
\DDsub{3}(M_1,m_2,M_3;p_i,p_j) &\NLLA&
\frac{1}{6}L^4
-\left( 3-\frac{2}{3}\Lrij
  + \frac{1}{3}\LrMI + \frac{1}{3}\LrMIII\right)L^3
,\nl
\DDsub{3}(0,M_2,M_3;p_i,p_j) &\NLLA&
\deltaz{i} \Biggl[
  -\frac{1}{3}L^3\Epsinv{1}
  - \frac{5}{12}L^4
  + \left( 2 - \Lrij + \LrMIII \right) L^2\Epsinv{1}
  + \biggl( \frac{1}{3} - \Lrij
\nn\\*&&{}
    + \frac{5}{3}\LrMIII \biggr) L^3
  \Biggr]
+ \deltat{i} \Biggl[
  \frac{1}{3}L^4
  - \left( \frac{11}{3} - \frac{4}{3}\Lrij + \LrMIII
    + \frac{1}{3}\Lrmi \right) L^3
  \Biggr]
,\nl
\DDsub{3}(M_1,M_2,0;p_i,p_j) &\NLLA&
-3 \deltaz{i} L\Epsinv{2}
+ \deltat{i} \left(
  \frac{9}{2}L^2\Epsinv{1} + \frac{7}{2}L^3
  \right)
+ \deltaz{j} \Biggl[
  -\frac{1}{3}L^3\Epsinv{1} - \frac{5}{12}L^4
\nn\\*&&{}
  - 3L\Epsinv{2}
  - \left(2 + \Lrij - \LrMI\right) L^2\Epsinv{1}
  + \left(\frac{2}{3} - \Lrij + \frac{5}{3}\LrMI\right) L^3
  \Biggr]
\nn\\&&{}
+ \deltat{j} \Biggl[
  \frac{1}{3}L^4
  + \frac{3}{2}L^2\Epsinv{1}
  + \left(\frac{5}{2} + \frac{4}{3}\Lrij - \LrMI
    - \frac{1}{3}\Lrmj\right) L^3
  \Biggr]
,
\eeqar
where the UV singularities 
\beqar\label{idiag3sing}%\refeq{idiag3sing}
\DDUV{3}(m_1,m_2,M_3;p_i,p_j) &\NLLA&
-3 L^{2}\Epsinv{1}
-2 L^{3}
,\nl
\DDUV{3}(m_1,m_2,0;p_i,p_j) &\NLLA&
-3 \left(\deltaz{i} + \deltaz{j}\right) \Epsinv{3}
+ \left(\deltat{i} + \deltat{j}\right)
  \left( 3L\Epsinv{2} + \frac{3}{2}L^2\Epsinv{1} + \frac{1}{2}L^3
  \right)
\nln
\eeqar
have been subtracted.

% Diagram 4
\subsubsection*{Diagram 4}
\vspace*{-3ex}
\beqar\label{diagram5}%\refeq{diagram5}
\nmel{2}{4,ij}
\eqdiagl
\vcenter{\hbox{
\unitlength 0.6pt \SetScale{0.6}
\diagV{$\leg{i}$}{$\leg{j}$}{$\scriptscriptstyle{V_1}$}{$\scriptscriptstyle{V_2}$}
{\factblob}
}}
%\eqdiagr
\NLLA
- \mel{0}{}
\sum_{V_1,V_2
=A,Z,W^\pm} 
I_i^{{V}_2}
I_i^{\bar{V}_2}
I_i^{{V}_1}
I_j^{\bar{V}_1}
\DD{4}(M_{V_1},M_{V_2};p_i,p_j)
,
\nn\\*[-4ex]
\eeqar
where the loop integral $\DD{4}$ is defined in \refeq{defint5} and yields
\beqar\label{idiag5sub}%\refeq{idiag5sub}
\DDsub{4}(M_1,m_2;p_i,p_j)&\NLLA&
\frac{1}{3}L^3
,\nl
\DDsub{4}(0,M_2;p_i,p_j)&\NLLA&
\left(\deltaz{i} + \deltaz{j}\right)
  \left(L\Epsinv{2} + L^2\Epsinv{1} + \frac{2}{3}L^3\right)
- \frac{1}{2} \deltat{i} \left(L^2\Epsinv{1} + L^3\right)
\nn\\*&&{}
- \deltat{j} \left(\frac{3}{2}L^2\Epsinv{1}
  + \frac{17}{6}L^3\right)
,\nl
\DDsub{4}(0,0;p_i,p_j)&\NLLA&
- \deltaz{i} \deltaz{j} \, \Epsinv{3}
+ \deltat{i} \deltaz{j} \left(
  \Epsinv{3} + 3L\Epsinv{2} + \frac{5}{2}L^2\Epsinv{1}
  + \frac{3}{2}L^2
  \right)
\nn\\*&&{}
+ \deltaz{i} \deltat{j} \left(
  -\frac{2}{3}\Epsinv{3} + \frac{1}{3}L\Epsinv{2}
  + \frac{1}{6}L^2\Epsinv{1} + \frac{1}{18}L^3
  \right)
\nn\\*&&{}
- \deltat{i} \deltat{j} \left(
  2L\Epsinv{2} + 8L^2\Epsinv{1} + \frac{38}{3}L^3
  \right)
.
\eeqar
Here the UV singularities 
\beqar\label{idiag5sing}%\refeq{idiag5sing}
\DDUV{4}(M_1,m_2;p_i,p_j) &\NLLA&
L^{2}\Epsinv{1}
+\frac{2}{3}L^{3}
,\nl
\DDUV{4}(0,m_2;p_i,p_j) &\NLLA&
\left(\deltaz{i} + \deltaz{j}\right) \Epsinv{3}
- \left(\deltat{i} + \deltat{j}\right) \left(
  L\Epsinv{2} + \frac{1}{2}L^2\Epsinv{1} + \frac{1}{6}L^3
  \right)
\qquad
\eeqar
have been subtracted.

Diagram~4 and the following diagram~5 are the only cases where
fermion masses in the numerator of the fermion line~$i$ contribute to
the result in NLL accuracy.  In principle, these fermion-mass terms
contribute with generators $I_i^V$ and $\hat I_i^V$, belonging to
representations with different chiralities [see \refeq{revchircoup}].
But we found that this happens only in the case when both gauge bosons
$V_1$ and $V_2$ are photons, such that $\hat I_i^A=I_i^A$.  Thus all
contributions to \refeq{diagram5} and \refeq{diagram7} can be
expressed in terms of the operators $I_i^V,I_j^V$, which belong to the
representations associated with the chiralities $\kappa_i,\kappa_j$ of
the external fermions.

% Diagram 5
\subsubsection*{Diagram 5}
\vspace*{-3ex}
\beqar\label{diagram7}%\refeq{diagram7}
\nmel{2}{5,ij}
\eqdiagl
\vcenter{\hbox{
\unitlength 0.6pt \SetScale{0.6}
\diagVII{$\leg{i}$}{$\leg{j}$}{$\scriptscriptstyle{V_1}$}{$\scriptscriptstyle{V_2}$}{\factblob}
}}
\NLLA
%\nl&=&
- \mel{0}{}
 \sum_{V_1,V_2
= A,Z,W^\pm} 
I_i^{{V}_2}
I_i^{{V}_1}
I_i^{\bar{V}_2}
I_j^{\bar{V}_1}
\DD{5}(M_{V_1},M_{V_2};p_i,p_j)
,
\nn\\*[-4ex]
\eeqar
where the loop integral $\DD{5}$ is defined in \refeq{defint7} and
to NLL accuracy is given by $\DD{4}$, up to a minus sign:
\beqar\label{idiag7sub}%\refeq{idiag7sub}
\DDsub{5}(m_1,m_2;p_i,p_j)&\NLLA&
-\DDsub{4}(m_1,m_2;p_i,p_j).
\eeqar
Note that this relation only holds if the fermion-mass terms in the
numerator along the line~$i$ are correctly taken into account.

% Diagram 6
\subsubsection*{Diagrams 6}
\vspace*{-3ex}
\label{sub:diagten}%\refse{sub:diagten}
\beqar\label{diagram10}%\refeq{diagram10}
\nmel{2}{6,ij}
\eqdiagl
\vcenter{\hbox{
\unitlength 0.6pt \SetScale{0.6}
\diagX{\factblob}
}}
\quad+\quad
\vcenter{\hbox{
\unitlength 0.6pt \SetScale{0.6}
\diagXI{\factblob}
}}
\nn\\*
&\NLLA&
\frac{1}{2} 
\frac{\gw^2}{e^2}
\mel{0}{}
\sum_{V_1,V_2,V_3,V_4
= A,Z,W^\pm} 
I_i^{\bar{V}_1}
I_j^{\bar{V}_4}\,
\teps^{V_1 \bar{V}_2 \bar{V}_3}
\teps^{V_4 V_2 V_3}\,
\DD{6}(M_{V_1},M_{V_2},M_{V_3},M_{V_4};p_i,p_j)
,\nl*[-1.7ex]
\eeqar
where the loop integral $\DD{6}$ is defined in \refeq{defint10} and yields
\beqar\label{idiag10sub}%\refeq{idiag10sub}
\DDsub{6}(M_1,m_2,m_3,M_4;p_i,p_j)
&\NLLA&
\frac{20}{9} L^3
,\nl
\DDsub{6}(0,M_2,M_3,M_4;p_i,p_j)
&\NLLA& 
\frac{20}{9} L^3
+ \frac{M_2^2+M_3^2}{{2M_4^2}}
\Biggl[
\left(\deltaz{i} + \deltaz{j}\right) \left(
  -8L\Epsinv{2}
  -4L^2\Epsinv{1}
  +\frac{8}{3}L^3
  \right)
\nn\\*&&{}
+ \left(\deltat{i} + \deltat{j}\right) \left(
  8L^2\Epsinv{1} + 12L^3
  \right)
\Biggr]
,\nl
\DDsub{6}(M_1,M_2,M_3,0;p_i,p_j)
&=& \DDsub{6}(0,M_2,M_3,M_1;p_i,p_j)
,\nl
\DDsub{6}(0,M_2,M_3,0;p_i,p_j)
&\NLLA& 
\left(\deltaz{i} + \deltaz{j}\right) \left(
  \frac{10}{3}L\Epsinv{2}
  + \frac{5}{3}L^2\Epsinv{1}
  \right)
\nn\\*&&{}
- \left(\deltat{i} + \deltat{j}\right) \left(
  \frac{10}{3}L^2\Epsinv{1} + \frac{35}{9}L^3
  \right)
.
\eeqar
Here the UV singularities 
\beqar\label{idiag10sing}%\refeq{idiag10sing}
\DDUV{6}(M_1,m_2,m_3,M_4;p_i,p_j) &\NLLA&
\frac{10}{3}L^{2}\Epsinv{1}
+\frac{20}{9}L^{3}
,\nl
\DDUV{6}(0,m_2,m_3,M_4;p_i,p_j) &\NLLA& 
\frac{10}{3}L^{2}\Epsinv{1}
+\frac{20}{9}L^{3}
+\frac{m_2^2+m_3^2}{{2M_4^2}}
  \Biggl[
  \left(\deltaz{i} + \deltaz{j}\right) \biggl(
  -8\Epsinv{3}
\nn\\*&&\hspace*{-0.5cm}{}
  +4L^2\Epsinv{1}
  +\frac{8}{3}L^3
  \biggr)
  + \left(\deltat{i} + \deltat{j}\right) \left(
    8L\Epsinv{2} + 8L^2\Epsinv{1} + 4L^3
  \right)
  \Biggr]
,\nl
\DDUV{6}(M_1,m_2,m_3,0;p_i,p_j)
&=& \DDUV{6}(0,m_2,m_3,M_1;p_i,p_j)
,\nl
\DDUV{6}(0,m_2,m_3,0;p_i,p_j) &\NLLA&
\frac{10}{3} \left(\deltaz{i} + \deltaz{j}\right)
  \Epsinv{3}
\nn\\*&&{}
- \left(\deltat{i} + \deltat{j}\right) \left(
  \frac{10}{3}L\Epsinv{2}
  + \frac{5}{3}L^2\Epsinv{1} + \frac{5}{9}L^3
  \right)
\eeqar
have been subtracted.
We observe that the loop integrals associated with $\FA$--$\FZ$
mixing-energy subdiagrams give rise to the contributions
\beqar\label{lineardep}%\refeq{lineardep}
\deDDsub{6}(0,\MW,\MW,\MZ;p_i,p_j)
&\NLLA&
\frac{\MW^2}{{\MZ^2}}
\Biggl[
\left(\deltaz{i} + \deltaz{j}\right) \left(
  -8L\Epsinv{2}
  -4L^2\Epsinv{1}
  +\frac{8}{3}L^3
  \right)
\nn\\*&&{}
+ \left(\deltat{i} + \deltat{j}\right) \left(
  8L^2\Epsinv{1} + 12L^3
  \right)
\Biggr]
,
\eeqar
which depend linearly on the ratio $\MW^2/\MZ^2$.  Similar terms
appear also in diagrams 7, 8, 9, and 10. These terms cancel when
adding all contributions owing to relations between these integrals
(see \refapp{se:looprelations}), which hold also in the presence of
massive external fermions.

% Diagram 7
\subsubsection*{Diagram 7}
\vspace*{-3ex}
\beqar\label{diagram16}%\refeq{diagram16}
\nmel{2}{7,ij}
\eqdiagl
\vcenter{\hbox{
\unitlength 0.6pt \SetScale{0.6}
\diagXVI{\factblob}
}}
\NLLA
- 
\frac{\gw^2}{e^2}
\mel{0}{}
\sum_{V_1,V_2,V_3
= A,Z,W^\pm} 
I_i^{\bar{V}_1}
I_j^{\bar{V}_3}
\sum_{V= A,Z,W^\pm} \teps^{V_1\bar{V_2}\bar{V}}\teps^{V_3{V_2}{V}}
\nl&&{}\times
(D-1) 
\DD{7}(M_{V_1},M_{V_2},M_{V_3};p_i,p_j)
,
\eeqar
where 
$D=4-2\veps$.
The loop integral $\DD{7}$ is defined in \refeq{defint16} and yields
\beqar\label{idiag16sub}%\refeq{idiag16sub}
\DDsub{7}(M_1,m_2,M_3;p_i,p_j)
&\NLLA&
0
,\nl
\DDsub{7}(0,M_2,M_3;p_i,p_j)
&\NLLA&
\frac{M_2^2}{{M_3^2}}
\Biggl[
\left(\deltaz{i} + \deltaz{j}\right) \left(
  -L\Epsinv{2}
  -\frac{1}{2}L^2\Epsinv{1}
  +\frac{1}{3}L^3
  \right)
\nn\\*&&{}
+ \left(\deltat{i} + \deltat{j}\right) \left(
  L^2\Epsinv{1} + \frac{3}{2}L^3
  \right)
\Biggr]
,\nl
\DDsub{7}(M_1,M_2,0;p_i,p_j)
&=& \DDsub{7}(0,M_2,M_1;p_i,p_j)
,\nl
\DDsub{7}(0,M_2,0;p_i,p_j)
&\NLLA&
0
,
\eeqar
where the UV singularities 
\beqar\label{idiag16sing}%\refeq{idiag16sing}
\DDUV{7}(0,M_2,M_3;p_i,p_j)
&\NLLA&
\frac{M_2^2}{{M_3^2}}
\Biggl[
\left(\deltaz{i} + \deltaz{j}\right) \left(
  -\Epsinv{3} + \frac{1}{2}L^2\Epsinv{1} + \frac{1}{3}L^3
  \right)
\nn\\*&&{}
+ \left(\deltat{i} + \deltat{j}\right) \left(
  L\Epsinv{2} + L^2\Epsinv{1} + \frac{1}{2}L^3
  \right)
\Biggr]
,\nl
\DDUV{7}(M_1,M_2,0;p_i,p_j)
&=& \DDUV{7}(0,M_2,M_1;p_i,p_j)
\eeqar
have been subtracted.

% Diagram 8
\subsubsection*{Diagram 8}
\vspace*{-3ex}
\beqar\label{diagram15}%\refeq{diagram15}
\nmel{2}{8,ij}
\eqdiagl
\vcenter{\hbox{
\unitlength 0.6pt \SetScale{0.6}
\diagXV{\factblob}
}}
\NLLA
- e^2 \vev^2 \mel{0}{}
\sum_{V_1,V_3,V_4
= A,Z,W^\pm} 
I_i^{\bar{V}_1}
I_j^{\bar{V}_4}
\sum_{\Phi_{2}=
H,\chi,\phi^\pm
}
\left\{
I^{{V}_1},
I^{\bar{V}_3}
\right\}_{H\Phi_{2}}
\nl&&{}\times
\left\{
I^{{V}_3},
I^{{V}_4}
\right\}_{\Phi_{2}H}
\DD{8}(M_{V_1},M_{\Phi_{2}},M_{V_3},M_{V_4};p_i,p_j)
,
\eeqar
where the curly brackets denote anticommutators and $\vev$ is the 
vacuum expectation value.
The loop integral $\DD{8}$ is defined in \refeq{defint15} and yields
\beqar\label{idiag15sub}%\refeq{idiag15sub}
\MW^2\DDsub{8}(M_1,M_2,m_3,M_4;p_i,p_j)
&\NLLA&
0
,\nl
\DDsub{8}(0,M_2,M_3,M_4;p_i,p_j)
&\NLLA&
\frac{1}{{M_4^2}}
\Biggl[
\left(\deltaz{i} + \deltaz{j}\right) \left(
  -L\Epsinv{2}
  -\frac{1}{2}L^2\Epsinv{1}
  +\frac{1}{3}L^3
  \right)
\nn\\*&&{}
+ \left(\deltat{i} + \deltat{j}\right) \left(
  L^2\Epsinv{1} + \frac{3}{2}L^3
  \right)
\Biggr]
,\nl
\DDsub{8}(M_1,M_2,M_3,0;p_i,p_j)
&=& \DDsub{8}(0,M_2,M_3,M_1;p_i,p_j)
,\nl
\MW^2
\DDsub{8}(0,M_2,M_3,0;p_i,p_j)
&\NLLA& 
0
.
\eeqar
Here the UV singularities 
\beqar\label{idiag15sing}%\refeq{idiag15sing}
\DDUV{8}(0,M_2,M_3,M_4;p_i,p_j) &\NLLA&
\frac{1}{M_4^2}
\Biggl[
\left(\deltaz{i} + \deltaz{j}\right) \left(
  -\Epsinv{3} + \frac{1}{2}L^2\Epsinv{1} + \frac{1}{3}L^3
  \right)
\nn\\*&&{}
+ \left(\deltat{i} + \deltat{j}\right) \left(
  L\Epsinv{2} + L^2\Epsinv{1} + \frac{1}{2}L^3
  \right)
\Biggr]
,\nl
\DDUV{8}(M_1,M_2,M_3,0;p_i,p_j)
&=& \DDUV{8}(0,M_2,M_3,M_1;p_i,p_j)
\eeqar
have been subtracted.

% Diagram 9
\subsubsection*{Diagram 9}
\vspace*{-3ex}
\beqar\label{diagram12}%\refeq{diagram12}
\nmel{2}{9,ij}
\eqdiagl
\vcenter{\hbox{
\unitlength 0.6pt \SetScale{0.6}
\diagXII{\factblob}
}}
%\eqdiagr
\NLLA
-\frac{1}{2} \mel{0}{}
\sum_{V_1,V_4
= A,Z,W^\pm} 
I_i^{\bar{V}_1}
I_j^{\bar{V}_4}
\sum_{\Phi_{2},\Phi_{3}
=H,\chi,\phi^\pm
}
I^{{V}_1}_{\Phi_{3}\Phi_{2}}
I^{{V}_4}_{\Phi_{2}\Phi_{3}}
\nn\\*&&{}\times
\DD{9}(M_{V_1},M_{\Phi_{2}},M_{\Phi_{3}},M_{V_4};p_i,p_j)
,
\eeqar
where the loop integral $\DD{9}$ is defined in \refeq{defint12} and yields
\beqar\label{massidiag12sub}%\refeq{massidiag12sub}
\DDsub{9}(M_1,M_2,M_3,M_4;p_i,p_j)
&\NLLA& 
\frac{2}{9}L^3
,\nl
\DDsub{9}(0,M_2,M_3,M_4;p_i,p_j)
&\NLLA& 
\frac{2}{9}L^3
+ \frac{M_2^2+M_3^2}{{2M_4^2}}
\Biggl[
\left(\deltaz{i} + \deltaz{j}\right) \left(
  2L\Epsinv{2}
  + L^2\Epsinv{1}
  - \frac{2}{3}L^3
  \right)
\nn\\*&&{}
- \left(\deltat{i} + \deltat{j}\right) \left(
  2L^2\Epsinv{1} + 3L^3
  \right)
\Biggr]
,\nl
\DDsub{9}(M_1,M_2,M_3,0;p_i,p_j)
&=& \DDsub{9}(0,M_2,M_3,M_1;p_i,p_j)
,\nl
\DDsub{9}(0,M_2,M_3,0;p_i,p_j)
&\NLLA&
\left(\deltaz{i} + \deltaz{j}\right) \left(
  \frac{1}{3}L\Epsinv{2}
  + \frac{1}{6}L^2\Epsinv{1}
  \right)
\nn\\*&&{}
- \left(\deltat{i} + \deltat{j}\right) \left(
  \frac{1}{3}L^2\Epsinv{1} + \frac{7}{18}L^3
  \right)
.
\eeqar
Here the UV singularities 
\beqar\label{idiag12sing}%\refeq{idiag12sing}
\DDUV{9}(M_1,M_2,M_3,M_4;p_i,p_j) &\NLLA&
\frac{1}{3}L^{2}\Epsinv{1}
+\frac{2}{9}L^{3}
,\nl
\DDUV{9}(0,M_2,M_3,M_4;p_i,p_j) &\NLLA& 
\frac{1}{3}L^{2}\Epsinv{1}
+\frac{2}{9}L^{3}
+\frac{M_2^2+M_3^2}{{2M_4^2}}
  \Biggl[
  \left(\deltaz{i} + \deltaz{j}\right) \biggl(
  2\Epsinv{3}
  -L^2\Epsinv{1}
\nn\\*&&{}
  -\frac{2}{3}L^3
  \biggr)
  - \left(\deltat{i} + \deltat{j}\right) \left(
    2L\Epsinv{2} + 2L^2\Epsinv{1} + L^3
  \right)
  \Biggr]
,\nl
\DDUV{9}(M_1,M_2,M_3,0;p_i,p_j)
&=& \DDUV{9}(0,M_2,M_3,M_1;p_i,p_j)
,\nl
\DDUV{9}(0,M_2,M_3,0;p_i,p_j) &\NLLA&
\frac{1}{3} \left(\deltaz{i} + \deltaz{j}\right)
  \Epsinv{3}
\nn\\*&&{}
- \left(\deltat{i} + \deltat{j}\right) \left(
  \frac{1}{3}L\Epsinv{2}
  + \frac{1}{6}L^2\Epsinv{1} + \frac{1}{18}L^3
  \right)
\eeqar
have been subtracted.

% Diagram 10
\subsubsection*{Diagram 10}
\vspace*{-3ex}
\beqar\label{diagram17}%\refeq{diagram17}
\nmel{2}{10,ij}
\eqdiagl
\vcenter{\hbox{
\unitlength 0.6pt \SetScale{0.6}
\diagXVII{\factblob}
}}
\NLLA
- \frac{1}{2}\mel{0}{}  
\sum_{V_1,V_3
= A,Z,W^\pm} 
I_i^{\bar{V}_1}
I_j^{\bar{V}_3}
\sum_{\Phi_{2}
=H,\chi,\phi^\pm
}
\left\{
I^{{V}_1}
,I^{{V}_3}
\right\}_{\Phi_{2}\Phi_{2}}
\nl&&{}\times
\DD{10}(M_{V_1},M_{\Phi_{2}},M_{V_3};p_i,p_j)
,
\eeqar
where
\beqar
\DD{10}\equiv \DD{7}.
\eeqar

% Diagram 11
\subsubsection*{Diagram 11}
\label{se:fermions}%\refeq{se:fermions}
For the diagrams involving fermionic self-energy subdiagrams we
consider the contributions of a generic fermionic doublet $\Psi$ with
components $\Psi_i=u,d$.  The sum over the three generations of
leptons and quarks is denoted by $\sum_\Psi$, and colour factors are
implicitly understood.  Assuming that all down-type fermions are
massless, $m_d=0$, and that the masses of up-type fermions are $m_u=
0$ or $\Mt$, we have
\beqar\label{diagram9}%\refeq{diagram9}
\nmel{2}{11,ij}
\eqdiagl
\vcenter{\hbox{
\unitlength 0.6pt \SetScale{0.6}
\diagIX{$\leg{i}$}{$\leg{j}$}{$\scriptscriptstyle{V_1}$}{$\scriptscriptstyle{V_4}$}{$\scriptscriptstyle{\Psi_{i_2}}$}{$\scriptscriptstyle{\Psi_{i_3}}$}{\factblob}
}}
\NLLA
-\frac{1}{2} \mel{0}{}
\sum_{V_1,V_4
= A,Z,W^\pm} 
I_i^{\bar{V}_1}
I_j^{\bar{V}_4}
\nl&&{}\times \sum_\Psi
\Biggl\{
\sum_{\Psi_{i_2},\Psi_{i_3}=u,d}
\;
\sum_{\kappa=\rR,\rL}
I^{{V}_1}_{\Psi_{i_3}^\kappa \Psi_{i_2}^\kappa}
I^{{V}_4}_{\Psi_{i_2}^\kappa \Psi_{i_3}^\kappa}
\,
\DD{11,0}(M_{V_1},m_{i_2},m_{i_3},M_{V_4};p_i,p_j)
\quad
\nl&&{}-
\left(
  I^{{V}_1}_{u^\rR u^\rR} I^{{V}_4}_{u^\rL u^\rL}
+ I^{{V}_1}_{u^\rL u^\rL} I^{{V}_4}_{u^\rR u^\rR}
\right)
m_u^2 \DD{11,m}(M_{V_1},m_u,m_u,M_{V_4};p_i,p_j)
\Biggr\}
,
\eeqar
where $\DD{11,m}\equiv -4 \DD{8}$ represents the contribution
associated with the $m_u$-terms in the numerator of the up-type
fermion propagators of the loop insertion, whereas the integral
$\DD{11,0}$, which is defined in \refeq{defint9}, accounts for the
remaining contributions.  This latter integral yields
\beqar\label{idiag90sub}%\refeq{idiag90sub}
\DDsub{11,0}(M_1,m_2,m_3,M_4;p_i,p_j)&\NLLA&
\frac{8}{9}L^{3}
,\nl
\DDsub{11,0}(0,m_2,m_3,M_4;p_i,p_j)&\NLLA& 
\frac{8}{9}L^{3}
+ \frac{m_2^2+m_3^2}{{2M_4^2}}
\Biggl[
\left(\deltaz{i} + \deltaz{j}\right) \left(
  4L\Epsinv{2}
  + 2L^2\Epsinv{1}
  -\frac{4}{3}L^3
  \right)
\nn\\*&&{}
- \left(\deltat{i} + \deltat{j}\right) \left(
  4L^2\Epsinv{1} + 6L^3
  \right)
\Biggr]
,\nl
\DDsub{11,0}(M_1,m_2,m_3,0;p_i,p_j)
&=& \DDsub{11,0}(0,m_2,m_3,M_1;p_i,p_j)
,\nl
\DDsub{11,0}(0,M_2,M_3,0;p_i,p_j)&\NLLA&
\left(\deltaz{i} + \deltaz{j}\right) \left(
  \frac{4}{3}L\Epsinv{2}
  + \frac{2}{3}L^2\Epsinv{1}
  \right)
\nn\\*&&{}
- \left(\deltat{i} + \deltat{j}\right) \left(
  \frac{4}{3}L^2\Epsinv{1} + \frac{14}{9}L^3
  \right)
,\nl
\DDsub{11,0}(0,0,0,0;p_i,p_j)&\NLLA&
- \left(\deltaz{i} + \deltaz{j}\right) \Epsinv{3}
+ \left(\deltat{i} + \deltat{j}\right) \left(
  \frac{2}{3}L\Epsinv{2} - \frac{2}{9}L^3
  \right)
,
\eeqar
where the UV singularities 
\beqar\label{idiag90sing}%\refeq{idiag90sing}
\DDUV{11,0}(M_1,m_2,m_3,M_4;p_i,p_j) &\NLLA&
\frac{4}{3}L^{2}\Epsinv{1}
+\frac{8}{9}L^{3}
,\nl
\DDUV{11,0}(0,m_2,m_3,M_4;p_i,p_j) &\NLLA& 
\frac{4}{3}L^{2}\Epsinv{1}
+\frac{8}{9}L^{3}
+\frac{m_2^2+m_3^2}{{2M_4^2}}
  \Biggl[
  \left(\deltaz{i} + \deltaz{j}\right) \biggl(
  4\Epsinv{3}
  -2 L^2 \Epsinv{1}
\nn\\*&&{}
  -\frac{4}{3}L^3
  \biggr)
  - \left(\deltat{i} + \deltat{j}\right) \left(
    4L\Epsinv{2} + 4L^2\Epsinv{1} + 2L^3
    \right)
  \Biggr]
,\nl
\DDUV{11,0}(M_1,m_2,m_3,0;p_i,p_j)
&=& \DDUV{11,0}(0,m_2,m_3,M_1;p_i,p_j)
,\nl
\DDUV{11,0}(0,m_2,m_3,0;p_i,p_j) &\NLLA& 
\frac{4}{3} \left(\deltaz{i} + \deltaz{j}\right) \Epsinv{3}
\nn\\*&&{}
- \left(\deltat{i} + \deltat{j}\right) \left(
  \frac{4}{3}L\Epsinv{2}
  + \frac{2}{3}L^2\Epsinv{1} + \frac{2}{9}L^3
  \right)
\eeqar
have been subtracted.
As a consequence of
\beqar
\deDDsub{11,0}(0,m_2,m_3,M_4;p_i,p_j)
&\NLLA&
\frac{m_2^2+m_3^2}{2} \,
\deDDsub{11,m}(0,m_2,m_3,M_4;p_i,p_j)
,\nl
\deDDsub{11,0}(M_1,m_2,m_3,0;p_i,p_j)
&\NLLA&
\frac{m_2^2+m_3^2}{2} \,
\deDDsub{11,m}(M_1,m_2,m_3,0;p_i,p_j)
,
\eeqar
all terms proportional to the fermion masses in
$\nmel{2}{11,ij}$ cancel.

% Diagram 12
\subsubsection*{Diagram 12}
\vspace*{-3ex}
\beqar\label{diagram20}%\refeq{diagram20}
\nmel{2}{12,ijk}
\eqdiagl
\vcenter{\hbox{
\unitlength 0.6pt \SetScale{0.6}
\diagXX{$\leg{j}$}{$\leg{i}$}{$\leg{k}$}{$\scriptscriptstyle{V_1}$}{$\scriptscriptstyle{V_2}$}{\factblob}
}}
\NLLA
\mel{0}{}
\sum_{V_1,V_2=A,Z,W^\pm} 
I_i^{\bar{V}_2}
I_i^{\bar{V}_1}
I_j^{{V}_1}
I_k^{{V}_2}
\DD{12}(M_{V_1},M_{V_2};p_i,p_j,p_k)
,
\nn\\*[-4ex]
\eeqar
where the loop integral $\DD{12}$ is defined in \refeq{defint1} and yields
{\newcommand{\shl}{\hspace*{-4cm}}%
\beqar\label{idiag20sub}%\refeq{idiag20sub}
\DDsub{12}(M_1,M_2;p_i,p_j,p_k)&\NLLA& 
\frac{1}{2}L^{4}
-\left(4 - 2\Lrik
  + \frac{4}{3}\LrMI+\frac{2}{3}\LrMII\right) L^{3}
,\nl
\DDsub{12}(0,M_2;p_i,p_j,p_k)&\NLLA&
\left(\deltaz{i} + \deltaz{j}\right)
  \Biggl[
  L^{2}\Epsinv{2}
  + L^{3}\Epsinv{1}
  + \frac{7}{12}L^{4}
\nn\\*&&\shl{}
  - \left(2-\Lrik\right) \left(
    2 L^{}\Epsinv{2}
    + L^{2}\Epsinv{1}
    + \frac{1}{3}L^{3}
    \right)
  - \LrMII\left(
    2L^{}\Epsinv{2}
    + 3L^{2}\Epsinv{1}
    + \frac{7}{3}L^{3}
    \right)
  \Biggr]
\nl&&\shl{}
+ \deltat{i} \Biggl[
  - \frac{2}{3}L^3\Epsinv{1} - \frac{5}{6}L^4
  + \left(2 - 2\Lrik + \LrMII + \Lrmi\right) L^2\Epsinv{1}
  + \left(\frac{2}{3} - 2\Lrik
    + \frac{5}{3}\LrMII + \frac{5}{3}\Lrmi\right) L^3
  \Biggr]
\nl&&\shl{}
+ \deltat{j} \Biggl[
  - \frac{4}{3}L^3\Epsinv{1} - \frac{7}{3}L^4
  + \left(6 - 2\Lrik - 2\Lrij + 3\LrMII + \Lrmj\right) L^2\Epsinv{1}
\nn\\*&&\shl{}
  + \left(\frac{26}{3} - 2\Lrik - \frac{14}{3}\Lrij
    + 7\LrMII + \frac{7}{3}\Lrmj\right) L^3
  \Biggr]
,\nl
\DDsub{12}(M_1,0;p_i,p_j,p_k)&\NLLA& 
\deltaz{k} \Biggl[
  -\frac{2}{3}L^{3}\Epsinv{1}
  - \frac{2}{3}L^{4}
  + \left(2-\Lrik\right) \left(
    2L^{2}\Epsinv{1}
    + \frac{4}{3}L^{3}
    \right)
\nn\\*&&\shl{}
  +\LrMI\left(
    2L^{2}\Epsinv{1}
    + \frac{8}{3}L^{3}
  \right)
  \Biggr]
+ \deltat{k} \Biggl[
  \frac{5}{6}L^4
  - \left(\frac{16}{3} - \frac{10}{3}\Lrik
    + \frac{8}{3}\LrMI + \frac{2}{3}\Lrmk\right) L^3  
  \Biggr]
,\nl
\DDsub{12}(0,0;p_i,p_j,p_k)&\NLLA&
\deltaz{i} \deltaz{j} \Biggl\{
\deltaz{k} \left[
  2\Epsinv{4}
  + \left(8 - 4\Lrik\right) \Epsinv{3}
  \right]
\nn\\*&&\shl\qquad{}
+ %\deltaz{i} \deltaz{j}
\deltat{k} \Biggl[
  \frac{2}{3}\Epsinv{4} - \frac{4}{3}L\Epsinv{3}
  - \frac{2}{3}L^2\Epsinv{2} - \frac{2}{9}L^3\Epsinv{1}
  - \frac{1}{18}L^4
  + \frac{4}{3}\left(2 - 2\Lrik + \Lrmk\right) \Epsinv{3}
\nn\\*&&\shl\qquad{}
  - \left(4 - \Lrik - \Lrmk\right) \left(
    \frac{4}{3}L\Epsinv{2} + \frac{2}{3}L^2\Epsinv{1}
    + \frac{2}{9}L^3 \right)
  \Biggr]
\Biggr\}
\nn\\&&\shl{}
+ \deltat{i} \deltaz{j} \Biggl\{
\deltaz{k} \Biggl[
  \frac{1}{2}\Epsinv{4} - L\Epsinv{3}
  + \frac{1}{3}L^3\Epsinv{1} + \frac{1}{4}L^4
  + \left(2 - 2\Lrik + \Lrmi\right) \Epsinv{3}
  - \left(4 - 2\Lrik\right) L\Epsinv{2}
\nn\\*&&\shl\qquad{}
  + \left(\Lrik - \Lrmi\right) L^2\Epsinv{1}
  + \left(\frac{4}{3} + \frac{1}{3}\Lrik - \Lrmi\right) L^3
  \Biggr]
\nn\\&&\shl\qquad{}
+ %\deltat{i} \deltaz{j}
\deltat{k} \Biggl[
  -L\Epsinv{3}
  - \frac{1}{3}L^3\Epsinv{1} - \frac{11}{12}L^4
  - \left(\Lrik - \frac{1}{2}\Lrmi - \frac{1}{2}\Lrmk\right)
    \Epsinv{3}
  - \left(4 - 2\Lrik\right) L\Epsinv{2}
\nn\\*&&\shl\qquad{}
  - \left(\Lrik - \Lrmi\right) L^2\Epsinv{1}
  - \left(\frac{4}{3} + 3\Lrik - 3\Lrmi - \frac{2}{3}\Lrmk\right) L^3
  \Biggr]
\Biggr\}
\nn\\&&\shl{}
+ \deltaz{i} \deltat{j} \Biggl\{
\deltaz{k} \Biggl[
  \frac{4}{3}\Epsinv{4} - \frac{2}{3}L\Epsinv{3}
  - \frac{1}{3}L^2\Epsinv{2} - \frac{1}{9}L^3\Epsinv{1}
  - \frac{1}{36}L^4
  + \biggl(\frac{16}{3} - 2\Lrik - \frac{4}{3}\Lrij
\nn\\*&&\shl\qquad{}
    + \frac{2}{3}\Lrmj\biggr) \Epsinv{3}
  - \left(\frac{4}{3} - \Lrik + \frac{2}{3}\Lrij -
    \frac{1}{3}\Lrmj\right)
    \left(2L\Epsinv{2} + L^2\Epsinv{1} + \frac{1}{3}L^3\right)
  \Biggr]
\nn\\&&\shl\qquad{}
+ %\deltaz{i} \deltat{j}
\deltat{k} \Biggl[
  \frac{1}{2}\Epsinv{4} - L\Epsinv{3}
  + \frac{1}{3}L^3\Epsinv{1} + \frac{1}{4}L^4
  + \left(2 - \frac{5}{3}\Lrik - \frac{1}{3}\Lrij
    + \frac{1}{6}\Lrmj + \frac{5}{6}\Lrmk\right) \Epsinv{3}
\nn\\*&&\shl\qquad{}
  - \left(4 - \frac{4}{3}\Lrik - \frac{2}{3}\Lrij
    + \frac{1}{3}\Lrmj - \frac{1}{3}\Lrmk\right) L\Epsinv{2}
  - \frac{1}{3} \left(\Lrik - 4\Lrij + 2\Lrmj + \Lrmk\right)
    L^2\Epsinv{1}
\nn\\*&&\shl\qquad{}
  + \left(\frac{4}{3} - \frac{7}{9}\Lrik + \frac{10}{9}\Lrij
    - \frac{5}{9}\Lrmj - \frac{4}{9}\Lrmk\right) L^3
  \Biggr]
\Biggr\}
\nn\\&&\shl{}
+ \deltat{i} \deltat{j} \Biggl\{
\deltaz{k} \Biggl[
  -L\Epsinv{3} + L^2\Epsinv{2}
  + 2L^3\Epsinv{1} + 2L^4
  - \left(\Lrij - \frac{1}{2}\Lrmi - \frac{1}{2}\Lrmj\right)
    \Epsinv{3}
\nn\\*&&\shl\qquad{}
  - \left(4 - 4\Lrik + 2\Lrmi\right) L\Epsinv{2}
  + \left(4 + 2\Lrik + 2\Lrij - 5\Lrmi - \Lrmj\right) L^2\Epsinv{1}
\nn\\*&&\shl\qquad{}
  + \left(8 + \frac{2}{3}\Lrik + \frac{10}{3}\Lrij
    - \frac{19}{3}\Lrmi - \frac{5}{3}\Lrmj\right) L^3
  \Biggr]
\nn\\&&\shl\qquad{}
+ %\deltat{i} \deltat{j}
\deltat{k} \Biggl[
  2L^2\Epsinv{2} + 2L^3\Epsinv{1} + \frac{7}{6}L^4
  + \left(2\Lrik + 2\Lrij - 2\Lrmi - \Lrmj - \Lrmk\right)
    L\Epsinv{2}
\nn\\*&&\shl\qquad{}
  + \left(8 - 2\Lrik + 4\Lrij - 3\Lrmi - 2\Lrmj - \Lrmk\right)
    L^2\Epsinv{1}
  + \left(8 - 4\Lrik + \frac{14}{3}\Lrij
    - \frac{7}{3}\Lrmi - \frac{7}{3}\Lrmj\right) L^3
  \Biggr]
\Biggr\}
.\nln
\eeqar}%
Here the UV singularities 
\beqar\label{idiag20sing}%\refeq{idiag20sing}
\DDUV{12}(M_1,m_2;p_i,p_j,p_k) &\NLLA&
-4 L^{2}\Epsinv{1}
-\frac{8}{3} L^{3}
,\nl
\DDUV{12}(0,m_2;p_i,p_j,p_k) &\NLLA&
-4 \left(\deltaz{i}+\deltaz{j}\right) \Epsinv{3}
+ \left(\deltat{i}+\deltat{j}\right) \left(
  4L\Epsinv{2} + 2L^2\Epsinv{1} + \frac{2}{3}L^3
  \right)
\nln
\eeqar
have been subtracted.
While the above diagram, to NLL accuracy, does not depend on
$r_{ij}$ and $r_{jk}$ for $p_j^2=0$, a dependency on $r_{ij}$ is
introduced for $p_j^2=\Mt^2$.

% Diagram 13
\subsubsection*{Diagram 13}
\vspace*{-3ex}
\beqar\label{diagram21}%\refeq{diagram21}
\nmel{2}{13,ijk}
\eqdiagl
\vcenter{\hbox{
\unitlength 0.6pt \SetScale{0.6}
\diagXXI{$\leg{j}$}{$\leg{i}$}{$\leg{k}$}{$\scriptscriptstyle{V_2}$}{$\scriptscriptstyle{V_1}$}{$\scriptscriptstyle{V_3}$}{\factblob}
}}
\NLLA
-\ri 
\frac{\gw}{e}
\mel{0}{}
\sum_{V_1,V_2,V_3
=A,Z,W^\pm} 
\teps^{V_1 V_2 V_3}
I_i^{\bar{V}_1}
I_j^{\bar{V}_2}
I_k^{\bar{V}_3}
\nn\\&&{}\times
\DD{13}(M_{V_1},M_{V_2},M_{V_3};p_i,p_j,p_k)
,
\eeqar
where the loop integral $\DD{13}$ is defined in \refeq{defint2}.
This integral is free of UV singularities and yields
\beqar\label{idiag21}%\refeq{idiag21}
\DD{13}(M_1,M_2,M_3;p_i,p_j,p_k)&\NLLA& 
0
,\nl
\DD{13}(0,M_2,M_3;p_i,p_j,p_k)&\NLLA& 
\left(l_{ij}-l_{ik}\right)
\left[
\deltaz{i} \left( L^2\Epsinv{1} +\frac{5}{3}L^3 \right)
-\frac{2}{3} \deltat{i} L^3
\right]
\nn\\*&&{}
+ \frac{1}{3} \deltat{i} \left(l_2 - l_3\right) L^3
,\nl
\DD{13}(M_1,0,M_3;p_i,p_j,p_k)
&=& \DD{13}(0,M_3,M_1;p_j,p_k,p_i)
,\nl
\DD{13}(M_1,M_2,0;p_i,p_j,p_k)
&=& \DD{13}(0,M_1,M_2;p_k,p_i,p_j)
.
\eeqar
When one of the gauge bosons is a photon, it couples to two
$\PW$~bosons with equal masses, so the terms with the mass-dependent
logarithms $l_{1}$, $l_{2}$, $l_{3}$ in~\refeq{idiag21} vanish in all
physically relevant cases.

% Diagram 14
\subsubsection*{Diagram 14}%
\vspace*{-3ex}
\beqar\label{diagram22}%\refeq{diagram22}
\nmel{2}{14,ijkl}
\eqdiagl
\vcenter{\hbox{
\unitlength 0.6pt \SetScale{0.6}
\diagXXII{$\leg{i}$}{$\leg{j}$}{$\leg{k}$}{$\leg{l}$}{$\scriptscriptstyle{V_1}$}{$\scriptscriptstyle{V_2}$}{\factblob}
}}
\NLLA
\mel{0}{}
\sum_{V_1,V_2
=A,Z,W^\pm} 
I_i^{\bar{V}_1}
I_j^{{V}_1}
I_k^{\bar{V}_2}
I_l^{{V}_2}
\DD{14}(M_{V_1},M_{V_2};p_i,p_j,p_k,p_l)
,
\nl*[-1.7ex]
\eeqar
where the loop integral $\DD{14}$ is simply given by the product of
one-loop integrals \refeq{idiag0subt},
\beqar\label{idiag22sub}%\refeq{idiag22sub}
\DDsub{14}(M_{V_1},M_{V_2};p_i,p_j,p_k,p_l)
&=&
\DDsub{0}(M_{V_1};p_i,p_j)\,
\DDsub{0}(M_{V_2};p_k,p_l)
.
\eeqar

\subsubsection*{Sum of two-loop diagrams involving gauge interactions}%

The complete contribution of all factorizable diagrams not involving
Yukawa contributions is obtained by inserting the above results into
\refeq{twoloopdiag3}.
For the case of massless external fermions, we have explained in
detail in App.~E of \citere{Denner:2006jr} how the factorizable
two-loop diagrams can be summed up to the total two-loop amplitude.
To this purpose we have used relations between the scalar loop
integrals which are listed in App.~B of \citere{Denner:2006jr} and are
valid in NLL approximation.
For the general case of diagrams with massive and massless fermions,
these relations receive only minor modifications which we
indicate in \refapp{se:looprelations} of the present paper.
Apart from that, the whole procedure remains exactly the same.
So here we only present the result and refer to \citere{Denner:2006jr}
for more details.

The factorizable two-loop contributions can be written in the form
\beqar\label{twoloopresultunr}%\refeq{twoloopresultunr}
\nmel{2}{\rF}&\NLLA&
\mel{0}{}
\Biggl\{
\frac{1}{2}\left[\FF{1}{\rF,\sew}\right]^2
+\FF{1}{\rF,\sew} \Delta \FF{1}{\rF,\elm}
+\FF{1}{\rF,\sew} \Delta \FF{1}{\rF,\PZ}
\nl&&{}
+\frac{1}{2}\left[\Delta \FF{1}{\rF,\elm}\right]^2
+\Delta \FF{1}{\rF,\PZ} \Delta \FF{1}{\rF,\elm}
+\GG{2}{\rF,\sew}
+\Delta \GG{2}{\rF,\elm}
\Biggr\}
,
\eeqar
where the one-loop terms are given in \refeq{onelooprepu}.
The additional two-loop terms read
\beqar\label{betatermsunr}%\refeq{betatermsunr}
e^2\GG{2}{\rF,\sew}
&=& \frac{1}{2}\sum_{i=1}^{n}
\left[
\betacoeff{1}^{(1)} \gb^2 \left(\frac{Y_i}{2}\right)^2
+\betacoeff{2}^{(1)} \gw^2 C_{i}
\right]
J(\veps,\MW,Q^2;p_i,p_i)
,\nl
\Delta \GG{2}{\rF,\elm}
&=&
\frac{1}{2}\sum_{i=1}^{n}
Q_i^2 \,
\Biggl\{
\betacoeff{e}^{(1)}
\left[
\Delta J(\veps,0,Q^2;p_i,p_i)
-\Delta J(\veps,0,\MW^2;p_i,p_i)
\right]
\nn\\*&&{}
+\betacoeff{\QED}^{(1)}\,
\Delta J(\veps,0,\MW^2;p_i,p_i)
\Biggr\}
,
\eeqar
with the one-loop $\be$-function coefficients \refeq{betacoeffres} and
\refeq{eq:betacoeffQED}, the SU(2) Casimir operator \refeq{casimir},
and the two-loop functions
\beqar\label{Jterms}%\refeq{Jterms}
J(\veps,m,\mu^2;p_i,p_j)&=&\frac{1}{\veps}\left[
I(2 \veps,m;p_i,p_j)
-\left(\frac{Q^2}{\mu^2}\right)^\veps
I(\veps,m;p_i,p_j)
\right]
,\nl
\Delta J(\veps,m,\mu^2;p_i,p_j)
&=&
J(\veps,m,\mu^2;p_i,p_j)-J(\veps,\MW,\mu^2;p_i,p_j)
,
\eeqar
for $m=\MW,\MZ,0$, which are combinations of one-loop functions $I$.
The expressions \refeq{betatermsunr} rely on the fact that the
$J$-function, to NLL accuracy, involves only the LL parts of the
$I$-function. In particular, no angular-dependent $\Lrij$-terms are
relevant for the $J$-function, so the identity \refeq{Iijsplit} yields
\beqar\label{Jijsplit}%\refeq{Jijsplit}
  J(\veps,m,\mu^2;p_i,p_j) \NLLA
  \frac{1}{2} \Bigl[
    J(\veps,m,\mu^2;p_i,p_i) + J(\veps,m,\mu^2;p_j,p_j)
  \Bigr] ,
\eeqar
and \refeq{chargeconservationsum} can be generalized to the
functions $J$ and $\Delta J$.

In order to combine the terms in
\refeq{betatermsunr} with \refeq{twoloopPR} we use
\beqar\label{twoloopPRJ}%\refeq{twoloopPRJ}
-\frac{1}{\veps} \left[\left(\frac{Q^2}{\muR^2}\right)^\veps-1\right]
\univfact{\veps}{m;p_i,p_j}
=
J(\veps,m,\muR^2;p_i,p_j) - J(\veps,m,Q^2;p_i,p_j)
\eeqar
and a corresponding relation between $\Delta I$ and $\Delta J$.

\subsection{Yukawa diagrams}
\label{se:Yukawatwoloop}%\refse{se:Yukawatwoloop}

Most diagrams with scalar bosons coupling to external fermions are
suppressed, as explained in \refse{se:NLLfromYukawa}.  The only
relevant Yukawa contributions from bare diagrams are presented in the
following.

% Yukawa diagram 1
\subsubsection*{Yukawa diagram 1}%
\vspace*{-3ex}
\beqar\label{diagram5Phi}%\refeq{diagram5Phi}
\nmel{2}{\mathrm{Y},1,ij}
\eqdiagl
\vcenter{\hbox{
\unitlength 0.6pt \SetScale{0.6}
\diagVYuk{$\leg{i}$}{$\leg{j}$}{$\scriptscriptstyle{V_1}$}{$\scriptscriptstyle{\Phi_{2}}$}
{\factblob}
}}
%\eqdiagr
\NLLA
- \frac{1}{e^2}\mel{0}{}
\sum_{V_1=A,Z,W^\pm} \,
\sum_{\Phi_{2}=H,\chi,\phi^\pm} 
\hat{G}_i^{\Phi_{2}^+}
G_i^{\Phi_{2}}
I_i^{{V}_1}
I_j^{\bar{V}_1}
\nn\\*&&{}\times
\DD{\mathrm{Y},1}(M_{V_1},M_{\Phi_{2}};p_i,p_j)
.%\nln
\eeqar
The loop integral $\DD{\mathrm{Y},1}$ is defined in \refeq{defint5Phi} and yields
\beqar\label{idiag5Phisub}%\refeq{idiag5Phisub}
\DDsub{\mathrm{Y},1}(m_1,M_2;p_i,p_j)&\NLLA&
\DDsub{\mathrm{Y}}(m_1;p_i,p_j)
\eeqar
with
\beqar\label{idiagPhisub}%\refeq{idiagPhisub}
\DDsub{\mathrm{Y}}(M_1;p_i,p_j) &\NLLA&
\frac{1}{6}L^3
,\nl
\DDsub{\mathrm{Y}}(0;p_i,p_j) &\NLLA&
 \left(\deltaz{i} + \deltaz{j}\right) \left(
  \frac{1}{2}L\Epsinv{2}
  + \frac{1}{2}L^2\Epsinv{1}
  + \frac{1}{3}L^3
  \right)
- \deltat{i} \left(
  \frac{1}{4}L^2\Epsinv{1} + \frac{1}{4}L^3
  \right)
\nn\\*&&{}
- \deltat{j} \left(
  \frac{3}{4}L^2\Epsinv{1} + \frac{17}{12}L^3
  \right)
,
\eeqar
where the UV singularities 
\beqar\label{idiagPhising}%\refeq{idiagPhising}
\DDUV{\mathrm{Y}}(M_1;p_i,p_j) &\NLLA&
 \frac{1}{2}L^{2}\Epsinv{1}
+\frac{1}{3}L^{3}
,\nl
\DDUV{\mathrm{Y}}(0;p_i,p_j) &\NLLA&
 \frac{1}{2} \left(\deltaz{i} + \deltaz{j}\right) \Epsinv{3}
- \left(\deltat{i} + \deltat{j}\right) \left(
  \frac{1}{2}L\Epsinv{2} + \frac{1}{4}L^2\Epsinv{1} + \frac{1}{12}L^3
  \right)
\nln
\eeqar
have been subtracted.

% Yukawa diagram 2
\subsubsection*{Yukawa diagram 2}%
\vspace*{-3ex}
\beqar\label{diagram7Phi}%\refeq{diagram7Phi}
\nmel{2}{\mathrm{Y},2,ij}
\eqdiagl
\vcenter{\hbox{
\unitlength 0.6pt \SetScale{0.6}
\diagVIIYuk{$\leg{i}$}{$\leg{j}$}{$\scriptscriptstyle{V_1}$}{$\scriptscriptstyle{\Phi_{2}}$}{\factblob}
}}
%\eqdiagr
\NLLA
- \frac{1}{e^2} \mel{0}{}
\sum_{V_1=A,Z,W^\pm} \,
\sum_{\Phi_{2}=H,\chi,\phi^\pm} 
\hat{G}_i^{\Phi_{2}^+}
\hat{I}_i^{{V}_1}
G_i^{\Phi_{2}}
I_j^{\bar{V}_1}
\nn\\*&&{}\times
\DD{\mathrm{Y},2}(M_{V_1},M_{\Phi_{2}};p_i,p_j)
.%\nln
\eeqar
The loop integral $\DD{\mathrm{Y},2}$ is defined in \refeq{defint7Phi} and yields
\beqar\label{idiag7Phisub}%\refeq{idiag7Phisub}
\DDsub{\mathrm{Y},2}(m_1,M_2;p_i,p_j)&\NLLA&
-\DDsub{\mathrm{Y}}(m_1;p_i,p_j)
\eeqar
with $\DDsub{\mathrm{Y}}(m_1;p_i,p_j)$ from~\refeq{idiagPhisub}.

% Yukawa diagram 3
\subsubsection*{Yukawa diagram 3}%
\vspace*{-3ex}
\beqar\label{diagram3Phi}%\refeq{diagram3Phi}
\nmel{2}{\mathrm{Y},3,ij}
\eqdiagl
\vcenter{\hbox{
\unitlength 0.6pt \SetScale{0.6}
\diagIIIYuka{$\leg{i}$}{$\leg{j}$}{$\scriptscriptstyle{\Phi_{1}}$}{$\scriptscriptstyle{V_3}$}{$\scriptscriptstyle{\Phi_{2}}$}{\factblob}
}}
%\eqdiagr
\NLLA
\frac{1}{e^2}
\mel{0}{}
\sum_{\Phi_{1},\Phi_{2}=H,\chi,\phi^\pm}
\hat{G}_i^{\Phi_{2}^+}
G_i^{\Phi_{1}}
\sum_{V_3=A,Z,W^\pm} 
I_j^{\bar{V}_3}
\,
I^{V_3}_{\Phi_{1} \Phi_{2}}
\nn\\*&&{}\times
\DD{\mathrm{Y},3}(M_{\Phi_{1}},M_{\Phi_{2}},M_{V_3};p_i,p_j)
.
%\nln
\eeqar
The loop integral $\DD{\mathrm{Y},3}$ is defined in \refeq{defint3Phi}
and yields
\beqar\label{idiag3Phisub}%\refeq{idiag3Phisub}
\DDsub{\mathrm{Y},3}(M_1,M_2,m_3;p_i,p_j) &\NLLA&
-\DDsub{\mathrm{Y}}(m_3;p_i,p_j)
\eeqar
with $\DDsub{\mathrm{Y}}(m_3;p_i,p_j)$ from~\refeq{idiagPhisub}.

\section{Definition of the loop integrals}\label{app:loops}%\refapp{app:loops}
\newcommand{\lmom}{l}
\newcommand{\slmom}{\lslash}
\newcommand{\mass}{m}
\newcommand{\linea}[1]{k_{#1}}
\newcommand{\lineb}[1]{q_{#1}}
\newcommand{\linec}[1]{r_{#1}}
\newcommand{\slinea}[1]{\ks_{#1}}
\newcommand{\slineb}[1]{\qs_{#1}}
\newcommand{\measure}[1]{\rd \tilde{\lmom}_{#1}}
\newcommand{\propagatorm}[2]{P(#1,#2)}
\newcommand{\propagator}[1]{P(#1)}

In this appendix, we list the explicit expressions for the Feynman
integrals that contribute to the one- and two-loop diagrams discussed
in \refapp{se:factcont}.  In order to keep our expressions as compact
as possible we define the momenta
\beqar\label{momentadef}
\linea{1}&=&p_i+\lmom_1
,\qquad
\linea{2}=p_i+\lmom_2
,\qquad
\linea{3}=p_i+\lmom_1+\lmom_2
,\nl
\lineb{1}&=&p_j-\lmom_1
,\qquad
\lineb{2}=p_j-\lmom_2
,\qquad
\lineb{3}=p_j-\lmom_1-\lmom_2
,\qquad
\lmom_{3}=-\lmom_1-\lmom_2
,\nl
\linec{1}&=&p_k-\lmom_1
,\qquad
\linec{2}=p_k-\lmom_2
,\qquad
\linec{3}=p_k-\lmom_1+\lmom_2
,\qquad
\lmom_{4}=\lmom_1-\lmom_2
.
\eeqar
For propagators with mass~$m$ we use the notation 
\beq\label{propagator}
\propagatorm{q}{m}=q^2-m^2+\ri 0
\eeq
and for triple gauge-boson couplings we write
\beq\label{YMvrtex}%\refeq{YMvrtex}
\Gamma^{\mu_1\mu_2\mu_3}(
\lmom_1,\lmom_2,\lmom_3)
=
g^{\mu_1\mu_2}(\lmom_1-\lmom_2)^{\mu_3}
+g^{\mu_2\mu_3}(\lmom_2-\lmom_3)^{\mu_1}
+g^{\mu_3\mu_1}(\lmom_3-\lmom_1)^{\mu_2}
.
\eeq
The normalization factors occurring in
\refeq{pertserie1a} 
are absorbed into the integration measure
\beq\label{measure}
\measure{i}=
{(4\pi)^2}
\left(\frac{4\pi\muD^2}{ \mathrm{e}^{\gamma_{\mathrm{E}}}Q^2}\right)^{D/2-2}
\muD^{4-D}
\frac{\rd^D \lmom_i}{\left(2\pi\right)^D}
=
\frac{1}{\pi^2} \,
\left(
 \mathrm{e}^{\gamma_{\mathrm{E}}}Q^2 \pi\right)^{2-D/2}
\,\rd^D \lmom_i
, 
\eeq
and for the projection introduced in \refeq{projectiondef2} we use the
equality
\newcommand{\traceproject}{\Pi} 
\beqar\label{traceproject}%\refeq{traceproject}
\traceproject_{ij}\left(\om_{\kappa_i} \Gamma\right)
&=&
\frac{1}{2} \traceproject_{ij}\left(\Gamma\right)
,
\eeqar
which holds if $\Gamma$ does not involve $\ga_5$
or $\om_{\rR,\rL}$, as it is the case in the following equations.
We have explicitly verified that the NLL contributions from the
projector  $\tilde\Pi_{ij}$ in \refeq{projectiondef3} cancel.

The integral functions $\DD{h}$ depend on the internal masses
$m_1,m_2,\ldots$ and, through the momenta $p_i,p_j,\ldots$, on the
kinematical invariants $r_{ij}$ and on the masses $m_i^2=p_i^2$ of the
external particles.
The definition of these integrals also involves the masses
$m'_i,m''_i,m'''_i$ of the particles
$\varphi'_i,\varphi''_i,\varphi'''_i$ along the fermionic line~$i$
after one, two or three interactions with gauge bosons or scalar
bosons, which possibly change the weak isospin of the external
particle~$\varphi_i$.  We have found, however, that the dependence of
the results on the masses along the fermionic lines is completely
fixed by the external masses~$m_i$, so we do not indicate the
additional masses $m'_i,\ldots$ in the arguments of the functions
$\DD{h}$.

With this notation we have
\beqar
\label{defint0}%\refeq{defint0}
&&\DD{0}(\mass_1;p_i,p_j) =
\int \measure{1}
\frac{
4 \ri
\linea{1}\lineb{1}
}{
\propagatorm{\lmom_1}{\mass_1}
\propagatorm{\linea{1}}{\mass'_i}
\propagatorm{\lineb{1}}{\mass'_j}
}
,\nl
\label{defint1}%\refeq{defint1}
&&\DD{1}(\mass_1,\mass_2;p_i,p_j) =
\int \measure{1}  \measure{2}
\frac{
-16
(\linea{1}\lineb{1})
(\linea{3} \lineb{3})
}{
\propagatorm{\lmom_1}{\mass_1}
\propagatorm{\lmom_2}{\mass_2}
\propagatorm{\linea{1}}{m'_i}
\propagatorm{\linea{3}}{m''_i}
\propagatorm{\lineb{1}}{m'_j}
\propagatorm{\lineb{3}}{m''_j}
}
,\nl
\label{defint2}%\refeq{defint2}
&&\DD{2}(\mass_1,\mass_2;p_i,p_j) =
\int \measure{1}  \measure{2}
\frac{
-16
(\linea{1}\lineb{3})
(\linea{3} \lineb{2})
}{
\propagatorm{\lmom_1}{\mass_1}
\propagatorm{\lmom_2}{\mass_2}
\propagatorm{\linea{1}}{m'_i}
\propagatorm{\linea{3}}{m''_i}
\propagatorm{\lineb{2}}{m'_j}
\propagatorm{\lineb{3}}{m''_j}
}
,\nl
\label{defint3}%\refeq{defint3}
&&\DD{3}(\mass_1,\mass_2,\mass_3;p_i,p_j) =
\nn\\*&&\qquad
\int \measure{1}  \measure{2}
\frac{
-2
\traceproject_{ij}\left[
  (\slinea{3} + m''_i)
  \gamma^{\mu_2}
  (\slinea{1} + m'_i)
  \gamma^{\mu_1}
\right]
\lineb{3}^{\mu_3}
\Gamma_{\mu_1\mu_2\mu_3}(
\lmom_1,\lmom_2,\lmom_3)
}{
\propagatorm{\lmom_1}{\mass_1}
\propagatorm{\lmom_2}{\mass_2}
\propagatorm{\lmom_3}{\mass_3}
\propagatorm{\linea{1}}{m'_i}
\propagatorm{\linea{3}}{m''_i}
\propagatorm{\lineb{3}}{m'_j}
}
,\nl
\label{defint5}%\refeq{defint5}
&&\DD{4}(\mass_1,\mass_2;p_i,p_j) =
%\nl&&\qquad
\int \measure{1}  \measure{2}
\frac{
2
\traceproject_{ij}\left[
  (\slinea{1} + m'_i)
  \gamma^{\mu_2}
  (\slinea{3} + m''_i)
  \gamma_{\mu_2}
  (\slinea{1} + m'_i)
  \slineb{1}
\right]
}{
\propagatorm{\lmom_1}{\mass_1}
\propagatorm{\lmom_2}{\mass_2}
\left[\propagatorm{\linea{1}}{m'_i}\right]^2
\propagatorm{\linea{3}}{m''_i}
\propagatorm{\lineb{1}}{m'_j}
}
,\nl
\label{defint7}%\refeq{defint7}
&&\DD{5}(\mass_1,\mass_2;p_i,p_j) =
\int \measure{1}  \measure{2}
\frac{
2
\traceproject_{ij}\left[
  (\slinea{1}+m'''_i)
  \gamma^{\mu_2}
  (\slinea{3}+m''_i)
  \slineb{1}
  (\slinea{2}+m'_i)
  \gamma_{\mu_2}
\right]
}{
\propagatorm{\lmom_1}{\mass_1}
\propagatorm{\lmom_2}{\mass_2}
\propagatorm{\linea{1}}{m'''_i}
\propagatorm{\linea{3}}{m''_i}
\propagatorm{\linea{2}}{m'_i}
\propagatorm{\lineb{1}}{m'_j}
}
,\nl
\label{defint10}%\refeq{defint10}
&&\DD{6}(\mass_1,\mass_2,\mass_3,\mass_4;p_i,p_j) =
\nn\\*&&\qquad
\int \measure{1}  \measure{2}
\frac{
-4 
\linea{1}^{\mu_1}\lineb{1 \mu_4}
\left[
\Gamma_{\mu_1\mu_2\mu_3}
(\lmom_1,\lmom_2,\lmom_3)
\Gamma^{\mu_4\mu_2\mu_3}
(\lmom_1,\lmom_2,\lmom_3)
+2\lmom_{2\mu_1}\lmom_{3}^{\mu_4}
\right]
}{
\propagatorm{\lmom_1}{\mass_1}
\propagatorm{\lmom_2}{\mass_2}
\propagatorm{\lmom_3}{\mass_3}
\propagatorm{\lmom_1}{\mass_4}
\propagatorm{\linea{1}}{m'_i}
\propagatorm{\lineb{1}}{m'_j}
}
,\nl
\label{defint16}%\refeq{defint16}
&&\DD{7}(\mass_1,\mass_2,\mass_3;p_i,p_j) =
\int \measure{1}  \measure{2}
\frac{
-4 
\linea{1}\lineb{1}
}{
\propagatorm{\lmom_1}{\mass_1}
\propagatorm{\lmom_2}{\mass_2}
\propagatorm{\lmom_1}{\mass_3}
\propagatorm{\linea{1}}{m'_i}
\propagatorm{\lineb{1}}{m'_j}
}
,\nl
\label{defint15}%\refeq{defint15}
&&\DD{8}(\mass_1,\mass_2,\mass_3,\mass_4;p_i,p_j) =
\nn\\*&&\qquad
\int \measure{1}  \measure{2}
\frac{
- 4 
\linea{1}\lineb{1}
}{
\propagatorm{\lmom_1}{\mass_1}
\propagatorm{\lmom_2}{\mass_2}
\propagatorm{\lmom_3}{\mass_3}
\propagatorm{\lmom_1}{\mass_4}
\propagatorm{\linea{1}}{m'_i}
\propagatorm{\lineb{1}}{m'_j}
},\nl
\label{defint12}%\refeq{defint12}
&&\DD{9}(\mass_1,\mass_2,\mass_3,\mass_4;p_i,p_j) =
\nn\\*&&\qquad
\int \measure{1}  \measure{2}
\frac{
4
\linea{1}^{\mu_1}\lineb{1}^{\mu_4}
(\lmom_{2}-\lmom_{3})_{\mu_1}
(\lmom_{2}-\lmom_{3})_{\mu_4}
}{
\propagatorm{\lmom_1}{\mass_1}
\propagatorm{\lmom_2}{\mass_2}
\propagatorm{\lmom_3}{\mass_3}
\propagatorm{\lmom_1}{\mass_4}
\propagatorm{\linea{1}}{m'_i}
\propagatorm{\lineb{1}}{m'_j}
}
,\nl
\label{defint17}%\refeq{defint17}
&&\DD{10}(\mass_1,\mass_2,\mass_3;p_i,p_j) =
\DD{7}(\mass_1,\mass_2,\mass_3;p_i,p_j)
,\nl
\label{defint9}%\refeq{defint9}
&&\DD{11,0}(\mass_1,\mass_2,\mass_3,\mass_4;p_i,p_j) =
\nn\\*&&\qquad
\int \measure{1}  \measure{2}
\frac{
4
\linea{1}^{\mu_1}\lineb{1}^{\mu_4}
\, \Tr\left(
\gamma_{\mu_1}
\slmom_2
\gamma_{\mu_4}
\slmom_3\right)
}{
\propagatorm{\lmom_1}{\mass_1}
\propagatorm{\lmom_2}{\mass_2}
\propagatorm{\lmom_3}{\mass_3}
\propagatorm{\lmom_1}{\mass_4}
\propagatorm{\linea{1}}{m'_i}
\propagatorm{\lineb{1}}{m'_j}
}
,\nl
\label{defint9m}%\refeq{defint9m}
&&\DD{11,m}(\mass_1,\mass_2,\mass_3,\mass_4;p_i,p_j) =
-4 \DD{8}(\mass_1,\mass_2,\mass_3,\mass_4;p_i,p_j)
,\nl
\label{defint20}%\refeq{defint20}
&&\DD{12}(\mass_1,\mass_2;p_i,p_j,p_k) =
\nn\\*&&\qquad
\int \measure{1}  \measure{2}
\frac{
-16
(\linea{1}\lineb{1})
(\linea{3} \linec{2})
}{
\propagatorm{\lmom_1}{\mass_1}
\propagatorm{\lmom_2}{\mass_2}
\propagatorm{\linea{1}}{m'_i}
\propagatorm{\linea{3}}{m''_i}
\propagatorm{\lineb{1}}{m'_j}
\propagatorm{\linec{2}}{m'_k}
}
,\nl
\label{defint21}%\refeq{defint21}
&&\DD{13}(\mass_1,\mass_2,\mass_3;p_i,p_j,p_k) =
\nn\\*&&\qquad
\int \measure{1}  \measure{2}
\frac{
8
\linea{1}^{\mu_1}
\lineb{2}^{\mu_2}
\linec{3}^{\mu_3}
\Gamma_{\mu_1\mu_2\mu_3}(
-\lmom_1,\lmom_2,\lmom_4)
}{
\propagatorm{\lmom_1}{\mass_1}
\propagatorm{\lmom_2}{\mass_2}
\propagatorm{\lmom_4}{\mass_3}
\propagatorm{\linea{1}}{m'_i}
\propagatorm{\lineb{2}}{m'_j}
\propagatorm{\linec{3}}{m'_k}
}
,\nl
&&\DD{14}(m_1,m_2;p_i,p_j,p_k,p_l) =
\DD{0}(m_1;p_i,p_j)
\DD{0}(m_2;p_k,p_l)
,\nl
\label{defint5Phi}%\refeq{defint5Phi}
&&\DD{\mathrm{Y},1}(\mass_1,\mass_2;p_i,p_j) =
%\nl&&\qquad
\int \measure{1}  \measure{2}
\frac{
-2
\traceproject_{ij}\left[
  (\slinea{1} + m'_i)
  (\slinea{3} + m''_i)
  (\slinea{1} + m'_i)
  \slineb{1}
\right]
}{
\propagatorm{\lmom_1}{\mass_1}
\propagatorm{\lmom_2}{\mass_2}
\left[\propagatorm{\linea{1}}{m'_i}\right]^2
\propagatorm{\linea{3}}{m''_i}
\propagatorm{\lineb{1}}{m'_j}
}
,\nl
\label{defint7Phi}%\refeq{defint7Phi}
&&\DD{\mathrm{Y},2}(\mass_1,\mass_2;p_i,p_j) =
\nn\\*&&\qquad
\int \measure{1}  \measure{2}
\frac{
-2
\traceproject_{ij}\left[
  (\slinea{1}+m'''_i)
  (\slinea{3}+m''_i)
  \slineb{1}
  (\slinea{2}+m'_i)
\right]
}{
\propagatorm{\lmom_1}{\mass_1}
\propagatorm{\lmom_2}{\mass_2}
\propagatorm{\linea{1}}{m'''_i}
\propagatorm{\linea{3}}{m''_i}
\propagatorm{\linea{2}}{m'_i}
\propagatorm{\lineb{1}}{m'_j}
}
,\nl
\label{defint3Phi}%\refeq{defint3Phi}
&&\DD{\mathrm{Y},3}(\mass_1,\mass_2,\mass_3;p_i,p_j) =
\nn\\*&&\qquad
\int \measure{1} \measure{2}
\frac{
2
\traceproject_{ij}\left[
  (\slinea{3} + m''_i)
  (\slinea{1} + m'_i)
\right]
\lineb{3} \lmom_4 %(\lmom_1 - \lmom_2)
}{
\propagatorm{\lmom_1}{\mass_1}
\propagatorm{\lmom_2}{\mass_2}
\propagatorm{\lmom_3}{\mass_3}
\propagatorm{\linea{1}}{m'_i}
\propagatorm{\linea{3}}{m''_i}
\propagatorm{\lineb{3}}{m'_j}
}
.
\eeqar
The previous definitions are valid for diagrams with incoming
fermions. In the case of an incoming antifermion~$\varphi_i$, all
masses $m_i,m'_i,\ldots$ along the fermionic line~$i$ have to be
multiplied by $(-1)$, in the integral definitions \refeq{defint0} as
well as in the projectors \refeq{projectiondef2}. But as mentioned in
\refses{se:gaugeandyuk} and \ref{se:NLLfromgauge}, the NLL results are
insensitive to this transformation.

The various coupling matrices $I_k^V$ and $G_k^{\Phi}$, which are
associated with the interactions along the fermionic lines
$k=i,j,\dots$, have been factorized from the loop integrals and can be
found in \refapp{se:factcont}.  Here a comment is in order since, in
principle, the fermion-mass terms in the numerator flip the chirality
of the fermions and give rise to coupling matrices $\hat I_k^V$ and
$\hat G_k^{\Phi}$ corresponding to opposite chirality states [see
\refeq{revchircoup}].  However, as discussed in \refapp{se:factcont}
for the case of diagrams 4 and 5 [see text after
\refeq{idiag5sing}], we have found that the fermion-mass terms in the
numerator are only relevant in the case of photon interactions, where
the representations of the generators are independent of the
chiralities, \ie $\hat I_k^A=I_k^A$.  Thus, all contributions can be
expressed in terms of the operators $I_k^V$, which belong to the
representations associated with the chiralities $\kappa_k$ of the
external fermions, and all coupling factors can be factorized as in
\refapp{se:factcont}.

\section{Relations between loop integrals in NLL approximation}
\label{se:looprelations}%\refapp{se:looprelations}

In order to combine the two-loop contributions of \refse{se:twoloop}
we use relations between the loop integrals.  These relations have
been obtained from the explicit results listed in \refses{se:oneloop}
and \ref{se:twoloop}.  They are valid after subtraction of the UV
singularities and in NLL approximation.

For the case of massless fermionic particles, the relevant relations
have been listed in App.~B of \citere{Denner:2006jr}.  We have found
that, after only small
modifications, they are all still valid for the case of massive
fermionic particles.  One trivial and obvious modification is the
following change of arguments in the integral functions~$\DDsub{h}$:
\beqar
  \DDsub{h}(\ldots;r_{ij}) \to \DDsub{h}(\ldots;p_i,p_j),
  \quad
  \mbox{for }
  \DDsub{1},\ldots,\DDsub{10},
  \DDsub{11,0}, \DDsub{11,m},
\eeqar
and similarly for the subtracted functions $\deDDsub{h}$ defined
in~\refeq{eq:subtractedintegral}.  Many relations do not need further
modifications, and we refer to App.~B of \citere{Denner:2006jr} for
them.

However, since the presence of fermion masses breaks the invariance of
some diagrams with respect to an exchange of external or internal
lines, certain relations obtained for massless fermions have to be
modified by an appropriate reordering of arguments in the
$\DDsub{h}$-functions.  We list these relations in the following.  As in
\refapp{se:factcont}, the symbols $m_i$ are used to denote generic
mass parameters, which can assume the values $m_i=\MW,\MZ,\Mt,\MH$ or
$m_i=0$, and the symbols $M_i$ are used to denote non-zero masses, \ie
$M_i=\MW,\MZ,\Mt,\MH$.

In the second line of (B.2) in \citere{Denner:2006jr} the arguments
$p_i,p_j$ have to be exchanged on the right-hand side.
This relation becomes
\beqar%\label{eq:2legintladder}%\refeq{eq:2legintladder}
\DDsub{2}(m_1,m_2;p_i,p_j) &=& \DDsub{2}(m_2,m_1;p_j,p_i).
\eeqar

In the relations (B.3) of \citere{Denner:2006jr} the order of the mass
parameters in $\DDsub{2}$ has to be reversed, and the order of the
momenta in the 3-leg integral $\DDsub{12}$ is now important:
\beqar%\label{eq:2legYM}%\refeq{eq:2legYM}
\DDsub{3}(M_1,m_2,m_3;p_i,p_j) &\NLLA&
\frac{1}{2}\DDsub{2}(M_1,m_3;p_i,p_j)
-\DDsub{4}(m_3,M_1;p_i,p_j)
\nl&&{}
-6\DDsub{9}(m_3,M_1,M_1,m_3;p_i,p_j),
\nl
\deDDsub{3}(\MW,\MW,m_1;p_i,p_j) &\NLLA&
\frac{1}{2}\deDDsub{12}(\MW,m_1;p_i,p_k,p_j)
-\deDDsub{4}(m_1,\MW;p_i,p_j)
\nl&&{}
-6\deDDsub{9}(m_1,\MW,\MW,m_1;p_i,p_j),
\nl
\deDDsub{3}(m_1,\MW,\MW;p_i,p_j) &\NLLA& \deDDsub{3}(\MW,m_1,\MW;p_i,p_j)
+\deDDsub{1}(\MW,m_1;p_i,p_j)
\nl&&{}
+\deDDsub{2}(m_1,\MW;p_i,p_j)+\deDDsub{4}(\MW,m_1;p_i,p_j)
\nl&&{}
-\frac{1}{2}\deDDsub{12}(\MW,m_1;p_j,p_k,p_i)
.
\eeqar
As in \citere{Denner:2006jr}, the first of these relations has been
\mla
verified and is needed only if at most one of the masses $m_2$ and
$m_3$ is zero.

\mda
The functions $J$ and $\De J$ defined in \refeq{Jterms} of this paper
now also depend on the external momenta $p_i,p_j$, and (B.6) and (B.7)
of \citere{Denner:2006jr} become
\mua
\beqar%\label{eq:2gse}%\refeq{eq:2gse}
3\DDsub{9}(M_1,M_2,M_3,M_4;p_i,p_j)&\NLLA& -J(\veps,\MW,Q^2;p_i,p_j),\nl
3\deDDsub{9}(0,M_2,M_3,0;p_i,p_j)&\NLLA& -\left[
\Delta J(\veps,0,Q^2;p_i,p_j)-\Delta J(\veps,0,\MW^2;p_i,p_j)
\right],
\nl
&&\hspace*{-5.5cm}
3\Bigl[\deDDsub{11,0}(0,0,0,0;p_i,p_j)
-\deDDsub{11,0}(0,M_2,M_3,0;p_i,p_j)\Bigr] \NLLA
  -4 \Delta J(\veps,0,\MW^2;p_i,p_j)
.\nln
\eeqar

\mda
For the 3-leg integrals the order of the external momenta $p_i,p_j,p_k$
becomes relevant and (B.8), (B.9) and (B.10)
of \citere{Denner:2006jr} generalize to 
\mua
\beqar%\label{eq:3legintfac}%\refeq{eq:3legintfac}
\DDsub{12}(m_1,m_2;p_i,p_j,p_k) + \DDsub{12}(m_2,m_1;p_i,p_k,p_j)
&\NLLA&
\DDsub{0}(m_1;p_i,p_j)\DDsub{0}(m_2;p_i,p_k)
,\nl
\sum_{\pi(i,j,k)}\sgn(\pi(i,j,k))\,
\DDsub{12}(M_1,M_2;p_i,p_j,p_k) 
&\NLLA& 0,
\eeqar
where the sum runs over all permutations $\pi(i,j,k)$ of $i,j,k$, with
sign $\sgn(\pi(i,j,k))$,
and
\beqar%\label{eq:3legYM}%\refeq{eq:3legYM}
\DDsub{13}(M_1,M_2,M_3;p_i,p_j,p_k) &\NLLA& 0,
\nl
2\deDDsub{13}(M_1,M_1,m_3;p_i,p_j,p_k) &=&
2\deDDsub{13}(m_3,M_1,M_1;p_k,p_i,p_j) 
\nl &=&
2\deDDsub{13}(M_1,m_3,M_1;p_j,p_k,p_i) 
\nl&\NLLA&
\deDDsub{12}(\MW,m_3;p_j,p_i,p_k) -\deDDsub{12}(\MW,m_3;p_i,p_j,p_k)
.\nln
\eeqar
When fermion masses are involved, the last relation is only true (and
only needed) if the two non-zero gauge-boson masses~$M_1$ on the
left-hand side are equal.

We have found that the relations from App.~B of \citere{Denner:2006jr},
together with the modifications presented above, are exactly the ones
needed to combine the two-loop contributions of \refse{se:twoloop}
into the complete amplitude (see end of \refapp{se:factcont}).

\section{Application to four-particle processes}
\label{se:2fproduction}%\refapp{se:2fproduction}

Here we apply our results to four-particle processes 
involving light fermions, heavy fermions and gluons.
We first examine four-fermion processes
\beqar\label{2fproduction}%\refeq{2fproduction}
  \varphi_1(p_1) \, \varphi_2(p_2) \to \varphi_3(-p_3) \, \varphi_4(-p_4)
,
\eeqar
where each of the $\varphi_i$ may be a fermion, $\varphi_i =
f^{\kappa_i}_{\sigma_i}$, or antifermion, $\varphi_i = \bar
f^{\kappa_i}_{\sigma_i}$, with the notations from
\refse{se:definitions}, provided that the number of fermions and
antifermions in the initial and final state is equal.
We exclude top quarks from the initial state, $\varphi_{1,2} \ne
t,\bar t$, but allow for bottom quarks there. The final state may
contain any combination of massless or massive fermions and
antifermions, including top and bottom quarks.

The scattering amplitudes for the processes~\refeq{2fproduction}
follow directly from our results for the generic $n \to 0$
process~\refeq{genproc} by crossing symmetry, and the Mandelstam
invariants are given by $s = r_{12} = r_{34}$, $t = r_{13} = r_{24}$,
and $u = r_{14} = r_{23}$ with $r_{ij} = (p_i+p_j)^2$.
In practice we can restrict ourselves to the calculation of
\mbox{$s$-channel} amplitudes $f^{\kappa_1}_{\sigma_1} \, \bar
f^{\kappa_2}_{\sigma_2} \to f^{\kappa_3}_{\sigma_3} \, \bar
f^{\kappa_4}_{\sigma_4}$, \ie amplitudes where the external fermion
lines are connected between $f^{\kappa_1}_{\sigma_1}$ and $\bar
f^{\kappa_2}_{\sigma_2}$ in the initial state and between
$f^{\kappa_3}_{\sigma_3}$ and $\bar f^{\kappa_4}_{\sigma_4}$ in the
final state.  All other scattering amplitudes needed for the
four-fermion processes can be obtained from the $s$-channel amplitudes
by crossing symmetry.

In \refse{se:4fNC} we treat neutral-current four-fermion amplitudes
where the particle pairs in the initial and final state are
antiparticles of each other.  \refse{se:4fCC} is devoted to
charged-current four-fermion amplitudes where the initial and final
state each consist of a pair of isospin partners.  Finally, in
\refse{se:ggff} we provide results for the annihilation of two gluons
into a fermion pair, $\Pg\,\Pg \to f^{\kappa}_{\sigma} \, \bar
f^{\kappa}_{\sigma}$.  All other four-particle processes involving two
gluons and two (anti)fermions, \ie $\Pg\,f^{\kappa}_{\sigma} \to \,
\Pg\,f^{\kappa}_{\sigma}$, $f^{\kappa}_{\sigma} \, \bar
f^{\kappa}_{\sigma}\to \Pg\,\Pg$, etc., are related to $\Pg\,\Pg \to
f^{\kappa}_{\sigma} \, \bar f^{\kappa}_{\sigma}$ by crossing symmetry.

In order to keep all results manifestly invariant with respect to
crossing symmetry, we keep the hard scale $Q^2$, which enters the
logarithms $L=\ln(Q^2/\MW^2)$, as a free parameter.  In practical
applications, $Q^2$ can be identified with the centre-of-mass energy
$s$ or, alternatively, with $|t|$ or $|u|$. This implies an ambiguity
of NNLL order, which corresponds to the intrinsic error of the NLL
approximation.

We present the results in the factorized form 
\refeq{factresult1f}, using the notation
\beqar\label{factresult4f}%\refeq{factresult4f}
\M_X&\NLLA&
\melQ{0,X}{}\, 
f_X^{\sew}\,
f_X^{\PZ}\,
f_X^{\elm},
\eeqar
where $X$ denotes a specific process.  In particular we separate the
finite parts of the corrections, $f_X^{\sew}$ and $f_X^{\PZ}$, from
the subtracted electromagnetic part $f_X^{\elm}$.  The latter contains
all soft/collinear $1/\veps$ poles that must be cancelled against
real photon emission or, in the case of initial-state singularities,
factorized.  In $f_X^{\sew}$ and $f_X^{\PZ}$ we will omit
contributions of $\order(\veps)$ and $\order(\veps^2)$, since such
terms are irrelevant after cancellation of the photonic soft/collinear
singularities.  The electromagnetic contributions $f_X^{\elm}$ have
the general form
\newcommand{\XF}[1]{f_X^{#1}}
\newcommand{\Xfem}{\Delta\Ff{1,X}{\elm}}
\newcommand{\Xgem}{\Delta\Gg{2,X}{\elm}}
\newcommand{\CXemz}{C_{1,X,0}^\elm}
\newcommand{\CXemt}{C_{1,X,\Pt}^\elm}
\newcommand{\CXadem}{C_{1,X}^{\mathrm{ad},\elm}}
\beqar\label{4Xresultem}%\refeq{4Xresultem}
  \XF{\elm} &\NLLA&
  1 + \frac{\alphaeps}{4\pi} \, \Xfem
  + \left(\frac{\alphaeps}{4\pi}\right)^2 \left[
    \frac{1}{2} \left(\Xfem\right)^2 + \Xgem
    \right]
\eeqar
with
\beqar
\label{4Xresultf1em}%\refeq{4Xresultf1em}
  \Xfem &\NLLA&
  - \left(
         2\Epsinv{2}
        -L^2
        -\frac{2}{3}L^3\Eps{}
        -\frac{1}{4}L^4\Eps{2}
        +3\Epsinv{1}
        +3L
        +\frac{3}{2}L^2\Eps{}
        +\frac{1}{2}L^3\Eps{2}
    \right)
    \CXemz
\nl&&{}
  + \left( 
      \Epsinv{1}
      +L + \frac{1}{2}L^2\Eps{} + \frac{1}{6}L^3\Eps{2}
    \right)
    \biggl[
      2 \left( L - 1 - \Lt \right) \CXemt
      - \CXadem
      \biggr]
  + \order(\veps^3)
,
\nl
\label{4Xresultg2em}%\refeq{4Xresultg2em}
  \Xgem &\NLLA&
  \biggl\{
    \LmuR \left[
      -2\Epsinv{2}
      -\left(2L-\LmuR\right)\Epsinv{1}
      +\LmuR L - \frac{1}{3}\LmuR^2
      \right]
      \betacoeff{e}^{(1)}
\nn\\*&&{}
  + \left(
      \frac{3}{2}\Epsinv{3} + 2L\Epsinv{2} + L^2\Epsinv{1}
    \right)
    \betacoeff{\QED}^{(1)}
    \biggr\} \,
  \CXemz
  + \biggl[
    \LmuR \left( 2L\Epsinv{1} + 4L^2 - \LmuR L \right)
      \betacoeff{e}^{(1)}
\nn\\*&&{}
    - \left( L\Epsinv{2} + 2L^2\Epsinv{1} + 2L^3 \right)
      \betacoeff{\QED}^{(1)}
  \biggr] \,
  \CXemt
  + \order(\veps)
,
\eeqar
and the factors $\CXemz, \CXemt, \CXadem$, which depend on the charges
and masses of the external particles, are given in the next sections.
The values for the $\be$-function coefficients $\betacoeff{e}^{(1)}$
and $\betacoeff{\QED}^{(1)}$ can be found in \refeq{betacoeffres} and
\refeq{eq:betacoeffQED}.

\subsection{Neutral-current four-fermion scattering}
\label{se:4fNC}%\refse{se:4fNC}

\newcommand{\NC}{\mathrm{NC}}
\newcommand{\NCMz}{\melQ{0,\NC}{}}
\newcommand{\NCamp}{\mathcal{A}_\NC}
\newcommand{\NCCz}{C_{0,\NC}}
\newcommand{\NCM}{\mel{\NC}{}}
\newcommand{\NCF}[1]{f_\NC^{#1}}
\newcommand{\NCCsew}{C_{1,\NC}^\sew}
\newcommand{\NCCad}{C_{1,\NC}^\mathrm{ad}}
\newcommand{\NCgsew}{g_{2,\NC}^\sew}
\newcommand{\NCfZ}{\Delta\Ff{1,\NC}{\PZ}}
\newcommand{\NCfem}{\Delta\Ff{1,\NC}{\elm}}
\newcommand{\NCgem}{\Delta\Gg{2,\NC}{\elm}}
\newcommand{\NCMfin}{\mel{\NC}{\mathrm{fin}}}
\newcommand{\NCUQED}{U_\NC^\QED}
\newcommand{\NCfQED}{\Ff{1,\NC}{\QED}}
\newcommand{\NCgQED}{\Gg{2,\NC}{\QED}}
\newcommand{\NCnMfin}[1]{\nmel{#1,\NC}{\mathrm{fin}}}
\newcommand{\Ptop}{\mathrm{top}}

\newcommand{\diaggaugeprocess}[5]{
\begin{picture}(150,70)(-75,-35)
\Photon(-30,0)(30,0){2.4}{4}\Text(0,8)[b]{#5}
\ArrowLine(-60,30)(-30,0)\Text(-65,30)[r]{#1}
\ArrowLine(-30,0)(-60,-30)\Text(-65,-30)[r]{#2}
\ArrowLine(30,0)(60,30)\Text(65,30)[l]{#3}
\ArrowLine(60,-30)(30,0)\Text(65,-30)[l]{#4}
\Vertex(-30,0){2}
\Vertex(30,0){2}
\end{picture}
}
\newcommand{\diagYukawaprocess}[5]{
\begin{picture}(150,70)(-75,-35)
\DashLine(-30,0)(30,0){3}\Text(0,5)[b]{#5}
\ArrowLine(-60,30)(-30,0)\Text(-65,30)[r]{#1}
\ArrowLine(-30,0)(-60,-30)\Text(-65,-30)[r]{#2}
\ArrowLine(30,0)(60,30)\Text(65,30)[l]{#3}
\ArrowLine(60,-30)(30,0)\Text(65,-30)[l]{#4}
\Vertex(-30,0){2}
\Vertex(30,0){2}
\end{picture}
}

\newcommand{\qt}{q_\Pt}
\newcommand{\qb}{q_\Pb}
\newcommand{\ytR}{y_{\Pt^\rR}}
\newcommand{\ytL}{y_{\Pt^\rL}}
\newcommand{\ybR}{y_{\Pb^\rR}}
\newcommand{\ybL}{y_{\Pb^\rL}}
\newcommand{\zYuktR}{\zYuk_{\Pt^\rR}}
\newcommand{\zYukbL}{\zYuk_{\Pb^\rL}}

This section deals with $s$-channel neutral-current four-fermion
amplitudes
\beqar\label{4fNCprocess}%\refeq{4fNCprocess}
  f^{\kappa}_{\sigma} \, \bar f^{\kappa}_{\sigma} \to
  f^{\kappa'}_{\sigma'} \, \bar f^{\kappa'}_{\sigma'}
\,,
\eeqar
where a fermion--antifermion pair annihilates and produces another
fermion--antifermion pair.
The Born diagram of such an amplitude is given by
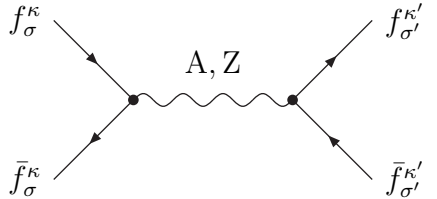
\[
\vcenter{\hbox{
\unitlength 1pt \SetScale{1}
\diaggaugeprocess{$f^{\kappa}_{\sigma}$}{$\bar f^{\kappa}_{\sigma}$}
  {$f^{\kappa'}_{\sigma'}$}{$\bar f^{\kappa'}_{\sigma'}$}
  {$\PA,\PZ$}
}}
\,.
\]
We allow for top quarks only in the final state.  Both particles of
the initial state must share the same chirality~$\kappa$, and both
particles of the final state must have the same chirality~$\kappa'$,
otherwise the amplitude is suppressed in NLL accuracy.  The
electromagnetic charge quantum numbers of the external particles are
given by
$q_f = q_{f^{\kappa}_{\sigma}} = -q_{\bar f^{\kappa}_{\sigma}}$
and $q_{f'} = q_{f^{\kappa'}_{\sigma'}} = -q_{\bar f^{\kappa'}_{\sigma'}}$,
the hypercharges by
$y_f = y_{f^{\kappa}_{\sigma}} = -y_{\bar f^{\kappa}_{\sigma}}$
and $y_{f'} = y_{f^{\kappa'}_{\sigma'}} = -y_{\bar f^{\kappa'}_{\sigma'}}$,
the isospin components by
$t^3_f = t^3_{f^{\kappa}_{\sigma}} = -t^3_{\bar f^{\kappa}_{\sigma}}$
and $t^3_{f'} = t^3_{f^{\kappa'}_{\sigma'}} =
  -t^3_{\bar f^{\kappa'}_{\sigma'}}$,
and the isospin by
$t_f = |t^3_f|$, $t_{f'} = |t^3_{f'}|$.
We use $\zYuk_f$ and $\zYuk_{f'}$ to denote the Yukawa factors
\refeq{Yukawafactors} of the initial- and final-state particles,
respectively.

The amplitude can be written in the factorized form \refeq{factresult1f},
\beqar
  \NCM \NLLA \NCMz\, \NCF{\sew}\, \NCF{\PZ}\, \NCF{\elm}
\,.
\eeqar
The Born amplitude combined with the (non-diagonal)
symmetric-electroweak contribution $\NCF{\sew}$~\refeq{fsew} reads
\beqar\label{4fresultsew}%\refeq{4fresultsew}
\lefteqn{
  \NCMz \, \NCF{\sew} \NLLA \frac{1}{s} \,
    \bar v(p_2,\kappa) \gamma^\mu u(p_1,\kappa) \,
    \bar u(-p_3,\kappa') \gamma_\mu v(-p_4,\kappa')
  \, \Biggl\{
  \NCCz
}
\nl&&{}
  + \frac{\alphaeps}{4\pi} \, \Biggl[
    -\NCCz \, \Biggl(
      \left( L^2 - 3L \right) \NCCsew
      + \LYuk \, \frac{\lambdat^2}{2e^2}
        \left(\zYuk_f + \zYuk_{f'}\right)
      \Biggr)
    + L \, \NCCad
    \Biggr]
\nl&&{}
  + \left(\frac{\alphaeps}{4\pi}\right)^2 \Biggl[
    \NCCz \, \Biggl(
      \left(\frac{1}{2} L^4 - 3L^3\right) \left(\NCCsew\right)^2
      + L^2 \LYuk \, \frac{\lambdat^2}{2e^2}
        \left(\zYuk_f + \zYuk_{f'}\right) \NCCsew
      + \NCgsew
      \Biggr)
\nn\\*&&\qquad
    - L^3 \, \NCCad \, \NCCsew
    \Biggr]
  \Biggr\}
  + \order(\veps)
,
\eeqar
where the Born term
\beqar\label{4fresultC0}%\refeq{4fresultC0}
  \NCCz = g_1^2(Q^2) \, \frac{y_f y_{f'}}{4} + g_2^2(Q^2) \, t^3_f t^3_{f'}
\eeqar
is written in terms of couplings $g_i(Q^2)$ renormalized at the
scale~$Q$, whereas $\alphaeps$ and the other couplings and mixing
angles in the loop corrections are renormalized at the scale~$\muR$.
The remaining terms in \refeq{4fresultsew} read
\beqar
\label{4fresultC1sew}%\refeq{4fresultC1sew}
  \NCCsew &=& \frac{g_1^2}{e^2} \, \frac{y_f^2}{4}
             + \frac{g_2^2}{e^2} \, t_f (t_f + 1)
             + (f \leftrightarrow f')
,
\nl
\label{4fresultC1ad}%\refeq{4fresultC1ad}
  \NCCad &=&
  \NCCz \left[
    4 \ln\left(\frac{u}{t}\right)
      \left( \frac{g_1^2}{e^2} \, \frac{y_f y_{f'}}{4}
        + \frac{g_2^2}{e^2} \, t^3_f t^3_{f'} \right)
    - 2 \ln\left(\frac{-s}{Q^2}\right) \NCCsew
    \right]
\nn\\*&&{}
  + 2 g_2^2(Q^2) \, \frac{g_2^2}{e^2} \left[
      \ln\left(\frac{u}{t}\right) t_f t_{f'}
    - \Biggl( \ln\left(\frac{t}{s}\right) +
        \ln\left(\frac{u}{s}\right) \Biggr) \, t^3_f t^3_{f'}
    \right]
,
\nl
\label{4fresultg2sew}%\refeq{4fresultg2sew}
  \NCgsew &\NLLA& \left(\frac{1}{3} L^3 - \LmuR L^2\right)
  \left[
    \betacoeff{1}^{(1)} \, \frac{g_1^2}{e^2} \, \frac{y_f^2}{4}
    + \betacoeff{2}^{(1)} \, \frac{g_2^2}{e^2} \, t_f (t_f + 1)
    + (f \leftrightarrow f')
  \right]
  + \order(\veps)
.\quad
\eeqar
The values for the $\be$-function coefficients $\betacoeff{1}^{(1)}$,
$\betacoeff{2}^{(1)}$ are given in \refeq{betacoeffres}.
The symmetric-electroweak result \refeq{4fresultsew} is
multiplied with the diagonal factors
\beqar\label{4fresultZ}%\refeq{4fresultZ}
  \NCF{\PZ} \NLLA 1 + \frac{\alphaeps}{4\pi} \,
  2 L \, \LMZW
  \left[
    \left(\frac{g_2}{e} \cw t^3_f - \frac{g_1}{e} \sw \frac{y_f}{2}
      \right)^2
    + (f \leftrightarrow f')
  \right]
  + \order(\veps)
\eeqar
and $\NCF{\elm}$. The latter is obtained from
\refeq{4Xresultem}--\refeq{4Xresultg2em} with
\newcommand{\CNCemz}{C_{1,\NC,0}^\elm}
\newcommand{\CNCemt}{C_{1,\NC,\Pt}^\elm}
\newcommand{\CNCadem}{C_{1,\NC}^{\mathrm{ad},\elm}}
\beqar\label{NCemfactors}%\refeq{NCemfactors}
\CNCemz&=& q_f^2 + \deltaz{f'} \, q_{f'}^2,
\qquad
\CNCemt= \deltat{f'} \, \qt^2,\nl
\CNCadem&=&  4 \ln\left(\frac{u}{t}\right) q_f q_{f'}
      - 2 \ln\left(\frac{-s}{Q^2}\right)
        \left( q_f^2 + q_{f'}^2 \right),
\eeqar
where $\qt=2/3$, and the symbols
\beqar
  \deltat{f'} = \left\{\barr{l}
    1, \; f_{\sigma'} = t\\
    0, \; f_{\sigma'} \ne t
  \earr\right\}
,\qquad
  \deltaz{f'} = 1 - \deltat{f'}
\eeqar
are used to distinguish between massive and massless fermions in the
final state.

Note that only the electromagnetic contributions $\NCF{\elm}$ depend
on the fermion masses.  The remaining contributions are simply
complemented by the Yukawa contributions proportional to $\lambdat^2$
in \refeq{4fresultsew}, otherwise they are equal to our results for
massless fermions presented in Sect.~8.4.1 of \citere{Denner:2006jr}.

\subsection{Charged-current four-fermion scattering}
\label{se:4fCC}%\refse{se:4fCC}

\newcommand{\CC}{\mathrm{CC}}
\newcommand{\CCMz}{\melQ{0,\CC}{}}
\newcommand{\CCamp}{\mathcal{A}_\CC}
\newcommand{\CCM}{\mel{\CC}{}}
\newcommand{\CCF}[1]{f_\CC^{#1}}
\newcommand{\CCCsew}{C_{1,\CC}^\sew}
\newcommand{\CCCad}{C_{1,\CC}^\mathrm{ad}}
\newcommand{\CCgsew}{g_{2,\CC}^\sew}
\newcommand{\CCfZ}{\Delta\Ff{1,\CC}{\PZ}}
\newcommand{\CCfem}{\Delta\Ff{1,\CC}{\elm}}
\newcommand{\CCgem}{\Delta\Gg{2,\CC}{\elm}}
\newcommand{\CCCemz}{C_{1,\CC,0}^\elm}
\newcommand{\CCCemt}{C_{1,\CC,\Pt}^\elm}
\newcommand{\CCCz}{\frac{g_2^2(Q^2)}{2}}
\newcommand{\CCCadem}{C_{1,\CC}^{\mathrm{ad},\elm}}

\newcommand{\CCY}{\mathrm{CCY}}
\newcommand{\CCYMz}{\melQ{0,\CCY}{}}
\newcommand{\CCYamp}{\mathcal{A}_\CCY}
\newcommand{\CCYM}{\mel{\CCY}{}}
\newcommand{\CCYF}[1]{f_\CCY^{#1}}
\newcommand{\CCYCsew}{C_{1,\CCY}^\sew}
\newcommand{\CCYCad}{C_{1,\CCY}^\mathrm{ad}}
\newcommand{\CCYgsew}{g_{2,\CCY}^\sew}
\newcommand{\CCYfZ}{\Delta\Ff{1,\CCY}{\PZ}}
\newcommand{\CCYfem}{\Delta\Ff{1,\CCY}{\elm}}
\newcommand{\CCYgem}{\Delta\Gg{2,\CCY}{\elm}}

In this section we treat \mbox{$s$-channel} charged-current
four-fermion amplitudes
$f^{\kappa}_{\sigma} \, \bar f^{\lambda}_{\rho} \to
f^{\kappa}_{\sigma'} \, \bar f^{\lambda}_{\rho'}
\,,$
where the fermions $f_{\sigma}$ and $f_{\sigma'}$
are the isospin partners of $f_{\rho}$ and $f_{\rho'}$,
respectively.
We allow for a top--antibottom or bottom--antitop pair both in
the initial and final state because the $s$-channel amplitude
$\Pt\,\bar\Pb \to \Pt\,\bar\Pb$ arises via crossing symmetry as a
contribution to the process $\Pb\,\bar\Pb \to \Pt\,\bar\Pt$.

For $s$-channel amplitudes with purely left-handed external fermions,
\beqar\label{4fCCgaugeprocess}%\refeq{4fCCgaugeprocess}
  f^{\rL}_{\sigma} \, \bar f^{\rL}_{\rho} \to
  f^{\rL}_{\sigma'} \, \bar f^{\rL}_{\rho'}
\,,
\eeqar
the Born diagram involves the exchange of a $\PW$ boson:
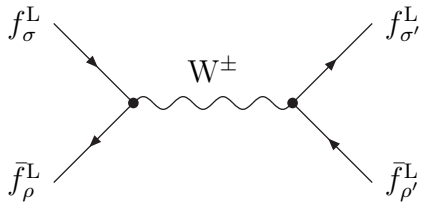
\[
\vcenter{\hbox{
\unitlength 1pt \SetScale{1}
\diaggaugeprocess{$f^{\rL}_{\sigma}$}{$\bar f^{\rL}_{\rho}$}
  {$f^{\rL}_{\sigma'}$}{$\bar f^{\rL}_{\rho'}$}
  {$\PW^\pm$}
}}
\,.
\]
Most of the other combinations of chiralities for the external
fermions yield contributions which are suppressed in NLL accuracy.
The only exception is the $s$-channel amplitude
\beqar\label{4fCCYukprocess}%\refeq{4fCCYukprocess}
  \Pt^{\rR} \, \bar\Pb^{\rL} \to
  \Pt^{\rR} \, \bar\Pb^{\rL}
\eeqar
or, via crossing symmetry,
$\Pb^{\rL} \, \bar\Pt^{\rR} \to \Pb^{\rL} \, \bar\Pt^{\rR}$,
where the exchange of a $\phi^\pm$ scalar boson in the Born diagram
produces a non-suppressed contribution with right-handed top quarks:
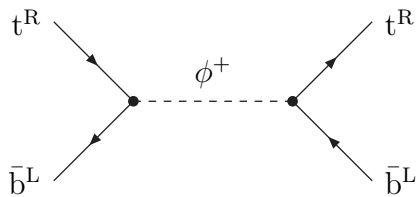
\[
\vcenter{\hbox{
\unitlength 1pt \SetScale{1}
\diagYukawaprocess{$\Pt^{\rR}$}{$\bar\Pb^{\rL}$}
  {$\Pt^{\rR}$}{$\bar\Pb^{\rL}$}
  {$\phi^+$}
}}
\,.
\]

We start with the fully left-handed amplitude \refeq{4fCCgaugeprocess}.
The hypercharge quantum numbers of the external
particles are given by
$y_f = y_{f^{\rL}_{\sigma}} = -y_{\bar f^{\rL}_{\rho}}$
and $y_{f'} = y_{f^{\rL}_{\sigma'}} = -y_{\bar f^{\rL}_{\rho'}}$,
the isospin components by
$t^3 = t^3_{f^{\rL}_{\sigma}} = t^3_{\bar f^{\rL}_{\rho}}
  = \smash{t^3_{f^{\rL}_{\sigma'}} = t^3_{\bar f^{\rL}_{\rho'}}}$,
and for left-handed fermions $|t^3| = {1}/{2}$.
The Yukawa factors \refeq{Yukawafactors} of the initial- and
final-state particles are denoted by 
$\zYuk_f =  \zYuk_{f^\rL_\sigma} =  \zYuk_{\bar f^\rL_\rho}$ and
$\zYuk_{f'} = \zYuk_{f^\rL_{\sigma'}} = \zYuk_{\bar f^\rL_{\rho'}}$,
respectively.

As in the previous section, the amplitude \refeq{4fCCgaugeprocess} is
written in the form
\beqar
  \CCM \NLLA \CCMz\, \CCF{\sew}\, \CCF{\PZ}\, \CCF{\elm}
\,.
\eeqar
Combining the Born amplitude with the (non-diagonal)
symmetric-electroweak contribution $\NCF{\sew}$~\refeq{fsew}, we find
\beqar\label{4fCCresultsew}%\refeq{4fCCresultsew}
\lefteqn{
  \CCMz \, \CCF{\sew} \NLLA \frac{1}{s} \,
    \bar v(p_2,\rL) \gamma^\mu u(p_1,\rL) \,
    \bar u(-p_3,\rL) \gamma_\mu v(-p_4,\rL)
  \, \Biggl\{
  \CCCz
}
\nl&&{}
  + \frac{\alphaeps}{4\pi} \, \Biggl[
    -\CCCz \, \Biggl(
      \left( L^2 - 3L \right) \CCCsew
      + \LYuk \, \frac{\lambdat^2}{2e^2}
        \left(\zYuk_f + \zYuk_{f'}\right)
      \Biggr)
    + L \, \CCCad
    \Biggr]
\nl&&{}
  + \left(\frac{\alphaeps}{4\pi}\right)^2 \Biggl[
    \CCCz \, \Biggl(
      \left(\frac{1}{2} L^4 - 3L^3\right) \left(\CCCsew\right)^2
      + L^2 \LYuk \, \frac{\lambdat^2}{2e^2}
        \left(\zYuk_f + \zYuk_{f'}\right) \CCCsew
      + \CCgsew
      \Biggr)
\nn\\*&&\qquad
    - L^3 \, \CCCad \, \CCCsew
    \Biggr]
  \Biggr\}
  + \order(\veps)
\eeqar
with
\beqar\label{4fCCresultC1sew}%\refeq{4fCCresultC1sew}
  \CCCsew &=& \frac{g_1^2}{e^2} \, \frac{y_f^2+y_{f'}^2}{4}
             + \frac{3}{2} \, \frac{g_2^2}{e^2}
\,,
\nl
\label{4fCCresultC1ad}%\refeq{4fCCresultC1ad}
  \CCCad &=&
  \CCCz \left[
    4 \ln\left(\frac{u}{t}\right)
      \frac{g_1^2}{e^2} \, \frac{y_f y_{f'}}{4}
    - 2 \,
      \Biggl( \ln\left(\frac{t}{s}\right) +
        \ln\left(\frac{u}{s}\right) \Biggr)
      \, \frac{g_2^2}{e^2}
    - 2 \ln\left(\frac{-s}{Q^2}\right) \CCCsew
    \right]
\nl &&{}
  + 2 \ln\left(\frac{u}{t}\right)
    g_1^2(Q^2) \, \frac{y_f y_{f'}}{4} \, \frac{g_2^2}{e^2}
\,,
\nl
\label{4fCCresultg2sew}%\refeq{4fCCresultg2sew}
  \CCgsew &\NLLA& \left(\frac{1}{3} L^3 - \LmuR L^2\right)
  \left(
    \betacoeff{1}^{(1)} \, \frac{g_1^2}{e^2} \, \frac{y_f^2+y_{f'}^2}{4}
    + \frac{3}{2} \, \betacoeff{2}^{(1)} \, \frac{g_2^2}{e^2}
  \right)
  + \order(\veps)
.
\eeqar
The $\MZ$-dependent correction factor reads
\beqar\label{4fCCresultZ}%\refeq{4fCCresultZ}
  \CCF{\PZ} \NLLA 1 + \frac{\alphaeps}{4\pi} \,
  L \, \LMZW
  \left(
    \frac{g_2^2}{e^2} \cw^2
    + 2 \, \frac{g_1^2}{e^2} \sw^2 \, \frac{y_f^2+y_{f'}^2}{4}
  \right)
  + \order(\veps),
\eeqar
and the electromagnetic factor $\CCF{\elm}$
is obtained from \refeq{4Xresultem}--\refeq{4Xresultg2em} with
\beqar
\label{4fCCresultC1em}%\refeq{4fCCresultC1em}
  \CCCemz &=&
    \frac{1}{2} \left[
      \deltaz{f} \left(
        q_{f^\rL_\sigma}^2 + q_{\bar f^\rL_\rho}^2 \right)
      + \deltaz{f'} \left(
        q_{f^\rL_{\sigma'}}^2 + q_{\bar f^\rL_{\rho'}}^2 \right)
      + \left( \deltat{f} + \deltat{f'} \right) \qb^2
      \right]
\nn\\* &=&
  \left[
    \deltaz{f} \left( \frac{g_1^2}{e^2} \cw^2 \, \frac{y_f^2}{4}
    + \frac{1}{4} \, \frac{g_2^2}{e^2} \sw^2 \right)
    + \frac{1}{2} \deltat{f} \, \qb^2
    + (f \leftrightarrow f')
    \right]
,
\nl
  \CCCemt &=&
    \frac{1}{2} \left( \deltat{f} + \deltat{f'} \right) \qt^2
,
\nl
\label{4fCCresultC1adem}%\refeq{4fCCresultC1adem}
  \CCCadem &=& 2 \, \Biggl[
    \ln\left(\frac{-s}{Q^2}\right) \left(
      q_{f^\rL_\sigma} q_{\bar f^\rL_\rho}
      + q_{f^\rL_{\sigma'}} q_{\bar f^\rL_{\rho'}} \right)
    - \ln\left(\frac{-t}{Q^2}\right) \left(
      q_{f^\rL_\sigma} q_{f^\rL_{\sigma'}}
      + q_{\bar f^\rL_\rho} q_{\bar f^\rL_{\rho'}} \right)
\nn\\*&&{}
    - \ln\left(\frac{-u}{Q^2}\right) \left(
      q_{f^\rL_\sigma} q_{\bar f^\rL_{\rho'}}
      + q_{f^\rL_{\sigma'}} q_{\bar f^\rL_\rho} \right)
    \Biggr]
\nl &=&
      4 \ln\left(\frac{u}{t}\right)
        \frac{g_1^2}{e^2} \cw^2 \, \frac{y_f y_{f'}}{4}
      - \Biggl( \ln\left(\frac{t}{s}\right)
          + \ln\left(\frac{u}{s}\right) \Biggr) \,
        \frac{g_2^2}{e^2} \sw^2
\nn\\*&&{}
      - 2 \ln\left(\frac{-s}{Q^2}\right)
        \left( \frac{g_1^2}{e^2} \cw^2 \, \frac{y_f^2+y_{f'}^2}{4}
        + \frac{1}{2} \, \frac{g_2^2}{e^2} \sw^2 \right)
,
\eeqar
where $\qt=2/3$, $\qb=-1/3$, and the symbols
\beqar
  \deltat{f} = \left\{\barr{l}
    1, \; f_{\sigma} = t \mbox{ or } f_{\rho} = t\\
    0, \; \mbox{otherwise}
  \earr\right\}
,\qquad
  \deltaz{f} = 1 - \deltat{f}
\eeqar
and similarly for $f \to f'$, $f_\sigma \to f_{\sigma'}$,
$f_\rho \to f_{\rho'}$
are used to distinguish between massive and massless fermions.

Again only the dependence of the electromagnetic contributions
$\CCF{\elm}$ on the fermion masses and the Yukawa terms in
\refeq{4fCCresultsew} are new compared with our results for massless
fermions in Sect.~8.4.2 of \citere{Denner:2006jr}.

Now we present the results for the $s$-channel amplitude
$\Pt^{\rR} \, \bar\Pb^{\rL} \to \Pt^{\rR} \, \bar\Pb^{\rL}$
\refeq{4fCCYukprocess} which contributes via crossing symmetry to the
process $\Pb^\rL \, \bar\Pb^\rL \to \Pt^\rR \, \bar\Pt^\rR$.
We need the hypercharges $\ytR=4/3$, $\ybL=1/3$
and the Yukawa factors $\zYuktR=2$, $\zYukbL=1$.
The amplitude reads
\beqar
  \CCYM \NLLA \CCYMz\, \CCYF{\sew}\, \CCYF{\PZ}\, \CCYF{\elm}
\eeqar
and is expressed through
\beqar\label{4fCCYresultsew}%\refeq{4fCCYresultsew}
\lefteqn{
  \CCYMz \, \CCYF{\sew} \NLLA
    -\frac{\lambdat^2(Q^2)}{s} \,
    \bar v(p_2,\rL) u(p_1,\rR) \,
    \bar u(-p_3,\rR) v(-p_4,\rL)
  \, \Biggl\{
  1
}
\nl&&{}
  + \frac{\alphaeps}{4\pi} \, \Biggl[
    - \left( L^2 - 3L \right) \CCYCsew
    - \LYuk \, \frac{\lambdat^2}{2e^2}
        \left(\zYuktR + \zYukbL\right)
    + L \, \CCYCad
    \Biggr]
\nl&&{}
  + \left(\frac{\alphaeps}{4\pi}\right)^2 \Biggl[
    \left(\frac{1}{2} L^4 - 3L^3\right) \left(\CCYCsew\right)^2
    + L^2 \LYuk \, \frac{\lambdat^2}{2e^2}
      \left(\zYuktR + \zYukbL\right) \CCYCsew
    + \CCYgsew
\nn\\*&&\qquad
    - L^3 \, \CCYCad \, \CCYCsew
    \Biggr]
  \Biggr\}
  + \order(\veps)
\eeqar
with
\beqar
\label{4fCCYresultC1sew}%\refeq{4fCCYresultC1sew}
  \CCYCsew &=&
    \frac{g_1^2}{e^2} \, \frac{\ytR^2 + \ybL^2}{4}
    + \frac{3}{4} \, \frac{g_2^2}{e^2}
\,,
\nl
\label{4fCCYresultC1ad}%\refeq{4fCCYresultC1ad}
  \CCYCad &=&
    4 \ln\left(\frac{u}{s}\right)
      \frac{g_1^2}{e^2} \, \frac{\ytR \ybL}{4}
    - 2 \ln\left(\frac{-t}{Q^2}\right) \CCYCsew
\,,
\nl
\label{4fCCYresultg2sew}%\refeq{4fCCYresultg2sew}
  \CCYgsew &\NLLA& \left(\frac{1}{3} L^3 - \LmuR L^2\right)
  \left(
    \betacoeff{1}^{(1)} \, \frac{g_1^2}{e^2} \,
      \frac{\ytR^2 + \ybL^2}{4}
    + \frac{3}{4} \, \betacoeff{2}^{(1)} \, \frac{g_2^2}{e^2}
  \right)
  + \order(\veps)
\eeqar
and the $\MZ$-dependent  factor
\beqar
\label{4fCCYresultZ}%\refeq{4fCCYresultZ}
  \CCYF{\PZ} &\NLLA& 1 + \frac{\alphaeps}{4\pi} \,
  2L \, \LMZW
  \left[
    \frac{g_1^2}{e^2} \sw^2 \, \frac{\ytR^2}{4}
    + \left( \frac{1}{2} \frac{g_2}{e} \cw
      + \frac{g_1}{e} \sw \frac{\ybL}{2} \right)^2 \,
  \right]
  + \order(\veps)
.
\eeqar
The electromagnetic correction factor $\CCYF{\elm}$
is obtained from \refeq{4Xresultem}--\refeq{4Xresultg2em} with
\newcommand{\CCYemz}{C_{1,\CCY,0}^\elm}
\newcommand{\CCYemt}{C_{1,\CCY,\Pt}^\elm}
\newcommand{\CCYadem}{C_{1,\CCY}^{\mathrm{ad},\elm}}
\beqar
\CCYemz&=& \qb^2,\qquad
\CCYemt= \qt^2,\nl
\CCYadem&=&  4 \ln\left(\frac{u}{s}\right) \qt\qb
      - 2 \ln\left(\frac{-t}{Q^2}\right)
        \left( \qt^2 +\qb^2 \right)
.
\eeqar

\subsection{Annihilation of two gluons into a fermion pair}
\label{se:ggff}%\refse{se:ggff}

\newcommand{\gf}{{\Pg f}}
\newcommand{\gfMz}{\mel{0,\gf}{}}
\newcommand{\gfCz}{C_{0,\gf}}
\newcommand{\gfM}{\mel{\gf}{}}
\newcommand{\gfF}[1]{f_\gf^{#1}}
\newcommand{\gfCsew}{C_{1,\gf}^\sew}
\newcommand{\gfCad}{C_{1,\gf}^\mathrm{ad}}
\newcommand{\gfgsew}{g_{2,\gf}^\sew}
\newcommand{\gffZ}{\Delta\Ff{1,\gf}{\PZ}}
\newcommand{\gffem}{\Delta\Ff{1,\gf}{\elm}}
\newcommand{\gfgem}{\Delta\Gg{2,\gf}{\elm}}
\newcommand{\gs}{g_{\mathrm{s}}}

\newcommand{\diaggfprocessS}[4]{
\begin{picture}(130,70)(-65,-35)
\Gluon(-20,0)(20,0){3}{4}
\Gluon(-50,30)(-20,0){3}{4}\Text(-55,30)[r]{#1}
\Gluon(-20,0)(-50,-30){3}{4}\Text(-55,-30)[r]{#2}
\ArrowLine(20,0)(50,30)\Text(55,30)[l]{#3}
\ArrowLine(50,-30)(20,0)\Text(55,-30)[l]{#4}
\Vertex(-20,0){2}
\Vertex(20,0){2}
\end{picture}
}
\newcommand{\diaggfprocessT}[4]{
\begin{picture}(110,70)(-55,-35)
\ArrowLine(0,-20)(0,20)
\Gluon(-40,30)(0,20){3}{4}\Text(-45,30)[r]{#1}
\Gluon(0,-20)(-40,-30){3}{4}\Text(-45,-30)[r]{#2}
\ArrowLine(0,20)(40,30)\Text(45,30)[l]{#3}
\ArrowLine(40,-30)(0,-20)\Text(45,-30)[l]{#4}
\Vertex(0,20){2}
\Vertex(0,-20){2}
\end{picture}
}
\newcommand{\diaggfprocessU}[4]{
\begin{picture}(110,70)(-55,-35)
\ArrowLine(0,-20)(0,20)
\Gluon(-40,30)(0,-20){3}{6}\Text(-45,30)[r]{#1}
\Gluon(0,20)(-13.2,3.5){-3}{2}\Gluon(-18,-2.5)(-40,-30){3}{4}
  \Text(-45,-30)[r]{#2}
\ArrowLine(0,20)(40,30)\Text(45,30)[l]{#3}
\ArrowLine(40,-30)(0,-20)\Text(45,-30)[l]{#4}
\Vertex(0,20){2}
\Vertex(0,-20){2}
\end{picture}
}

This section completes the four-particle processes by amplitudes with
two gluons in the initial state.  The process
\beqar\label{gfprocess}%\refeq{gfprocess}
%  \Pg_a \, \Pg_b \to
  \Pg \, \Pg \to
  f^{\kappa'}_{\sigma'} \, \bar f^{\kappa'}_{\sigma'}
\eeqar
involves three diagrams at Born level,
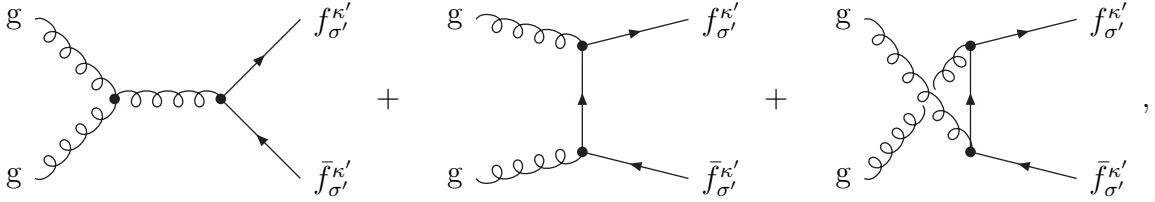
\[
\unitlength 1pt \SetScale{1}
\vcenter{\hbox{
%\diaggfprocessS{$\Pg_a$}{$\Pg_b$}
\diaggfprocessS{$\Pg$}{$\Pg$}
  {$f^{\kappa'}_{\sigma'}$}{$\bar f^{\kappa'}_{\sigma'}$}
}}
\; + \;
\vcenter{\hbox{
%\diaggfprocessT{$\Pg_a$}{$\Pg_b$}
\diaggfprocessT{$\Pg$}{$\Pg$}
  {$f^{\kappa'}_{\sigma'}$}{$\bar f^{\kappa'}_{\sigma'}$}
}}
\; + \;
\vcenter{\hbox{
%\diaggfprocessU{$\Pg_a$}{$\Pg_b$}
\diaggfprocessU{$\Pg$}{$\Pg$}
  {$f^{\kappa'}_{\sigma'}$}{$\bar f^{\kappa'}_{\sigma'}$}
}}
\,,
\]
and the corresponding Born amplitude in the high-energy limit reads
\beqar\label{gfBorn}%\refeq{gfBorn}
  \gfMz &=& - \gs^2 \,
  \varepsilon_\mu(p_1) \varepsilon_\nu(p_2) \,
  \bar u(-p_3,\kappa')
  \times
\nn\\*&&{}
  \, \Biggl\{
    \frac{\ri}{s} \, f^{abc} t^c
      \, \Bigl[ g^{\mu\nu} (p_1-p_2)^\rho + g^{\nu\rho} (p_1+2p_2)^\mu
      - g^{\rho\mu} (2p_1+p_2)^\nu \Bigr] \,
      \gamma_\rho
\nn\\&&{}
    + \frac{1}{t} \, t^a t^b \,
      \gamma^\mu (\ps_2+\ps_4) \gamma^\nu
    + \frac{1}{u} \, t^b t^a \,
      \gamma^\nu (\ps_1+\ps_4) \gamma^\mu
    \Biggr\} \,
    v(-p_4,\kappa')
,
\eeqar
where $\gs$ is the strong coupling, $\varepsilon_\mu(p_{1,2})$ are the
polarization vectors of the initial-state gluons with colour indices
$a$ and $b$, $t^a$ are the generators of the QCD SU(3) gauge group,
and $f^{abc}$ are the corresponding structure constants.

As discussed in \refse{se:disc}, our NLL results for $n$-fermion
processes can be applied also to QCD processes involving fermions and
gluons.  The NLL amplitude for the process \refeq{gfprocess} assumes
the usual factorized form,
\beqar
  \gfM \NLLA \gfMz\, \gfF{\sew}\, \gfF{\PZ}\, \gfF{\elm}
\,,
\eeqar
and the presence of the gluons affects only the factorized Born
amplitude $\gfMz$.  The NLL correction factors $\gfF{\sew}$,
$\gfF{\PZ}$, and $\gfF{\elm}$ are obtained from the general results of
\refse{se:res1loop} and \refse{se:res2loop} by treating the reaction
\refeq{gfprocess} as a $0\to 2$ process, in the sense that the NLL
correction factors receive contributions only from the final-state
fermions.

The symmetric-electroweak operator $\gfF{\sew}$ is diagonal for this
process,
\beqar\label{gfresultsew}%\refeq{gfresultsew}
  \gfF{\sew} &\NLLA& 1
  - \frac{\alphaeps}{4\pi} \, \Biggl\{
    \left[ L^2 - 3L
      + 2L \ln\left(\frac{-s}{Q^2}\right)
      \right] \gfCsew
    + \LYuk \, \frac{\lambdat^2}{2e^2} \, \zYuk_{f'}
    \Biggr\}
\nl&&{}
  + \left(\frac{\alphaeps}{4\pi}\right)^2 \Biggl\{
    \left[ \frac{1}{2} L^4 - 3L^3
      + 2L^3 \ln\left(\frac{-s}{Q^2}\right)
      \right] \left(\gfCsew\right)^2
    + L^2 \LYuk \, \frac{\lambdat^2}{2e^2} \, \zYuk_{f'} \, \gfCsew
    + \gfgsew
    \Biggr\}
\nl&&{}
  + \order(\veps)
,
\eeqar
and simply multiplies the Born amplitude \refeq{gfBorn}.  The terms
$\gfCsew$ and $\gfgsew$ used here as well as the two other correction
factors $\gfF{\PZ}$ and $\gfF{\elm}$ can be obtained from the
corresponding results \refeq{4fresultC1sew}--\refeq{NCemfactors} of
the neutral-current four-fermion amplitude in \refse{se:4fNC} by
setting all electroweak quantum numbers of the initial-state fermions
($y_f$, $t^3_f$, $t_f$, $q_f$) to zero, keeping only the final-state
quantum numbers $y_{f'}$, $t^3_{f'}$, $t_{f'}$, $q_{f'}$.

\subsection{Comparison with effective field-theory results}
\newcommand{\SCETEW}{\ensuremath{\mathrm{SCET}_{\mathrm{EW}}}}
\newcommand{\SCETg}{\ensuremath{\mathrm{SCET}_{\gamma}}}

The four-fermion amplitudes presented in this appendix can be compared
with the results of \citeres{Chiu:2007yn,Chiu:2008vv} based on
soft--collinear effective theory (SCET).  To this end we have to use
the results in Sect.~VII of \citere{Chiu:2008vv}, omitting QCD
contributions.  We have found agreement at NLL accuracy for the
symmetric-electroweak and $\MZ$-dependent parts of our results.

In the SCET framework, the full electroweak theory is matched at the
high scale $\mu=Q$ to an effective theory \SCETEW{} where the degrees
of freedom above the scale~$Q$ are integrated out.  In NLL accuracy,
only the tree-level expressions of the corresponding matching
coefficients are relevant.  These matching coefficients are evolved
from the scale $\mu=Q$ down to the scale $\mu=\MW$ using
anomalous-dimension matrices calculated in \SCETEW{}.  This step
yields LL and NLL contributions which agree with the
symmetric-electroweak parts $\NCF{\sew}$ \refeq{4fresultsew} and
\mla
$\CCF{\sew}$ \refeq{4fCCresultsew} of the neutral- and charged-current
amplitudes presented here.  At the low scale $\mu=\MW$, \SCETEW{} is
matched to another effective theory \SCETg{} where the massive gauge
\mla
bosons are integrated out.  The loop corrections resulting from this
second matching agree with the factors $\NCF{\PZ}$~\refeq{4fresultZ}
and $\CCF{\PZ}$~\refeq{4fCCresultZ} arising in our calculation from
the difference in the $\PW$- and $\PZ$-boson masses.

Finally, the matching coefficients in \SCETg{} are evolved down to
\mda
some finite scale $\mu_0 < \MW$ ($\mu_0 = 30\GeV$ for the numerics in
\citere{Chiu:2008vv}), which acts as a cut-off for the singular
contributions due to soft and collinear photons. Due to the different
regularization employed in our calculation these contributions cannot
directly be compared with our electromagnetic factors $\NCF{\elm}$ and
$\CCF{\elm}$.
\mua

\mda
In addition to the results of \citere{Chiu:2008vv}, our analysis in
\refse{se:4fCC} also includes the charged-current $t$-channel
amplitude for the process $\Pb \, \bar\Pb \to \Pt \, \bar\Pt$ and, in
particular, the tree-level exchange of a scalar boson in the case of
right-handed top quarks. 
\mua

Our results concerning four-particle processes with two gluons and
two quarks in \refse{se:ggff} are in agreement with the comments on these
reactions in \citere{Chiu:2008vv}.

\end{appendix}

\subsection*{Acknowledgements}
This work is supported in part by the European Community's Marie-Curie
Research Training Network under contract MRTN-CT-2006-035505 ``Tools
and Precision Calculations for Physics Discoveries at Colliders''.
We thank the Galileo Galilei Institute for Theoretical
Physics in Florence for the hospitality and the INFN for partial
support during some weeks in 2007.

\end{document}